\documentclass[prd,aps,showpacs,superscriptaddress,nofootinbib, preprint,tightenlines]{revtex4-1}
\usepackage{amsmath}
\usepackage{epsfig}
\usepackage{bm}  
\usepackage{ulem}
\usepackage[sans]{dsfont}  
\usepackage{slashed}

\allowdisplaybreaks

\begin{document}
\title{Mass Dependence of Higgs Production\\
 at Large Transverse Momentum}

\author{Eric Braaten}
\email{braaten@mps.ohio-state.edu}
\affiliation{Department of Physics,
                  The Ohio State University,
                   Columbus, OH 43210, USA}
                                      
\author{Hong Zhang}
\email{zhang.5676@osu.edu}
\affiliation{Department of Physics,
                   The Ohio State University,  
                   Columbus, OH 43210, USA}

\author{Jia-Wei Zhang}
\email{jwzhang@cqust.edu.cn}
\affiliation{Department of Physics,
                   The Ohio State University,  
                   Columbus, OH 43210, USA}
\affiliation{Department of Physics,
                   Chongqing University of Science and Technology,
                   Chongqing 401331, P.R. China}

\date{\today}
                                  
\begin{abstract}
The transverse momentum distribution of the Higgs at large  $P_T$ is complicated by its dependence  on three important energy scales: $P_T$, the top quark mass $m_t$, and the Higgs mass $m_H$. A strategy for simplifying the calculation of the cross section at large $P_T$  is to calculate only the leading terms in its expansion in  $m_t^2/P_T^2$ and/or $m_H^2/P_T^2$. The expansion of the cross section in inverse powers of $P_T$ is complicated by logarithms of $P_T$ and by mass singularities. In this paper, we consider the top-quark loop contribution to the subprocess $q\bar{q}\to H+g$ at leading order  in $\alpha_s$. We show that the leading power of $1/P_T^2$ can be expressed in the form of  a factorization formula that separates the large scale $P_T$ from the scale of the masses. All the  dependence on $m_t$ and  $m_H$ can be factorized into a distribution amplitude for $t \bar t$ in the Higgs,  a distribution amplitude for $t \bar t$ in a real gluon, and an endpoint contribution. The factorization formula can be used to simplify calculations of the $P_T$ distribution at large $P_T$ to next-to-leading order in $\alpha_s$.
 
\end{abstract}

\maketitle

\section{Introduction}
\label{sec:intro}

The discovery of the Higgs boson in the year 2012 completed the Standard Model (SM) of particle physics \cite{Aad:2012tfa, Chatrchyan:2012ufa}. Since then, many of the properties of the Higgs have been measured and compared with the theoretical predictions of the SM. As the experimental precision improves with the collection of more and  more data at the Large Hadron Collider (LHC), it is very important that theoretical uncertainties in the SM predictions are under control. The most straightforward way to reduce the theoretical uncertainties is to carry out calculations to higher orders in perturbation theory, and to resum to all orders those terms (usually logarithms) that spoil the perturbative expansion in certain kinematic regions. Calculations to higher orders are increasingly difficult, but they can sometimes be simplified by separating scales. An important example is the Higgs Effective Field Theory (HEFT), in which the top quark mass $m_t$ is taken to be much larger than all other scales and the top quark is integrated out of the theory. Using HEFT, the  total coss section for Higgs production has been calculated to next-to-leading order (NLO)  in $\alpha_s$ \cite{Dawson:1990zj, Djouadi:1991tka, Spira:1995rr}, to next-to-next-to-leading order (N$^2$LO) \cite{Harlander:2002wh, Anastasiou:2002yz, Ravindran:2003um}, and finally to the impressive precision of N$^3$LO \cite{Anastasiou:2015ema, Anastasiou:2016cez}. The accuracy has been further improved by the resummation of threshold logarithms  \cite{Ahrens:2008nc, Bonvini:2014joa, Li:2014afw, Bonvini:2014tea, Schmidt:2015cea, Bonvini:2016frm}. HEFT has also been used to calculate the cross section for Higgs plus one jet to $\text{N}^\text{2}\text{LO}$  \cite{Boughezal:2013uia, Chen:2014gva, Boughezal:2015dra, Boughezal:2015aha} and the cross section for Higgs plus two or more jets to NLO \cite{Campbell:2006xx, Campbell:2010cz, Cullen:2013saa}.

HEFT breaks down for Higgs produced with large transverse momentum $P_T$ of order  $m_t$,   because the large momentum transfer resolves the top quark loop that is integrated out in HEFT.  The effect of the top quark mass is only at the percent level for the total cross section for Higgs production at the LHC, since the Higgs is produced dominantly with $P_T\ll m_t$ \cite{Harlander:2009mq, Pak:2009dg, Harlander:2009bw, Harlander:2009my, Marzani:2008az, Pak:2009bx}. However, the effect of the top quark mass is much larger for the Higgs $P_T$ distribution, especially at large $P_T$. The Higgs $P_T$ distribution is particularly important for searches for new physics beyond the SM. For example, new physics that modifies the top-quark Yukawa coupling and also introduces new heavy colored particles may mimic the SM in the total cross section for Higgs production, but the deviation from the SM is manifest in the Higgs $P_T$ distribution when $P_T \gtrsim 250$ GeV \cite{Schlaffer:2014osa, Dawson:2015gka}. Higgs production at large $P_T$  has also been applied to the search for new particles in other scenarios beyond the SM \cite{Schlaffer:2014osa, Dawson:2015gka, Bagnaschi:2015qta, Grazzini:2016paz}. With more data being collected in the present and future runs of the LHC, the production of Higgs  at large $P_T$ is a promising channel to search for new physics.

The effect of the top quark mass must be considered in predictions of Higgs production at large $P_T$. Predictions for the production of Higgs at large $P_T$ without final-state top quarks is only available with full $m_t$ dependence at leading order (LO) in $\alpha_s$ \cite{Ellis:1987xu, Baur:1989cm}. At next-to-leading order (NLO), there are real and virtual contributions. The real NLO contribution, which is the same as $H+ 2$ jets at LO, has been calculated with full $m_t$ dependence \cite{Delduca:2001eu, DelDuca:2001fn}. The virtual NLO contribution with full $m_t$ dependence  is still not available. There have been efforts to develop approximations that include some effects of the top quark mass. One approach is to take into account dimension-7 operators in HEFT (for example, Ref.~\cite{Dawson:2014ora}). Another approach is to multiply the LO result with a {\it K-factor}  given by the NLO/LO ratio from HEFT (for example, Ref.~\cite{Neumann:2016dny}). Numerical studies show that these approaches  improve the accuracy at intermediate $P_T$, but the accuracy becomes worse at  large $P_T$. As a result of the unsystematic treatment of the large $P_T$ region, the uncertainties are out of control, making it impossible to estimate the errors introduced. 

A  new approach based on factorization was proposed in Ref.~\cite{Braaten:2015ppa}. At large $P_T$, it is reasonable to expand the cross section in powers of $M^2/Q^2$, where $M$ is a mass scale and $Q$ is a large kinematic scale. The expansion is straightforward for terms that are  analytic in $M^2/Q^2$, but there are also terms that are nonanalytic in $M^2/Q^2$, such as logarithms. Ref.~\cite{Braaten:2015ppa}   showed how factorization theorems could be used to factor the nonanalytic terms into fragmentation functions, allowing the expansion in $M^2/Q^2$. In Ref.~\cite{Braaten:2015ppa}, this  procedure was illustrated with the subprocess $q\bar{q}\to H +t \bar{t}$ at LO. The mass scales are $M \sim m_H,m_t$, where $m_H$ is the Higgs mass, and the kinematic scales are  $Q\sim P_T,\sqrt{\hat s}$, where $\sqrt{\hat s}$ is the center-of-mass energy of the colliding partons. It was shown analytically that the factorization formula reproduces the full LO result up to corrections of order $M^2/Q^2$. Thus the numerical error decreases rapidly as $1/P_T^2$ as $P_T$ increases, indicating  that the errors are under control. 

In addition to better control of the theoretical errors, there are other advantages of the factorization approach. First, the  different energy scales are separated into different pieces in the factorization formula. Consequently,  fewer scales need to be considered in each piece. For example, in the subprocess $q\bar{q}\to H +t \bar{t}$ at LO, the hard-scattering  cross sections are free of the mass scales $m_t$ and $m_H$, and the fragmentation functions are free of the kinematic scales $P_T$ and $\sqrt{\hat s}$ \cite{Braaten:2015ppa}. The calculation of each piece  at higher order would therefore be much simpler. Second, some  pieces in the factorization formula may be directly used in other subprocesses. For example, the fragmentation function for $t^*\to H+t$ is the same for $q\bar{q}\to H +t \bar{t}$ and for $gg\to H+t\bar{t}$. Finally, the factorization formula makes it possible to sum large logarithms to all orders. For example, in the subprocess $q\bar{q}\to H+ b\bar{b}$, logarithms of $m_b^2/p_T^2$ can be summed by solving evolution equations for the fragmentation functions.

In this work, we demonstrate that  the factorization approach can also be used to simplify calculations of virtual NLO contributions to Higgs production at large $P_T$. We consider as a specific example the subprocess $q\bar{q}\to H g$  at LO, which proceeds through a top quark loop. We choose the soft scale to be $M\sim m_H, m_t$ and the hard scale to be $Q\sim P_T, \sqrt{\hat s}$. We express the leading power in the expansion of the amplitude in powers of $M^2/Q^2$ in the form of a factorization formula in which the scales $M$ and $Q$ are separated. The factorization formula involves distribution amplitudes for a $t \bar t$ pair in the Higgs and for  a $t \bar t$ pair in a real gluon. Our factorization formula provides an approximation with errors of order $M^2/Q^2$ that go to zero as the kinematic scale $Q$ increases.

The method we  present in this paper can be used to simplify NLO calculations of the top-quark loop contribution to the Higgs $P_T$ distribution. The method can also be applied to the bottom-quark-loop contribution and to other processes, including the production of $HZ$. Expressing the amplitude in the form of a factorization formula may facilitate the resummation of large logarithms of $P_T^2$. The method can be applied more generally to exclusive processes for the production at large $P_T$ of other elementary particles besides the Higgs boson.

This paper is organized as follows.  In Section~\ref{sec:FormFactor}, we introduce the form factor that determines the cross section for $q\bar{q}\to H g$. We define  the leading-power (LP) form factor to be the leading term in the expansion of the form factor in powers of $M^2/Q^2$. In Section~\ref{sec:AnalReg}, we calculate the LP form factor in the limit $m_H=0$ using analytic regularization to regularize rapidity divergences. In Sections~\ref{sec:FactorHcoll} and \ref{sec:Factorgcoll}, we separate the scales $Q$ and $M$ in the Higgs collinear and gluon collinear contributions to the LP form factor. Each of these contributions is  expressed  as an integral over a relative longitudinal momentum fraction of the product of a   hard  form factor that depends on the scale $Q$ and a distribution amplitude that depends on the scale $M$. In Section~\ref{sec:RapidReg}, we recalculate the LP form factor in the limit $m_H=0$ using rapidity regularization, which makes  rapidity divergences appear as ultraviolet divergences. In Section~\ref{sec:Fragmentation}, we   simplify the  calculations of the Higgs collinear and gluon collinear contributions by calculating the   hard  form factors  and the  distribution amplitudes directly from diagrams. Readers who are not interested in the technical details of factorization can skip Sections~\ref{sec:AnalReg} to \ref{sec:Fragmentation}. In Section~\ref{sec:Factorization}, we renormalize all the ultraviolet divergences to obtain a finite factorization formula for the LP form factor. We present an improvement in the factorization formula that includes all dependence on $m_t$ that is not suppressed by $m_H^2/Q^2$, so that the errors are reduced from order $m_t^2/Q^2$ to order $m_H^2/Q^2$. We show that the  improved factorization formula gives a good approximation to the full form factor whose error decreases to 0 rapidly as $P_T$ increases. We discuss the prospects for extending our approach to NLO in $\alpha_s$ in Section~\ref{sec:Discuss}. In the Appendix, we  calculate a function that  appears in the distribution amplitude for $t \bar t$ in the Higgs using analytic regularization and using rapidity  regularization.

\section{Higgs production by $\bm{q \bar q \to H+g}$}
\label{sec:FormFactor}

In this Section, we define the form factor that determines the cross section for $q \bar q\to H + g$ at leading order in $\alpha_s$. We give the leading power in the expansion of the form factor in powers of $M^2/Q^2$. We also present the schematic form of a factorization formula for the leading-power form factor.

\subsection{Form factor for $\bm{g^* \to H+g}$}

\begin{figure}
\includegraphics[width=10cm]{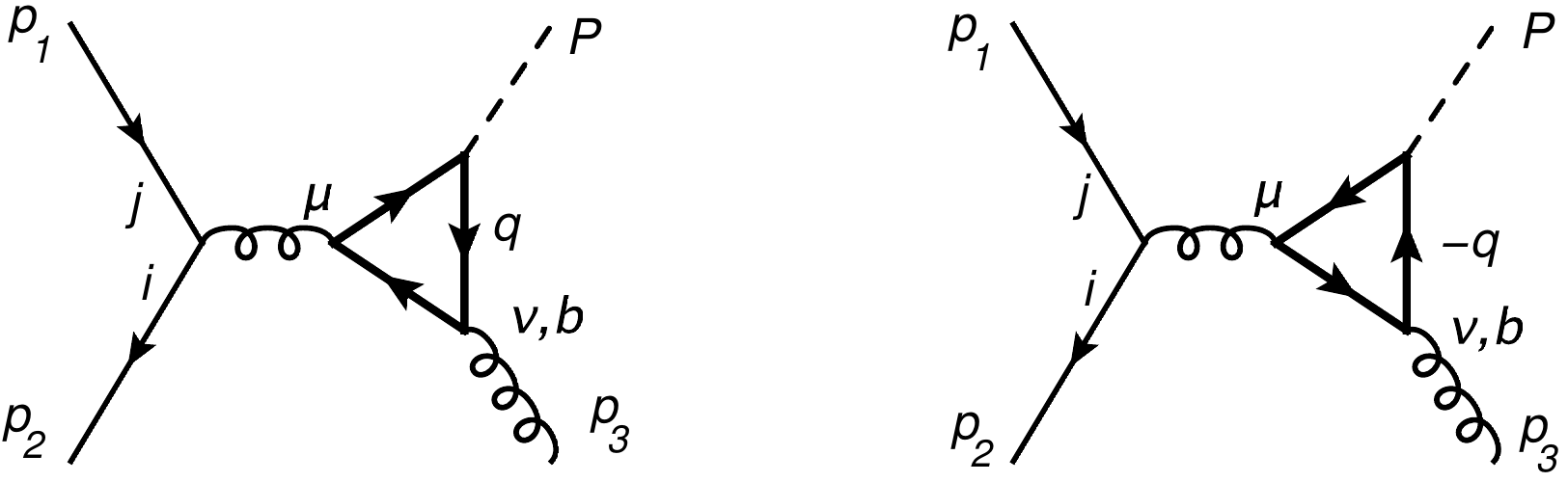}
\caption{Feynman diagrams for $q\bar{q}\to H+g$ at LO.
\label{fig:diagrams}}
\centering
\end{figure}

The reaction $q \bar q\to H + g$ proceeds at leading order (LO) in the QCD coupling constant $g_s$ through the two one-loop Feynman diagrams in Fig.~\ref{fig:diagrams}. The dominant contribution comes from the top-quark loop because of the large Yukawa coupling constant $y_t$. The matrix element  for $q(p_1) \bar q(p_2)\to H(P) + g(p_3)$ at LO has the form
\begin{equation}
\mathcal{M}=  
\frac{g_s}{2\hat s} \, T^b_{ij}\,\bar v_2 \gamma_\mu u_1\, \mathcal{T}^{\mu \nu}(P,p_3)\, \varepsilon^*_{3\nu},
\label{eq:M}
\end{equation}
where $T^b_{ij}$ is the color factor, $\bar v_2$ and  $u_1$ are the Dirac spinors for $\bar q$ and $q$, and $\varepsilon_3$ is the polarization vector for the final-state gluon. The $q\bar{q}$ invariant mass $\hat{s}=(p_1+p_2)^2$ is also the invariant mass of the Higgs and the final-state gluon. The amplitude $\mathcal{T}^{\mu \nu}$ for $g^* \to H + g$ is
\begin{equation}
\mathcal{T}^{\mu\nu} (P,p_3)= i g_s^2 y_t \int_q 
\frac{\text{Tr}\big[ ( \slash \!\!\!q + \slash \!\!\!\!P + m_t) \gamma^\mu 
(\slash \!\!\!q- \slash \!\!\!p_3 + m_t) \gamma^\nu (\slash \!\!\!q +m_t)\big]  - (m_t \to -m_t)}
{ [(q \!+\! P)^2 \!-\! m_t^2  \!+\! i \epsilon]\,  [q^2 \!-\! m_t^2  \!+\! i \epsilon]\,  
[(q \!-\! p_3)^2 \!-\! m_t^2  \!+\! i \epsilon]},
\label{eq:Tdef}
\end{equation}
where the integration measure is $\int_q = \int d^4q/(2\pi)^4$. A color trace tr($T^a T^b$), which is diagonal in the color indices of the virtual gluon and the real gluon, has been absorbed into the prefactor of $ \mathcal{T}^{\mu \nu}$ in Eq.~\eqref{eq:M}. The explicit Dirac trace in Eq.~\eqref{eq:Tdef} comes from the first diagram in Fig.~\ref{fig:diagrams}. Since the only nonzero terms in the trace are proportional to $m_t$ or $m_t^3$, the two diagrams are equal.

The tensor structure of $ \mathcal{T}^{\mu \nu}$ is constrained by the Ward identities $(P+p_3)_\mu \mathcal{T}^{\mu \nu} = 0$ and $p_{3\nu} \mathcal{T}^{\mu \nu} = 0$
to have the form
\begin{eqnarray}
\mathcal{T}^{\mu \nu}(P,p_3) &=& 
4 m_t \, \mathcal{F}(\hat s, m_t^2, m_H^2) 
\left( g^{\mu\nu} - \frac{p_3^\mu (P+p_3)^\nu}{P.p_3} \right)
\nonumber\\
&&+4 m_t\,  \mathcal{G}(\hat s, m_t^2, m_H^2) 
\frac{\big[ P.p_3\, P^\mu - (P.p_3 + m_H^2)p_3^\mu\big]p_3^\nu}{(P.p_3)^2} ,
\label{eq:Ttensor}
\end{eqnarray}
where the form factors $\mathcal{F}$ and $\mathcal{G}$ are dimensionless functions of the $q \bar q$ invariant mass $\hat{s}$  and the masses $m_t$ and $m_H$. The form factor $\mathcal{G}$ does not contribute to the matrix element $\mathcal{M}$ in Eq.~\eqref{eq:M}, because the tensor it multiplies in Eq.~\eqref{eq:Ttensor} is orthogonal to the polarization vector $ \varepsilon^*_{3\nu}$ of the real gluon. The form factor $\mathcal{F}$ can be expressed as
\begin{equation}
\mathcal{F}(\hat s, m_t^2, m_H^2) = 
\frac{1}{(D-2)4 m_t}  \left( g_{\mu\nu} - \frac{p_{3\mu} (P+p_3)_\nu}{P.p_3} \right) \mathcal{T}^{\mu\nu} (P,p_3),
\label{eq:Fdef}
\end{equation}
where $D=4$ is the number of space-time dimensions. The form factor can be expressed as an integral over a loop momentum:
\begin{equation}
\mathcal{F}(\hat s, m_t^2, m_H^2) =  ig_s^2 y_t\int_q 
\frac{q^2 + 2 p_3.q + 2 P.p_3 + 3 m_t^2 - 4(P+p_3).q\, p_3.q/P.p_3}
{ [(q \!+\! P)^2 \!-\! m_t^2  \!+\! i \epsilon]\, [q^2 \!-\! m_t^2 \!+\! i \epsilon]\,
[(q \!-\! p_3)^2-m_t^2  \!+\! i \epsilon]}.
\label{eq:Fint4}
\end{equation}
The square of the matrix element $\mathcal{M}$ for $q \bar q\to H + g$ summed over spins  and colors is proportional to $|\mathcal{F}|^2$: 
\begin{equation}
\frac{1}{4N_c^2} \sum |\mathcal{M}|^2 =
\frac{2(N_c^2-1)g_s^2m_t^2}{N_c^2}\,
\frac{\hat{t}^2+\hat{u}^2}{\hat s(\hat{s}-m_H^2)^2}
\, |\mathcal{F}(\hat s, m_t^2, m_H^2)|^2,
\label{eq:MM-F}
\end{equation}
where $\hat t$ and $\hat u$ are Mandelstam variables that satisfy $\hat s + \hat t +\hat u = m_H^2$. The cross section for $q \bar q\to H + g$ at LO was first calculated  in Refs.~\cite{Ellis:1987xu,Baur:1989cm}. In Ref.~\cite{Keung:2009bs}, $\mathcal{F}$ is expressed compactly in terms of the finite parts of simple scalar one-loop integrals.

The matrix elements for $g\,  q \to H + q$ and  $g \, \bar q \to H + \bar q$ at LO can be expressed in terms of the same function  $\mathcal{F}$  as the form factor for $q \bar q\to H + g$, but with the positive Mandelstam variable $\hat s$ replaced by a negative Mandelstam variable $\hat t$. If the form factor $\mathcal{F}$  for $q \bar q\to H + g$ is expressed in terms of the complex variable $\hat s + i \epsilon$, it can be applied to  $g\,  q \to H + q$ and $g\, \bar q \to H + \bar q$ by analytic continuation.

\subsection{Simple approximations}
\label{sec:LPFF}

The form factor $\mathcal{F}$ is a function of the three energy scales $\hat s^{1/2}$, $m_t$, and $m_H$, which satisfy the inequalities $m_H \le \sqrt{\hat s}$ and $m_H < 2m_t$. Analytic expressions for $\mathcal{F}$ are given in Refs.~\cite{Ellis:1987xu,Baur:1989cm}. There are three limits in which the analytic expression for $\mathcal{F}$ can be simplified. One such limit is $m_H,\hat s^{1/2} \ll m_t$. In this limit,  $\mathcal{F}$ can be expanded in powers of $\hat s/m_t^2$ and $m_H^2/m_t^2$. The leading term in the expansion is
\begin{equation}
\mathcal{F}^\text{HEFT}(\hat s, m_t^2,m_H^2) =
\frac{g_s^2 y_t}{48\pi^2 m_t^2} \left( \hat s - m_H^2 \right).
\label{eq:FHEFT}
\end{equation}

This can be derived more directly using Higgs Effective Field Theory (HEFT). The expression in HEFT for the amplitude $\mathcal{T}^{\mu\nu}$ defined by Eq.~\eqref{eq:M} is
\begin{equation}
\mathcal{T}^{\mu\nu} (P,p_3) = 
\frac{g^2 y_t}{6 \pi^2 m_t}  \big[ P.p_3 \, g^{\mu\nu} - p_3^\mu (P+p_3)^\nu \big].
\label{eq:THEFT}
\end{equation}
The Lorentz contractions in Eq.~\eqref{eq:Fdef} give the form factor in Eq.~\eqref{eq:FHEFT}.

Another limit in which the form factor can be simplified is $m_H \ll\hat s^{1/2} ,m_t$. The leading term in this limit can be obtained by setting $m_H=0$ in the full form factor. The form factor reduces  to
\begin{equation}
\mathcal{F}(\hat s, m_t^2, 0) =
\frac{g_s^2 y_t}{16\pi^2}\left\{
2\, \frac{\hat s + 4 m_t^2}{\hat s} \arcsin^2 z + 4 \frac{\sqrt{1-z^2}}{z}\arcsin z - 6\right\},
\label{eq:FmH=0}
\end{equation}
where $z$ is defined as
\begin{equation}
z=\sqrt{\frac{\hat s+i\epsilon}{4m_t^2}}.
\label{eq:zdef}
\end{equation}

The third limit in which  the form factor can be simplified is $m_H,m_t \ll\hat s^{1/2} $. In this limit,  $\mathcal{F}$ can be expanded in powers of $m_t^2/\hat s$ and $m_H^2/\hat s$. The expansion can be interpreted as an expansion around $\hat s = \infty$ or as an expansion around $m_t=m_H =0$. The expansion in powers of $1/\hat s$ is complicated by terms that are not analytic in $\hat s$, such as $\log(\hat s/m_t^2)$. The expansion in powers of $m_t$ and $m_H$ is complicated by {\it mass singularities}. We define mass singularities to be terms that either diverge in the limits $m_t \to 0$ and $m_H \to 0$ or else depend on the order in which the two limits are taken. The logarithm $\log(\hat s/m_t^2)$ is a mass singularity. Any function of the ratio $m_H/m_t$ that is not suppressed by a factor of $m_t^2/\hat s$ and $m_H^2/\hat s$ is also a mass singularity. We refer to the leading term in the expansion of the form factor in powers of $m_t^2/\hat s$ and $m_H^2/\hat s$ as the {\it leading-power (LP) form factor}. The LP form factor can be derived from the full  form factor in Refs.~\cite{Ellis:1987xu,Baur:1989cm}:
\begin{eqnarray}
\mathcal{F}^\text{LP}(\hat s, m_t^2, m_H^2) &=& 
\frac{g_s^2 y_t}{16\pi^2} \left( - \frac12 \log^2 \frac{-\hat s - i \epsilon}{m_t^2}  + 2\log \frac{-\hat s - i \epsilon}{m_t^2} \right.
\nonumber\\
&& \hspace{2cm}\left.
- 2 \arcsin^2 r - \frac{4\sqrt{1-r^2}}{r}  \arcsin r -  2 \right),
\label{eq:FFLP}
\end{eqnarray}
where $r$ is the mass ratio defined by
\begin{equation}
r \equiv m_H/(2m_t) = 0.36.
\label{eq:r}
\end{equation}
The mass singularities in Eq.~\eqref{eq:FFLP} are the single and double  logarithms of $\hat s/m_t^2$, which diverge as $m_t \to 0$, and the functions of $r$, whose limits as $m_H \to 0$ and $m_t \to 0$ depend on the order of the limits. The only term in Eq.~\eqref{eq:FFLP}  that is not a mass singularity  is the last term $-2$ inside the parentheses.

\begin{figure}
\includegraphics[width=16cm]{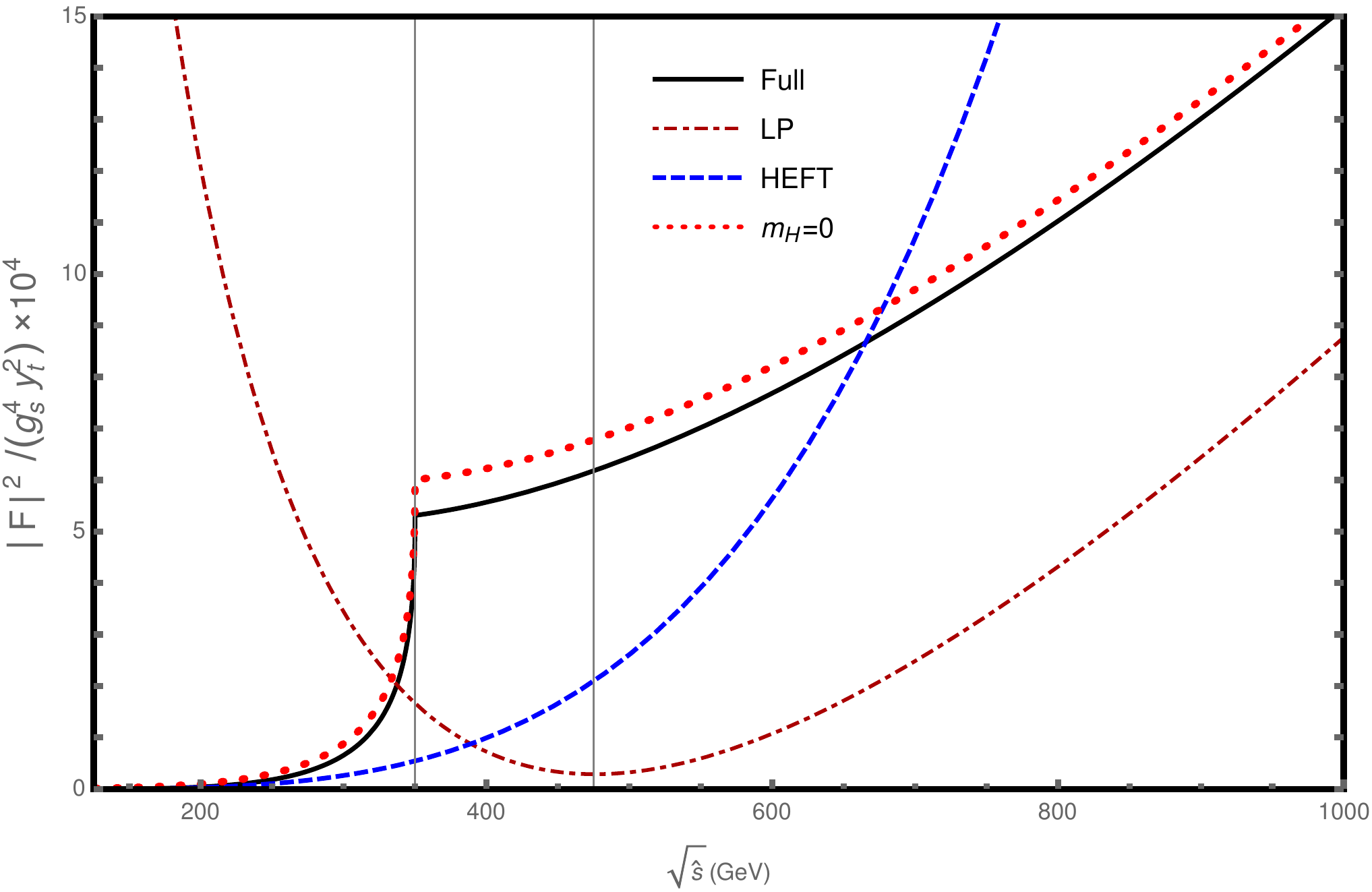}
\caption{Form factors for $q\bar{q}\to H+g$ as functions of the center-of-mass energy $\sqrt{\hat{s}}$: the full form factor $|\mathcal{F}|^2$ (solid curve), the HEFT form factor (dashed curve), the $m_H=0$ form factor (dotted curve), and the LP form factor  (dot-dashed curve). The two vertical lines mark the $t\bar{t}$ threshold $2m_t$ and the $t\bar{t}H$ threshold $2m_t+m_H$.
\label{fig:Fullvsapprox}}
\centering
\end{figure}

In Fig.~\ref{fig:Fullvsapprox}, we compare  the simple approximations described above to the full form factor $\mathcal{F}(\hat s, m_t^2,m_H^2)$ given in Refs.~\cite{Ellis:1987xu,Baur:1989cm}. The three approximations are
\begin{itemize}
\item 
the {\it HEFT form factor} $\mathcal{F}^\text{HEFT}(\hat s, m_t^2,m_H^2)$ in Eq.~\eqref{eq:FHEFT}, which can be obtained by taking $m_t \gg \sqrt{\hat s},m_H$ in the full form factor,
\item 
the {\it $m_H=0$ form factor} $\mathcal{F}(\hat s, m_t^2, 0)$ in Eq.~\eqref{eq:FmH=0}, which is obtained by setting $m_H=0$ in the full form factor,
\item 
the {\it LP form factor} $\mathcal{F}^\text{LP}(\hat s, m_t^2, m_H^2)$ in Eq.~\eqref{eq:FFLP}, which is the leading power in the expansion in $m_t^2/\hat s$ and $m_H^2/\hat s$.
\end{itemize}
We set $m_H=125$~GeV and $m_t=175$~GeV. The squares of the absolute values of the form factors are shown as functions of the center-of-mass energy $\sqrt{\hat{s}}$, which
ranges  from the threshold $m_H$ for producing the Higgs to 1~TeV. The full form factor $|\mathcal{F}|^2$ is zero at the Higgs threshold $m_H$, and it begins increasing quadratically in $\sqrt{\hat{s}}-m_H$. It increases sharply as $\sqrt{\hat{s}}$ approaches the $t \bar t$ threshold, where it has a discontinuity in slope. The discontinuity arises  from the onset of an imaginary part of the form factor from producing $t$ and $\bar t$ on shell. At larger $\sqrt{\hat{s}}$, $|\mathcal{F}|^2$ continues to increase. It increases asymptotically as $\log^4(\hat s/m_t^2)$. The HEFT form factor in Eq.~\eqref{eq:FHEFT} provides a good approximation near the Higgs threshold, but it breaks down before the $t \bar t $ threshold. The $m_H=0$ form factor  has the same qualitative behavior as the full form factor. It  seems to provide a reasonably good  approximation to the full form factor over the range of $\sqrt{\hat{s}}$ shown in Fig.~\ref{fig:Fullvsapprox}. The  absolute error in $|\mathcal{F}(\hat{s}, m_t^2, 0)|^2$ is largest at the $t \bar t $ threshold, where the percentage error is about $14\%$. The LP form factor has a completely different qualitative behavior from the full form factor. It is very large at the Higgs threshold, decreases smoothly  to a minimum near the $t \bar t H$ threshold, and then increases monotonically.  It must be a good approximation to the full form factor at very large $\hat s$, because the error decreases to 0 as $\hat s$ increases. However it provides a very poor approximation in the range of $\sqrt{\hat{s}}$ shown in  Fig.~\ref{fig:Fullvsapprox}. In Section~\ref{sec:massdep}, we will find that a simple modification of the LP form factor provides an approximation that is significantly better than the $m_H=0$ form factor.

\subsection{Leading-power factorization formula}

In order to understand the mass singularities in the leading-power  form factor in Eq.~\eqref{eq:FFLP}, it is necessary to separate the  dependence on $\hat s$ from the dependence on the masses $m_H$ and $m_t$. We refer to the kinematic scale $Q=\hat s^{1/2}$ as the {\it hard scale}. We refer to the scale $M$ provided by the masses $m_H$ and $m_t$ as the {\it soft scale}. We will find that there are four contributions to the LP form factor:
\begin{itemize}
\item
{\it direct production} of $H+g$, in which the Higgs $H$ and the real gluon $g$ are  produced   by the  process $g^* \to H+g$ at the  hard  scale $Q$,
\item
$t \bar t$ {\it fragmentation} into $H$, in which  a nearly collinear $t \bar t$ pair and the real gluon are created by the process  $g^* \to t \bar t+g$ at the  hard  scale $Q$, and the Higgs is produced by the subsequent transition $t \bar t \to H$ at the soft scale $M$,
\item
$t \bar t$ {\it fragmentation} into $g$, in which a nearly collinear $t \bar t$ pair and the Higgs are created by   the  process  $g^* \to H + t \bar t$ at the  hard  scale $Q$, and the real gluon is produced by the subsequent transition $t \bar t \to g$ at the  soft  scale $M$,
\item
{\it endpoint production} of $H+g$, in which  a $t$ and $\bar t$ are created  by the process $g^* \to t + \bar t$  at the  hard  scale $Q$, and the Higgs and the real gluon are produced by the subsequent transition $t \bar t \to H + g$ at the  soft  scale $M$.
\end{itemize}
In the $t \bar t$ fragmentation processes, the collinear $t$ and $ \bar t$  are created with longitudinal momenta that add up to the momentum of the $t \bar t$ pair. We denote the longitudinal momentum fractions of the $t$ and $\bar t$ by $(1+\zeta)/2$ and $(1-\zeta)/2$, respectively. The range of the momentum fraction variable $\zeta$ is  $-1 \le \zeta \le +1$.

We will show that the separation of the hard scale  $Q$   from the soft scale  $M$   in the LP  form factor for $g^* \to H + g$ at LO can be expressed in terms of a factorization formula that has the schematic form
\begin{eqnarray}
\mathcal{F}^\text{LP}[H+g] &=&  \widetilde{\mathcal{F}} [H+g]   
+ \widetilde{\mathcal{F}}[t \bar t_{1V}+g]  \otimes d[t \bar t_{1V} \to H]
\nonumber\\
&& \hspace{2cm}
+ \widetilde{\mathcal{F}} [H + t \bar t_{8T}]  \otimes d[t \bar t_{8T} \to g]
+ \mathcal{F}_\text{endpt}[H+g].
\label{eq:Ffact}
\end{eqnarray}
The  terms on the right side correspond to the  four contributions itemized above. The subscripts on $t \bar t$ indicate the color channel, which can be color-singlet (1) or color-octet (8), and the Lorentz channel, which can be vector ($V$) or tensor ($T$). The $\otimes$ represents an integral over the momentum fraction variable $\zeta$. The factors represented  by $\widetilde{\mathcal{F}}$ are {\it hard form factors} that depend only on the scale $Q$. The factors represented  by $d$ are {\it distribution amplitudes} that depend only on the scale $M$. Regularized expressions for the terms in the factorization formula in Eq.~\eqref{eq:Ffact} will be obtained in Sections~\ref{sec:AnalReg}, \ref{sec:FactorHcoll}, and \ref{sec:Factorgcoll} using analytic regularization and in Sections~\ref{sec:RapidReg}  and \ref{sec:Fragmentation} using rapidity regularization. Renormalized expressions for the terms in the factorization formula will be given in Section~\ref{sec:Factorization}.

\section{LP Form Factor using Analytic Regularization}
\label{sec:AnalReg}

In this Section, we identify the regions of the loop momentum that contribute to the LP form factor for $g^* \to H+g$  at LO. We calculate the LP form factor using analytic regularization in conjunction with dimensional regularization to separate the contributions from the various regions. We set $m_H=0$ in this section to simplify the calculations.

\subsection{Analytic regularization}

The form factor $\mathcal{F}$ in Eq.~\eqref{eq:Fdef} is finite, but we will decompose it into contributions that have ultraviolet divergences and  infrared divergences. The divergences cancel when all the  contributions are added. Some of the divergences can be regularized using {\it dimensional regularization} of the integral in Eq.~\eqref{eq:Tdef} with $D=4-2\epsilon$ space-time dimensions. These divergences appear as poles in $\epsilon$. There are additional infrared divergences called {\it rapidity divergences}  that require some other regularization procedure. They can be regularized using analytic regularization \cite{Beneke:2003pa}, in which the following substitution is applied to appropriate propagators:
\begin{equation}
\frac{1}{p^2 - m^2+i\epsilon}  \longrightarrow 
\frac{(e^{i \pi}\nu^2)^{\delta}}{(p^2 - m^2+i\epsilon)^{1 + \delta}},
\label{eq:analreg}
\end{equation}
where $\delta$ is the analytic regularization parameter and $\nu$ is an arbitrary momentum scale. The phase factor $e^{i \pi \delta}$ is introduced to cancel a phase that arises from the Wick rotation of a loop momentum. The limit $ \delta \to 0$ should be taken before the limit $\epsilon \to 0$. The propagators in Eq.~\eqref{eq:Tdef} that produce rapidity divergences and therefore  require analytic regularization are those with momenta $q+P$ and $q-p_3$. If they are regularized with different parameters $\delta_1$ and $\delta_3$, the  rapidity divergences appear as poles in $\delta_1-\delta_3$. We can regularize the rapidity divergences by applying analytic regularization to either the propagator with momentum $q+P$ or the propagator with momentum $q-p_3$ or both. We choose to apply  analytic regularization to both and to also apply  analytic regularization with parameter $\delta_2$ to the propagator with momentum $q$.

The dimensionally and analytically regularized expression for the amplitude for $g^* \to H + g$ at LO in Eq.~\eqref{eq:Tdef}  is
\begin{equation}
\mathcal{T}^{\mu\nu}(P,p_3) = i g_s^2 y_t \int_q 
\frac{\text{Tr}\big[ ( \slash \!\!\!q + \slash \!\!\!\!P + m_t) \gamma^\mu 
(\slash \!\!\!q- \slash \!\!\!p_3 + m_t) \gamma^\nu (\slash \!\!\!q +m_t)\big]  - (m_t \to -m_t)}
{ [(q \!+\! P)^2 \!-\! m_t^2  \!+\! i \epsilon]^{1 + \delta_1}  [q^2 \!-\! m_t^2  \!+\! i \epsilon]^{1 + \delta_2} 
[(q \!-\! p_3)^2 \!-\! m_t^2  \!+\! i \epsilon]^{1 + \delta_3}}.
\label{eq:Tdefreg}
\end{equation}
The measure of the momentum integral is
\begin{equation}
\int_q \equiv \, (e^{i \pi}\nu^2)^{\delta_1 + \delta_3 + \delta_2} \mu^{2\epsilon} 
\frac{(4 \pi)^{-\epsilon}}{\Gamma(1+\epsilon)} \int \frac{d^{4-2\epsilon}q}{(2 \pi)^{4-2\epsilon}}.
\label{eq:intreg}
\end{equation}
The powers of the analytic regularization scale $\nu$ and the dimensional regularization scale $\mu$ ensure that the dimension of the integral is the same as in 4 dimensions. The factor $(4 \pi)^{- \epsilon}/\Gamma(1+\epsilon)$ has been included in the measure in order to simplify the analytic expressions for loop integrals. With this measure, the minimal subtraction of poles in $\epsilon$ from ultraviolet divergences corresponds to the $\overline{\text{MS}}$ renormalization scheme.

\subsection{Leading-power regions}

The $g^* \to H+g$ form factor at LO with dimensional regularization and analytic regularization is obtained by contracting the tensor $\mathcal{T}^{\mu\nu}$ in Eq.~\eqref{eq:Tdefreg} with the tensor in Eq.~\eqref{eq:Fdef}. After evaluating the Dirac trace in Eq.~\eqref{eq:Tdefreg}, the form factor reduces to
\begin{eqnarray}
\mathcal{F}(\hat s, m_t^2, m_H^2) &=& \frac{2 ig_s^2 y_t}{D-2} \int_q 
\frac{1}
{ [(q \!+\! P)^2 \!-\! m_t^2  \!+\! i \epsilon]^{1 + \delta_1} [q^2 \!-\! m_t^2 \!+\! i \epsilon]^{1 + \delta_2}
[(q \!-\! p_3)^2-m_t^2  \!+\! i \epsilon]^{1 + \delta_3}}.
\nonumber\\
&&\hspace{-2cm}\times \left(
(5-D)q^2 - 4\frac{(P+p_3).q\, p_3.q}{P.p_3}
+  2(D-3)p_3.q + (D-2)P.p_3 + (D-1) m_t^2
\right).
\label{eq:Fint}
\end{eqnarray}

Below we will calculate the LP contribution to the form factor from various regions of the loop integral over the momentum $q$ using the {\it method of regions} \cite{Beneke:1997zp, Smirnov:2002pj}. The regions are 
\begin{itemize}
\item
the {\it hard} region, in which $q^\mu$ is order $Q$, so $q^2$, $P.q$, and  $p_3.q$ are all order $Q^2$,
\item
the {\it Higgs collinear} region, in which $p_3.q$ is order $Q^2$, but $q^2$ and $P.q$ are order $M^2$,
\item
the {\it gluon collinear} region, in which $P.q$ is order $Q^2$, but $q^2$ and $p_3.q$ are order $M^2$,
\item
the {\it soft} region, in which $q^\mu$ is order $M$, so $q^2$ is order $M^2$ but $P.q$  and $p_3.q$ are order $MQ$.
\end{itemize}
The contributions to the LP form factor from each of the regions itemized above can be obtained from the expression in Eq.~\eqref{eq:Fint} by keeping only the leading terms in the numerator and the leading terms in each of the denominators. Analytic regularization is Lorentz invariant. This ensures that the only kinematic variable that the contribution to the LP form factor from each region can depend on is $\hat s$.

\subsection{Hard contribution}
\label{sec:Fhard0}

The contribution to the  LP form factor from the hard region in which $q^\mu$ is order $\hat s^{1/2}$ is 
\begin{eqnarray}
\mathcal{F}^{\text{LP}}_\text{hard}(\hat s) &=& 
 \frac{2 ig_s^2 y_t}{D-2} \int_q \frac{1}{[(q + \tilde P)^2  + i \epsilon]\,  [q^2 + i \epsilon]\, [(q - p_3)^2 + i \epsilon]}
\nonumber\\
&&\hspace{1cm}\times 
\left( (5-D)q^2 - 4\frac{(\tilde P+p_3).q\, p_3.q}{\tilde P.p_3}+  2(D-3)p_3.q + (D-2)\tilde P.p_3 \right).
\label{eq:F0hardint}
\end{eqnarray}
Since there are no divergences as $\delta_1$, $\delta_2$, and $\delta_3$ approach 0 with $\epsilon$ fixed, we have set the three analytic regularization parameters to 0. The 4-momentum $P$ of the Higgs has been replaced by a light-like 4-vector $\tilde P$ whose 3-vector component is collinear to $\bm{P}$ and whose normalization is given by  $2\tilde{P}.p_3=\hat{s}$. The hard contribution does not depend on the masses $m_t$ and $m_H$.

The integral in Eq.~\eqref{eq:F0hardint} can be calculated analytically: 
\begin{equation}
\mathcal{F}^{\text{LP}}_\text{hard}(\hat s) = 
- \frac{g_s^2 y_t}{16 \pi^2} \left[\frac{-\hat s - i \epsilon}{\mu^2}\right]^{-  \epsilon}
\frac{(1)_{-\epsilon}(1)_{-\epsilon}}{(1)_{-2\epsilon}}
\left( \frac{1}{\epsilon^2} + \frac{2}{\epsilon(1-\epsilon)(1-2\epsilon)}    \right).
\label{eq:F0hard}
\end{equation}
We have expressed a factor involving gamma functions in a compact form using the Pochhammer symbol:
\begin{equation}
(n)_z = \frac{\Gamma(n+z)}{\Gamma(n)}.
\label{eq:Pochhammer}
\end{equation}
 The Taylor expansion of  the Pochhammer symbol $(1)_z=\Gamma(1+z)$  can be conveniently expressed in an exponentiated form:
\begin{equation}
(1)_z = \exp\left( - \gamma z + \frac{\pi^2}{12} z^2 + \ldots\right),
\label{eq:(1)_z}
\end{equation}
where $\gamma$ is Euler's constant. This expression makes it easy to expand a combination of Pochhammer symbols like that in Eq.~\eqref{eq:F0hard} in powers of $\epsilon$, especially if the sum of the subscripts in the numerator is equal to  the sum of the subscripts in the denominator. The single pole in $\epsilon$ in Eq.~\eqref{eq:F0hard} is an ultraviolet divergence and the double pole is an infrared divergence. The Laurent expansion in $\epsilon$ of the scale-free factor gives
\begin{equation}
\mathcal{F}^{\text{LP}}_\text{hard}(\hat s) = 
- \frac{g_s^2 y_t}{16 \pi^2} \left[\frac{-\hat s - i \epsilon}{\mu^2}\right]^{-\epsilon}
\left(  \frac{1}{\epsilon^2} +\frac{2}{\epsilon} + 6 - \frac{\pi^2}{6}   \right).
\label{eq:F0hardeps}
\end{equation}

\subsection{Higgs collinear contribution}
\label{sec:FHcoll0}

The contribution to the  LP form factor from the Higgs collinear region in which $p_3.q$ is order $\hat s$ but $q^2$ and $P.q$ are order $m_t^2$ is
\begin{equation}
\mathcal{F}^{\text{LP}}_{H\,\text{coll}}(\hat s, m_t^2,m_H^2) = 
 \frac{2 ig_s^2 y_t}{D-2} \int_q 
\frac{ - 4 (p_3.q)^2/P.p_3 + 2(D-3)p_3.q + (D-2)P.p_3}
{[(q+P)^2 - m_t^2  + i \epsilon]^{1+\delta_1} [q^2 - m_t^2  + i \epsilon]^{1+\delta_2} [-2p_3.q  + i \epsilon]^{1+\delta_3}}.
\label{eq:F0intHcoll}
\end{equation}
If $m_H\neq 0$, the dependence on $m_H$ enters through the  4-momentum $P$ of the Higgs, which satisfies $P^2 = m_H^2$. The Higgs collinear contribution is the only leading-power contribution that depends on $m_H$. In this Section, we simplify it by setting $m_H=0$.

The integral over the loop momentum in Eq.~\eqref{eq:F0intHcoll} can be evaluated analytically:
\begin{eqnarray}
\mathcal{F}^{\text{LP}}_{H\,\text{coll}}(\hat s, m_t^2,0) &=& 
- \frac{g_s^2 y_t}{16 \pi^2} \left[ \frac{\mu^2}{m_t^2} \right]^{\epsilon} 
\left[ \frac{\nu^2}{m_t^2} \right]^{\delta_1+ \delta_2} \left[\frac{-\hat s - i \epsilon}{\nu^2}\right]^{-\delta_3}
\frac{(1)_{\epsilon+\delta_1+\delta_2}(1)_{\delta_1-\delta_3}}{(1)_\epsilon (1)_{\delta_1}(1)_{\delta_1+\delta_2-\delta_3}}\nonumber\\
&&\times  
\frac{1}{(1-\epsilon)(\epsilon+\delta_1+\delta_2)}  
\left( \frac{1-\epsilon}{\delta_1-\delta_3}  - \frac{1-2\epsilon}{1+\delta_1+\delta_2-\delta_3} \right.
\nonumber\\
&&\hspace{4cm}
\left.  
  - \frac{2(1+\delta_1-\delta_3)}{(1+\delta_1+\delta_2-\delta_3)(2+\delta_1+\delta_2-\delta_3)}\right) .
\label{eq:F0Hcoll}
\end{eqnarray}
This contribution has an ultraviolet divergence in the form of a pole in $\epsilon+\delta_1+\delta_2$ and a rapidity divergence in the form of a pole in $\delta_1-\delta_3$. The rapidity divergence comes from the region where $p_3.q \to 0$.

\subsection{Gluon collinear contribution}

The contribution to the  LP form factor from the gluon collinear region in which $P.q$ is order $\hat s$ but $q^2$ and $p_3.q$ are order $m_t^2$ is
\begin{equation}
\mathcal{F}^{\text{LP}}_{g\,\text{coll}}(\hat s, m_t^2) = 
2i g_s^2 y_t \int_q
\frac{\tilde P.p_3}
{ [2\tilde P.q   + i \epsilon]^{1+\delta_1}[q^2 - m_t^2  + i \epsilon]^{1+\delta_2} [(q-p_3)^2 - m_t^2  + i \epsilon]^{1+\delta_3}}.
\label{eq:F0intgcoll}
\end{equation}
The  4-momentum $P$ of the Higgs has been replaced by the light-like 4-vector $\tilde P$. The gluon collinear contribution does not depend on $m_H$.

The integral over the loop momentum in Eq.~\eqref{eq:F0intgcoll} can be evaluated analytically:
\begin{eqnarray}
\mathcal{F}^{\text{LP}}_{g\,\text{coll}}(\hat s, m_t^2) = 
- \frac{g_s^2 y_t}{16 \pi^2} \left[ \frac{\mu^2}{m_t^2} \right]^{\epsilon} 
\left[ \frac{\nu^2}{m_t^2} \right]^{\delta_2+\delta_3}  \left[\frac{-\hat s - i \epsilon}{\nu^2}\right]^{-\delta_1}
\frac{(1)_{\epsilon+\delta_3+\delta_2}(1)_{\delta_3-\delta_1}}{(1)_\epsilon (1)_{\delta_3}(1)_{\delta_3+\delta_2-\delta_1}}  
\nonumber\\
\times 
\frac{1}{(\epsilon+\delta_3+\delta_2)(\delta_3-\delta_1)}  .
\label{eq:F0gcoll}
\end{eqnarray}
This contribution has  an ultraviolet divergence in the form of a pole in $\epsilon+\delta_3+\delta_2$ and a rapidity divergence in the form of a pole in $\delta_1-\delta_3$. The rapidity divergence comes from the region where $\tilde P.q \to 0$.

\subsection{Soft contribution}
\label{sec:Fgcoll0}

The contribution to the  LP form factor from the soft region in which $q^\mu$ is order $m_t$ is
\begin{equation}
\mathcal{F}^{\text{LP}}_\text{soft}(\hat s, m_t^2) = 
2i  g_s^2 y_t  \int_q
\frac{\tilde P.p_3}
{[2 \tilde P.q  + i \epsilon]^{1+\delta_1} [q^2 - m_t^2  + i \epsilon]^{1+\delta_2} [-2p_3.q  + i \epsilon]^{1+\delta_3}}.
\label{eq:F0intsoft}
\end{equation}
The  4-momentum $P$ of the Higgs has been replaced by the light-like 4-vector $\tilde P$. The soft contribution does not depend on $m_H$.

The denominators $2\tilde{P}.q$ and $-2p_3.q$ can be combined using Feynman parameters $x$ and $1-x$. The resulting denominator and the second denominator can be combined using Feynman parameters $y$ and $1-y$. After integrating over $q$ and changing variables to $w = y/(1-y)$, the soft contribution is
\begin{eqnarray}
\mathcal{F}^{\text{LP}}_\text{soft}(\hat s, m_t^2) =
\frac{g_s^2 y_t \hat s}{16 \pi^2} (\nu^2)^{\delta_1 + \delta_2+\delta_3}  (\mu^2)^{\epsilon} 
\frac{  (1)_{\epsilon+\delta_1+\delta_3+\delta_2}  }{  (1)_\epsilon (1)_{\delta_1} (1)_{\delta_3}(1)_{\delta_2}  }
 \int _0^1dx\,  x^{\delta_3} (1-x)^{\delta_1}
\nonumber\\
 \times 
\int_0^\infty \!\!dw\,  w^{1+\delta_1+\delta_3}
\left[ m_t^2 - w^2 x (1-x)  \hat s - i \epsilon \right]^{-1-\epsilon-\delta_1-\delta_3-\delta_2}.
\label{eq:F0intsoft2}
\end{eqnarray}
The integral over  $w$ can be evaluated analytically. The subsequent integral over $x$ is
\begin{eqnarray}
 \int _0^1dx\,  x^{-1+(\delta_3-\delta_1)/2} (1-x)^{-1+(\delta_1-\delta_3)/2} \big[(1-x)+x \big] =
2 \,(1)_{(\delta_3-\delta_1)/2}  (1)_{(\delta_1-\delta_3)/2}
\nonumber\\
\times \left(  \frac{1}{\delta_3-\delta_1} + \frac{1}{\delta_1-\delta_3}\right).
\label{eq:xint0}
\end{eqnarray}
We have separated the integral into two terms by inserting a factor of $(1-x)+x$ into the integrand. The two terms have poles in $\delta_1-\delta_3$ that come from the $x=0$ and $x=1$ endpoints of the integral, respectively. The two terms cancel, so the soft contribution is zero. If the  contributions from the two  terms are made explicit, the soft contribution  can be expressed as
\begin{eqnarray}
\mathcal{F}^{\text{LP}}_\text{soft}(\hat s, m_t^2) &=& 
-\frac{g_s^2 y_t}{16 \pi^2} \left[ \frac{\mu^2}{m_t^2} \right]^{\epsilon} 
 \left[ \frac{\nu^2}{m_t^2} \right]^{(\delta_1+\delta_3)/2+ \delta_2} 
\left[\frac{-\hat s - i \epsilon}{\nu^2}\right]^{-  (\delta_1+\delta_3)/2}  (1)_{(\delta_3-\delta_1)/2} (1)_{(\delta_1-\delta_3)/2} 
\nonumber\\
&&\hspace{0cm}\times 
\frac{  (1)_{(\delta_1+\delta_3)/2}(1)_{\epsilon+(\delta_1+\delta_3)/2+\delta_2}  }
{  (1)_\epsilon (1)_{\delta_1} (1)_{\delta_3} (1)_{\delta_2}  }
\frac{1}{\epsilon+\tfrac12(\delta_1+\delta_3)+\delta_2}
\left( \frac{1}{\delta_3-\delta_1} + \frac{1}{\delta_1-\delta_3}  \right) .
\label{eq:F0soft}
\end{eqnarray}
This contribution has  an ultraviolet divergence in the form of a pole in $\epsilon+(\delta_1+\delta_3)/2+\delta_2$ and rapidity divergences in the form of poles in $\delta_1-\delta_3$. Note that the separation of the soft contribution into two terms with rapidity divergences that cancel is not unique. Another way to separate the soft contribution into two such terms that does not depend on the choice of Feynman parameters is to multiply the integrand in Eq.~\eqref{eq:F0intsoft} by $[\tilde P.q  -p_3.q]/(\tilde P -p_3).q$. The resulting Feynman parameter integrals are more difficult to evaluate.

\subsection{LP form factor for massless Higgs}
\label{sec:LPF0}

The complete LP form factor is obtained by adding the hard contribution in Eq.~\eqref{eq:F0hardeps}, the Higgs collinear contribution in Eq.~\eqref{eq:F0Hcoll}, the gluon collinear contribution in Eq.~\eqref{eq:F0gcoll}, and the soft contribution in Eq.~\eqref{eq:F0soft} (which is equal to 0). The poles in $\delta_1-\delta_3$ cancel between the Higgs collinear contribution and the gluon collinear contribution. In the sum of those contributions, we can take the limit as the analytic regularization parameters approach zero, and then do a Laurent expansion in $\epsilon$:
\begin{eqnarray}
\mathcal{F}^{\text{LP}}_{H\,\text{coll}}(\hat s, m_t^2,0) 
+ \mathcal{F}^{\text{LP}}_{g\,\text{coll}}(\hat s, m_t^2) = 
\frac{g_s^2 y_t}{16 \pi^2} \left[ \frac{\mu^2}{m_t^2} \right]^{\epsilon} 
\left\{  \frac{1}{\epsilon^2} -\frac{1}{\epsilon}   \left( \log\frac{-\hat s - i \epsilon}{m_t^2} -2 \right)
 - \frac{\pi^2}{6}\right\}.~~~~
\label{eq:F0H+gcoll}
\end{eqnarray}
The double and single poles in $\epsilon$  are canceled by the hard contribution in Eq.~\eqref{eq:F0hardeps}. The final result for the LP form factor with $m_H=0$ is
\begin{equation}
\mathcal{F}^{\text{LP}}(\hat s, m_t^2,0) = 
-\frac{g_s^2 y_t}{16 \pi^2}
\left(\frac12 \log^2\frac{-\hat s - i \epsilon}{m_t^2} - 2 \log\frac{-\hat s - i \epsilon}{m_t^2} +6 \right).
\label{eq:F0LP}
\end{equation}
This agrees with the LP form factor in Eq.~\eqref{eq:FFLP} in the limit $m_H \to 0$.

\subsection{Simple choice of analytic regularization parameters}
\label{sec:FsoftmH}

The contributions to the LP form factor for $m_H=0$  from the different regions were calculated  using different analytic regularization parameters $\delta_1$, $\delta_2$, and $\delta_3$ for the three top-quark propagators. The soft contribution is 0 and the poles in $\delta_1-\delta_3$ cancel between the Higgs collinear  and gluon collinear contributions. The cancellation of these rapidity divergences can alternatively be regarded as cancellations between collinear contributions and soft contributions. The pole in $\delta_1-\delta_3$ in the Higgs collinear contribution  in Eq.~\eqref{eq:F0Hcoll} is canceled by the first pole in the soft contribution  in Eq.~\eqref{eq:F0soft}. The pole in $\delta_1-\delta_3$ in the gluon collinear contribution  in Eq.~\eqref{eq:F0gcoll} is canceled by the second pole in the soft contribution. The poles in $\delta_1-\delta_3$ come from endpoints of Feynman parameter integrals in which the coefficient of one of the propagators goes to zero. We can exploit this by using different analytic regularization parameters for the two endpoints.

In each collinear region, it will prove to be convenient to treat the two propagators whose momenta are nearly collinear symmetrically by using the same analytic regularization parameter for both propagators. The simplest possibility is to set the analytic regularization parameter for the third propagator equal to 0. In the Higgs collinear contribution and in the first term of the soft contribution, we choose to set $\delta_2 = \delta_1$ and $\delta_3=0$. In the gluon collinear contribution and in the second term of the soft contribution, we choose to set $\delta_2= \delta_3$ and $\delta_1=0$. Since each of the resulting terms depends on a single analytic regularization parameter, it can be simplified by a Laurent expansion in that parameter followed by a Laurent expansion in $\epsilon$. The Higgs collinear and gluon collinear contributions in Eqs.~\eqref{eq:F0Hcoll} and \eqref{eq:F0gcoll}  reduce to
\begin{subequations}
\begin{eqnarray}
\mathcal{F}^{\text{LP}}_{H\,\text{coll}}(\hat s, m_t^2, 0) &=& 
- \frac{g_s^2 y_t}{16 \pi^2} \left[ \frac{\mu^2}{m_t^2} \right]^{\epsilon} 
\left[ \frac{\nu^2}{m_t^2} \right]^{2\delta_1} 
\left(  \frac{1}{\epsilon \delta_1}  - \frac{2}{\epsilon^2} - \frac{2}{\epsilon}  + \frac{\pi^2}{3} \right) ,
\label{eq:F0Hcoll1}
\\
\mathcal{F}^{\text{LP}}_{g\,\text{coll}}(\hat s, m_t^2) &=& 
- \frac{g_s^2 y_t}{16 \pi^2} \left[ \frac{\mu^2}{m_t^2} \right]^{\epsilon} 
\left[ \frac{\nu^2}{m_t^2} \right]^{2\delta_3} 
\left(  \frac{1}{\epsilon \delta_3}  - \frac{2}{\epsilon^2} + \frac{\pi^2}{3} \right) .
\label{eq:F0gcoll1}
\end{eqnarray}
\label{eq:F0coll1}%
\end{subequations}
The soft contribution in Eq.~\eqref{eq:F0soft} reduces to 
\begin{eqnarray}
\mathcal{F}^{\text{LP}}_{\text{soft}}(\hat s, m_t^2) =
\frac{g_s^2 y_t}{16 \pi^2} \left[ \frac{\mu^2}{m_t^2} \right]^{\epsilon} 
\bigg\{
 \left[ \frac{\nu^2}{m_t^2} \right]^{3\delta_1/2} 
\left[\frac{-\hat s - i \epsilon}{\nu^2}\right]^{-\delta_1/2} 
\left( \frac{1}{\epsilon\delta_1}  - \frac{3}{2 \epsilon^2}+\frac{\pi^2}{4} \right)
\nonumber\\
 + (\delta_1 \to \delta_3)\bigg\}.
\label{eq:F0soft1}
\end{eqnarray}
Note that this contribution is no longer 0. The sum of the two collinear contributions in Eqs.~\eqref{eq:F0coll1} and the soft contribution in Eq.~\eqref{eq:F0soft1} agrees with the sum of the two collinear contributions in Eq.~\eqref{eq:F0H+gcoll}.

\section{Factorization of Higgs Collinear Contribution}
\label{sec:FactorHcoll}

In this section, we  separate the scales $Q$ and $M$ in the  Higgs collinear contribution to the LP form factor using analytic regularization. The resulting expression has the schematic form of the $t \bar t _{1V}$ term in the factorization formula in Eq.~\eqref{eq:Ffact}. We keep $m_H$ nonzero in this section, and use the results to complete the calculation of the LP form factor.

\subsection{Higgs collinear region}
\label{sec:THcoll}

In order to separate the hard scale $Q$ from the soft scale $M$ in the Higgs collinear region, it is convenient to shift the loop momentum $q$ in Eq.~\eqref{eq:Tdefreg} so that the momenta of the collinear $t$ and $\bar t$ that form the Higgs are $\tfrac12 P+q$ and $\tfrac12 P-q$, respectively. For the two diagrams in Fig.~\ref{fig:diagrams}, the appropriate shifts in the loop momentum are $q \to q- \tfrac12 P$ and $q \to -q- \tfrac12 P$, respectively. The resulting expression for the regularized amplitude for $g^* \to H+g$  is
\begin{eqnarray}
\mathcal{T}^{\mu \nu}(P,p_3)&=&i g_s^2 y_t  \int_q
\bigg( \frac{1}{[(p_3 + \tfrac12 P -q)^2-m_t^2  + i \epsilon]^{1 + \delta_3} }
\nonumber\\
&&\hspace{2.5cm} \times
\frac{\text{Tr}\big[ \gamma^\mu ( \slash \!\!\!p_3 + \tfrac12 \slash \!\!\!\!P  - \slash \!\!\!q - m_t) \gamma^\nu 
(\tfrac12 \slash \!\!\!\!P - \slash \!\!\!q - m_t)  (\tfrac12 \slash \!\!\!\!P + \slash \!\!\!q + m_t)\big]}
{[(\tfrac12 P-q)^2 - m_t^2  + i \epsilon]^{1+\delta_1} [(\tfrac12 P+q)^2-m_t^2  + i \epsilon]^{1+\delta_1} }
\nonumber\\
&&\hspace{2cm} 
- (m_t \to -m_t,q \to -q) \bigg).
\label{eq:THcoll}
\end{eqnarray}
We have chosen the  same analytic regularization parameter $\delta_1$ for the propagators with momenta $\tfrac12P \pm q$. The measure for the integral over $q$ is therefore given by Eq.~\eqref{eq:intreg} with $\delta_2=\delta_1$. The shift in $q$ does  not change the power counting in the Higgs collinear region: $p_3.q$ is order $Q^2$, but $q^2$ and $P.q$ are order $M^2$. The hard scale $\hat s$ enters  into the integral in  Eq.~\eqref{eq:THcoll}  only through the denominator that depends on $p_3$ and through the factor in the trace that depends on $p_3$.  

\subsection{Fierz decomposition and tensor decomposition}

A Fierz identity can be used to express the trace in Eq.~\eqref{eq:THcoll} in terms of traces that involve only the collinear momenta $\tfrac12 P\pm q$ and traces that involve $p_3$. A convenient  basis for matrices acting on Dirac spinors in an arbitrarily large  even number $D$ of space time dimensions is the unit matrix $\mathds{1}$, the Dirac matrices $\gamma^\mu$, and the completely antisymmetrized products $\gamma^{[\mu_1} \gamma^{\mu_2} \cdots \gamma^{\mu_n]}$ of $n \ge 2$ Dirac matrices. The Fierz identity for the tensor product of two unit matrices is particularly simple. The coefficients depend on $D$ only through an overall multiplicative factor determined by the trace of the unit matrix.  If we choose Tr$(\mathds{1}) =4$, the Fierz identity for the tensor product of two unit matrices is
\begin{equation}
\mathds{1} _{ij} \mathds{1} _{kl} = \frac14
\Big[\mathds{1} _{il} \mathds{1}_{kj} + (\gamma^\alpha)_{il}  (\gamma_\alpha)_{kj} 
+ \tfrac12 (\sigma^{\alpha\beta})_{il} (\sigma_{\alpha\beta})_{kj}
+ \ldots  \Big].
\label{eq:Fierz}
\end{equation}
where $\sigma^{\alpha\beta} = \tfrac{i}2 (\gamma^\alpha \gamma^\beta - \gamma^\beta \gamma^\alpha)$. We refer to the three terms shown explicitly  as the {\it scalar} ($S$), {\it vector} ($V$), and {\it tensor} ($T$) terms.

The Fierz identity in Eq.~\eqref{eq:Fierz} can be used to separate the collinear factors $\tfrac12 \slash \!\!\!\!P  \pm( \slash \!\!\!q +  m_t)$ in the trace in Eq.~\eqref{eq:THcoll} from the other factors. The only nonzero contributions come from the $\mathds{1} \otimes \mathds{1}$, $\gamma^\alpha \otimes\gamma_\alpha$, and $\sigma^{\alpha\beta} \otimes \gamma_{\alpha\beta}$ terms in the Fierz identity. The trace in Eq.~\eqref{eq:THcoll} is decomposed into the sum of products of a hard trace and a collinear trace. We label the three terms in the sum $1S$, $1V$, and $1T$. The 1 indicates that the collinear $t$ and $\bar t$ that form  the Higgs must be in a color-singlet state. After the contraction of Lorentz indices in Eq.~\eqref{eq:Fdef} that defines the form factor, the only term with a leading-power contribution from the Higgs collinear region is the $1V$ term. We therefore drop the $1S$ and $1T$ terms.

After using the Fierz identity, the scales $Q$ and $M$  are not yet separated, because the hard trace and the collinear trace both depend on the relative momentum $q$ of the virtual $t$ and $\bar t$ that form the Higgs. In the Higgs collinear region, $q$ has a large longitudinal component along the direction of the Higgs momentum $P$. Its large components can be expressed as $q^\lambda \approx \tfrac12\zeta P^\lambda$, where $\zeta \equiv 2q.p_3/P.p_3$.
The separation of the scales $Q$ and $M$ can be facilitated by inserting an integral over $\zeta$ into the integral in Eq.~\eqref{eq:THcoll}:
\begin{equation}
\int_{-1}^{+1}\!\!\! d\zeta \, \delta(\zeta - 2q.p_3/P.p_3) = 1.
\label{eq:intzetaH}
\end{equation}
Since we wish to keep only the LP terms in the amplitude  in  Eq.~\eqref{eq:THcoll}, the 4-momentum $q^\lambda$  can be replaced by $\tfrac12 \zeta P^\lambda$ in the first denominator and in the hard trace. The first denominator and the hard trace can then be pulled outside the integral over $q$. The $1V$ term  from the Fierz transformation  reduces to
\begin{eqnarray}
\mathcal{T}^{\mu \nu}_{1V}(P,p_3)&=& \frac{i g_s^2 y_t }{4} 
\int_{-1}^{+1}\!\!\! d\zeta \, \left( 
\frac{\text{Tr}\big[ \gamma^\mu (\slash \!\!\!p_3 + \tfrac12(1-\zeta) \slash \!\!\!\!P - m_t) \gamma^\nu \gamma_\alpha \big]}
{[  (1-\zeta)P.p_3 + i \epsilon]^{1 + \delta_3}} - (m_t \to -m_t,\zeta \to -\zeta) \right)
\nonumber\\
&&\hspace{2cm} 
\times \int_q
\frac{ \delta(\zeta - 2q.p_3/P.p_3) \, 
\text{Tr}\big[  (\tfrac12 \slash \!\!\!\!P - \slash \!\!\!q - m_t)  (\tfrac12 \slash \!\!\!\!P + \slash \!\!\!q + m_t)\gamma^\alpha \big]}
{[(\tfrac12 P-q)^2 - m_t^2  + i \epsilon]^{1+\delta_1} [(\tfrac12 P+q)^2-m_t^2  + i \epsilon]^{1+\delta_1} }.
\label{eq:THcoll1}
\end{eqnarray}
The integrand of the integral over $q$ is invariant under $m_t \to -m_t$, $q \to -q$, and $\zeta \to -\zeta$. The integral over $q$ in Eq.~\eqref{eq:THcoll1} defines a Lorentz vector function of $P$ and $p_3$ with index $\alpha$.  Since the integrand is homogeneous in $p_3$ with degree 0, the leading power is  in the term proportional to $P^\alpha$. That term can be isolated by replacing $\gamma^\alpha$ in the collinear trace by $P^\alpha\, \slash \!\!\!p_3/P.p_3$. The factor $P^\alpha$ can then be moved into the hard trace.  In the first denominator, in the hard trace, and in the factor $1/P.p_3$, $P$ can be replaced by the light-like 4-vector $\tilde P$ whose 3-vector component is collinear to $\bm{P}$ and whose normalization is given by  $2\tilde{P}.p_3=\hat{s}$. Since the term proportional to $m_t$ in the hard trace is traceless, we can set $m_t=0$ in the hard trace. The tensor in Eq.~\eqref{eq:THcoll1}  therefore reduces to
\begin{eqnarray}
\mathcal{T}^{\mu \nu}_{1V}(P,p_3)&=& \frac{i g_s^2 y_t }{4 \tilde P.p_3} 
\int_{-1}^{+1}\!\!\! d\zeta \, \left( 
\frac{\text{Tr}\big[ \gamma^\mu (\slash \!\!\!p_3 + \tfrac12(1-\zeta) \slash \!\!\!\!\tilde P) \gamma^\nu  \slash \!\!\!\!\tilde P \big]}
{[  (1-\zeta)\tilde P.p_3 + i \epsilon]^{1 + \delta_3}} - (\zeta \to -\zeta) \right)
\nonumber\\
&&\hspace{2cm} 
\times 
\frac{ \delta(\zeta - 2q.p_3/P.p_3) \, 
\text{Tr}\big[  (\tfrac12 \slash \!\!\!\!P - \slash \!\!\!q - m_t)  (\tfrac12 \slash \!\!\!\!P + \slash \!\!\!q + m_t)\slash \!\!\!p_3 \big]}
{[(\tfrac12 P-q)^2 - m_t^2  + i \epsilon]^{1+\delta_1} [(\tfrac12 P+q)^2-m_t^2  + i \epsilon]^{1+\delta_1} }.
\label{eq:THcoll2}
\end{eqnarray}
After evaluating the traces, the $1V$ term in the LP contribution to the tensor reduces to
\begin{eqnarray}
\mathcal{T}_{1V} ^{\mu \nu}(P,p_3) &=& 
-\frac{4g_s^2 y_t m_t}{\tilde P.p_3} \left[\frac{\tilde P.p_3+ i \epsilon}{ e^{i \pi}\nu^2} \right]^{- \delta_3} \int_{-1}^{+1}\!\!\! d\zeta
\bigg( \frac{\tilde P.p_3 g^{\mu\nu} -(\tilde P^\mu p_3^\nu + p_3^\mu \tilde P^\nu) -(1-\zeta)\tilde P^\mu \tilde P^\nu}
{(1-\zeta)^{1 + \delta_3}} 
\nonumber\\
&&\hspace{6.5cm}
- (\zeta \to -\zeta) \bigg)  \zeta\, d(\zeta) ,
\label{eq:THcoll3}
\end{eqnarray}
where the function $d(\zeta)$ is
\begin{equation}
d(\zeta) = -i \int_q
\frac{\delta(\zeta - 2q.p_3/P.p_3)}
{[(\tfrac12 P + q)^2 - m_t^2  + i \epsilon]^{1+\delta_1} [(\tfrac12 P - q)^2-m_t^2  + i \epsilon]^{1+\delta_1} }.
 \label{eq:dzetadef}
\end{equation}
The measure for the  integral over $q$ is given by Eq.~\eqref{eq:intreg} with $\delta_2=\delta_1$ and $\delta_3=0$. The integral in Eq.~\eqref{eq:dzetadef} defines a Lorentz scalar function of $P$ and $p_3$ that is a homogeneous function of $p_3$ with degree 0. Since a homogeneous function of $p_3$ with degree 0 cannot be formed from the Lorentz scalars $P^2=m_H^2$, $p_3^2=0$, and $P.p_3$, the integral must actually be independent of  $p_3$. The dimensionless function $d(\zeta)$ depends only on $\zeta$ and on ratios of the masses $m_t$ and $m_H$ and the regularization scales $\mu$ and $\nu$.

\subsection{Form factor}

A factorized expression for the Higgs collinear contribution to the LP form factor in which the scales $Q$ and $M$ are separated can be obtained by contracting the tensor $\mathcal{T}_{1V}^{\mu \nu}$ in Eq.~\eqref{eq:THcoll3} with  the tensor in Eq.~\eqref{eq:Fdef}:
\begin{eqnarray}
\mathcal{F}_{H\, \text{coll}}^\text{LP}(\hat s, m_t^2, m_H^2) &=& 
- \frac{g_s^2 y_t}{D-2} \left[\frac{ \hat s+ i \epsilon}{2 e^{i \pi} \nu^2} \right]^{-\delta_3}\int_{-1}^{+1}\!\!\! d\zeta
\left( \frac{D-1-\zeta}{(1-\zeta)^{1 + \delta_3} } - \frac{D-1+\zeta}{(1+\zeta)^{1 + \delta_3} } \right)
\zeta\, d(\zeta),~~~
\label{eq:FHcollfact}
\end{eqnarray}
where $d(\zeta)$ is defined by the momentum integral in Eq.~\eqref{eq:dzetadef}. This function is calculated using analytic regularization in Appendix~\ref{app:FragAmp} and is given in Eq.~\eqref{eq:dint3}: 
\begin{equation}
d(\zeta) = \frac{1}{32 \pi^2}
\left[ \frac{\mu^2}{m_t^2} \right]^{\epsilon} \left[ \frac{\nu^2}{m_t^2} \right]^{2\delta_1}
\frac{(1)_{\epsilon+2\delta_1}}{(1)_{\epsilon}(1)_{\delta_1}(1)_{\delta_1}} \frac{1}{\epsilon+2\delta_1}
\left( \frac{1-\zeta^2}{4} \right)^{\delta_1}
\big[ 1 - (1-\zeta^2) r^2 \big]^{-\epsilon-2\delta_1},
 \label{eq:d-zeta}
\end{equation}
where $r = m_H/2m_t$. The subsequent integral over $\zeta$ in Eq.~\eqref{eq:FHcollfact} produces a pole in $\delta_1-\delta_3$.

The scales $Q$ and $M$ are separated in Eq.~\eqref{eq:FHcollfact}. All the dependence on $\hat s$ is in the prefactor   factor  $ \hat s^{-\delta_3}$. All the dependence on $m_t$ and $m_H$ is in the function $d(\zeta)$ in the integrand. The expression in Eq.~\eqref{eq:FHcollfact} has the schematic form $\widetilde{\mathcal{F}}[t \bar t_{1V}+g]  \otimes d[t \bar t_{1V} \to H]$, which corresponds to one of the terms in the factorization formula in Eq.~\eqref{eq:Ffact}. The notation $t \bar t_{1V}$ represents a collinear $t \bar t$ pair in the color-singlet Lorentz-vector channel. The symbol $\otimes$ represents the  integral over $\zeta$ in Eq.~\eqref{eq:FHcollfact}.

The factorized expression for the Higgs collinear contribution to the LP form factor in Eq.~\eqref{eq:FHcollfact} can be simplified by choosing $\delta_3= 0$.  The rapidity divergence is now a pole in $\delta_1$. The contribution to the LP form factor  reduces to
\begin{equation}
\mathcal{F}_{H\, \text{coll}}^\text{LP}(\hat s, m_t^2, m_H^2) = 
-2g_s^2y_t \int_{-1}^{+1}\!\!\! d\zeta \, \zeta^2 \, \frac{d(\zeta)}{1-\zeta^2}.
\label{eq:FfactHcoll}
\end{equation}
One advantage of this choice of  analytic regularization parameters is that the Higgs collinear contribution no longer depends on $\hat s$. Another advantage is that the poles in $\delta_1$ and $\epsilon$ can be extracted before the integration over $\zeta$. The result is derived in the Appendix and given in  Eq.~\eqref{eq:dLaurent}:
\begin{eqnarray}
\frac{d(\zeta)}{1-\zeta^2} =
\frac{1}{32 \pi^2} \left[ \frac{\mu^2}{m_t^2} \right]^{\epsilon} \left[ \frac{\nu^2}{m_t^2} \right]^{2 \delta_1}
\Bigg\{
 \left(\frac{1}{\epsilon \delta_1}  - \frac{2}{\epsilon^2} +  \frac{\pi^2}{3} \right)  \, \delta\big(1 - \zeta^2\big) 
 + \frac{1}{\epsilon} \frac{1}{(1 - \zeta^2)_+}
\nonumber \\
- \frac{\log\big(1- (1 - \zeta^2) r^2 \big)}{1 - \zeta^2} \Bigg\}.
\label{eq:dLaurentr}
\end{eqnarray}
The plus distribution is defined in Eq.~\eqref{eq:intplus}.

The integral over $\zeta$ in the  Higgs collinear contribution in Eq.~\eqref{eq:FfactHcoll} can be evaluated analytically:
\begin{eqnarray}
\mathcal{F}_{H\, \text{coll}}^\text{LP}(\hat s, m_t^2, m_H^2)  &= &
-\frac{g_s^2 y_t}{16 \pi^2} \left[ \frac{\mu^2}{m_t^2} \right]^{\epsilon}  \left[ \frac{\nu^2}{m_t^2} \right]^{2 \delta_1}
\Bigg\{\frac{1}{\epsilon \delta_1}  - \frac{2}{\epsilon^2}   - \frac{2}{\epsilon}+  \frac{\pi^2}{3} 
\nonumber \\
&&\hspace{4cm} 
-  \int_{-1}^{+1}\!\!\! d\zeta \, \zeta^2 \frac{\log\big(1- (1 - \zeta^2) r^2 \big)}{1 - \zeta^2} \Bigg\}.
\label{eq:FHcollint1}
\end{eqnarray}
The remaining integral over $\zeta$ is
\begin{eqnarray}
\int_{-1}^{+1} \!\!\!\!d\zeta\, \frac{\zeta^2}{1 - \zeta^2} \log\big(1 - (1-\zeta^2) r^2\big)  =
-2 \arcsin^2r - \frac{4\sqrt{1-r^2}}{r} \arcsin r+4.
\label{eq:intzeta}
\end{eqnarray}
The only difference between the $m_H$-dependent Higgs collinear contribution to the LP form factor in Eq.~\eqref{eq:FHcollint1} and the contribution with $m_H=0$ in Eq.~\eqref{eq:F0Hcoll1} is the terms from the integral over $\zeta$ in Eq.~\eqref{eq:intzeta}. Adding those terms to the complete LP form factor with $m_H=0$ in Eq.~\eqref{eq:F0LP}, we obtain the complete LP form factor with nonzero $m_H$ in Eq.~\eqref{eq:FFLP}.

\section{Factorization of Gluon Collinear Contribution}
\label{sec:Factorgcoll}

In this section, we separate the scales $Q$ and $M$ in the regularized gluon collinear contribution to the LP form factor. The resulting expression has the schematic form of the $t \bar t _{8T}$ term in the factorization formula in Eq.~\eqref{eq:Ffact}.

\subsection{Gluon collinear region}
\label{sec:Tgcoll}

In order to separate  the hard scale $Q$ from the soft scale $M$ in the gluon collinear region, it is convenient to shift the loop momentum $q$ in Eq.~\eqref{eq:Tdefreg} so that the momenta of the collinear $t$ and $\bar t$ that form the gluon are $\tfrac12 p_3+q$ and $\tfrac12 p_3-q$, respectively. For the two diagrams in Fig.~\ref{fig:diagrams}, the appropriate shifts in the loop momentum are $q \to q+ \tfrac12 p_3$ and $q \to -q+ \tfrac12 p_3$, respectively. The resulting expression for the regularized amplitude for $g^* \to H+g$ is
\begin{eqnarray}
\mathcal{T}^{\mu \nu}(P,p_3) &=& -i g_s^2 y_t  \int_q
\bigg( \frac{1}{[(P + \tfrac12 p_3  + q)^2-m_t^2  + i \epsilon]^{1 + \delta_1} }
\nonumber\\
&&\hspace{2.5cm} \times
\frac{\text{Tr}\big[ (\slash \!\!\!\!P + \tfrac12 \slash \!\!\!p_3 + \slash \!\!\!q + m_t) \gamma^\mu 
(\tfrac12 \slash \!\!\!p_3 - \slash \!\!\!q - m_t)  \gamma^\nu (\tfrac12 \slash \!\!\!p_3 + \slash \!\!\!q + m_t)\big]}
{[(\tfrac12 p_3-q)^2 - m_t^2  + i \epsilon] ^{1+\delta_3} [(\tfrac12 p_3 +q)^2-m_t^2  + i \epsilon]^{1+\delta_3} }
\nonumber\\
&&\hspace{2cm} 
- (m_t \to -m_t, q \to -q) \bigg).
\label{eq:Tgcoll}
\end{eqnarray}
We have chosen the  same analytic regularization parameter $\delta_3$ for the propagators with momenta $\tfrac12p_3 \pm q$. The measure for the integral over $q$ is therefore given by Eq.~\eqref{eq:intreg} with $\delta_2=\delta_3$. The shift of $q$ does  not change the power counting in the gluon collinear region: $P.q$ is order $Q^2$, but $q^2$ and $p_3.q$ are order $M^2$. The hard scale $\hat s$ enters  into the integral in  Eq.~\eqref{eq:Tgcoll} only through the denominator that depends on $P$ and through the factor in the trace that depends on $P$.  

\subsection{Fierz decomposition and tensor decomposition}

The Fierz identity in Eq.~\eqref{eq:Fierz} can be used to separate the collinear factors $\tfrac12 \slash \!\!\!p_3\pm (\slash \!\!\!q + m_t)$ in the trace in Eq.~\eqref{eq:Tgcoll} from the other factors. The only nonzero contributions come from the $\mathds{1} \otimes \mathds{1} $, $\gamma^\alpha \otimes\gamma_\alpha$, and $\sigma^{\alpha\beta} \otimes \gamma_{\alpha\beta}$ terms in the Fierz identity. The trace in Eq.~\eqref{eq:Tgcoll} is decomposed into the sum of products of a hard trace and a collinear trace. We label the three terms in the sum $8S$, $8V$, and $8T$. The 8 indicates that the collinear $t$ and $\bar t$ that form  the real gluon must be in a color-octet state. After the contraction of Lorentz indices in Eq.~\eqref{eq:Fdef} that defines the form factor, the only term with a leading-power contribution from the gluon collinear region is the $8T$ term. We therefore drop the $8S$ and $8V$ terms.

After using the Fierz identity, the scales $Q$ and $M$  are not yet separated, because the hard trace and the collinear trace both depend on the relative momentum $q$ of the virtual $t$ and $\bar t$ that form the gluon. In the gluon collinear region, $q$ has a large longitudinal component along the direction of the gluon momentum $p_3$. Its large components can be expressed as $q^\lambda \approx \tfrac12\zeta p_3^\lambda$, where  $\zeta = 2q.\tilde P/p_3.\tilde P$ and $\tilde P$ is the light-like 4-vector whose 3-vector component is collinear to $\bm{P}$ and whose normalization is given by $2\tilde{P}.p_3=\hat{s}$.
The separation of the scales $Q$ and $M$ can be facilitated by inserting an integral over $\zeta$ into the integral in Eq.~\eqref{eq:Tgcoll}:
\begin{equation}
\int_{-1}^{+1}\!\!\! d\zeta \, \delta(\zeta - 2q.\tilde P/p_3.\tilde P) = 1.
\label{eq:intzetag}
\end{equation}
Since we wish to keep only the LP terms in the amplitude  in  Eq.~\eqref{eq:Tgcoll}, the 4-momenta $q^\lambda$ and $P^\lambda$  can be replaced by $\tfrac12 \zeta p_3^\lambda$ and $\tilde P^\lambda$  in the first denominator and in the hard trace. These factors  can then be pulled outside the integral over $q$. The $8T$ term from the Fierz transformation reduces to 
\begin{eqnarray}
\mathcal{T}_{8T}^{\mu \nu}(P,p_3) &=& 
-\frac{i g_s^2 y_t}{8} \int_{-1}^{+1}\!\!\! d\zeta \, \left( 
\frac{\text{Tr}\big[ (\slash \!\!\!\!\tilde P + \tfrac12(1+\zeta) \slash \!\!\!p_3  + m_t) \gamma^\mu \sigma_{\alpha\beta} \big]}
{[(1+\zeta) \tilde P.p_3 + i \epsilon]^{1 + \delta_1} } + (m_t \to -m_t, \zeta \to -\zeta) \right)
\nonumber\\
&&\hspace{1.5cm} \times  \int_q
\frac{\delta(\zeta - 2q.\tilde P/p_3.\tilde P)\, 
\text{Tr}\big[(\tfrac12 \slash \!\!\!p_3 - \slash \!\!\!q - m_t)  \gamma^\nu 
(\tfrac12 \slash \!\!\!p_3 + \slash \!\!\!q + m_t)  \sigma^{\alpha\beta} \big]}
{[(\tfrac12 p_3-q)^2 - m_t^2  + i \epsilon] ^{1+\delta_3} [(\tfrac12 p_3 +q)^2-m_t^2  + i \epsilon]^{1+\delta_3} }.
\label{eq:Tgcoll1}
\end{eqnarray}
The integrand of the integral over $q$ changes sign under $m_t \to -m_t$, $q \to -q$, and $\zeta \to -\zeta$. The integral over $q$ in Eq.~\eqref{eq:Tgcoll1} defines a Lorentz tensor function of $\tilde P$ and $p_3$ with indices $\nu$, $\alpha$, and $\beta$.  Since the integrand is homogeneous in  $\tilde P$ with degree 0, the leading power has the maximum number of indices carried by the 4-vector $p_3$. In particular, one of the indices $\alpha$ and $\beta$ must be carried by $p_3$. This can be exploited to reduce the number of free indices in the hard trace and in the collinear trace. The matrix $\sigma^{\alpha\beta}$ in the collinear trace can be replaced by $-(p_3^\alpha \sigma^{\beta\gamma} - p_3^\beta \sigma^{\alpha\gamma})\tilde P_\gamma/\tilde P.p_3$. The 4-momentum $p_3$ can be moved to the hard trace. Since the term proportional to $m_t$ in the hard trace is traceless, we can set $m_t=0$ in the hard trace. The tensor in Eq.~\eqref{eq:Tgcoll1} can therefore be expressed as
\begin{eqnarray}
\mathcal{T}_{8T}^{\mu \nu}(P,p_3) &=& 
\frac{i g_s^2 y_t}{16\tilde P.p_3}\! \int_{-1}^{+1}\!\!\!\!\! d\zeta \, \bigg( 
\frac{\text{Tr}\big[ (\slash \!\!\!\!\tilde P + \tfrac12(1+\zeta) \slash \!\!\!p_3 ) \gamma^\mu [\slash \!\!\!p_3, \gamma_\alpha] \big]}
{[(1+\zeta) \tilde P.p_3 + i \epsilon]^{1 + \delta_1} } + (\zeta \to -\zeta) \bigg)
\nonumber\\
&&\hspace{0.0cm} \times   \int_q
\frac{\delta(\zeta - 2q.\tilde P/p_3.\tilde P)\, 
\text{Tr}\big[(\tfrac12 \slash \!\!\!p_3 - \slash \!\!\!q - m_t)  \gamma^\nu 
(\tfrac12 \slash \!\!\!p_3 + \slash \!\!\!q + m_t) [\slash \!\!\!\! \tilde P,\gamma^\alpha] \big]}
{[(\tfrac12 p_3-q)^2 - m_t^2  + i \epsilon] ^{1+\delta_3} [(\tfrac12 p_3 +q)^2-m_t^2  + i \epsilon]^{1+\delta_3} }.
\label{eq:Tgcoll2}
\end{eqnarray}
After evaluating the traces, the $8T$ term in the LP contribution to the tensor reduces to
\begin{eqnarray}
\mathcal{T}_{8T} ^{\mu \nu}(P,p_3) &=& 
-\frac{2g_s^2 y_t m_t}{\tilde P.p_3}\left[ \frac{\tilde P.p_3 + i \epsilon}{e^{i \pi} \nu^2} \right]^{-\delta_1} \int_{-1}^{+1}\!\!\! d\zeta
\bigg( \frac{2 \tilde P.p_3  g^{\mu\nu}  -  2 p_3^\mu  \tilde P^\nu - (1+\zeta) p_3^\mu p_3^\nu}
{(1+\zeta)^{1 + \delta_1} }
\nonumber\\
&& \hspace{6.5cm} 
+ (\zeta \to -\zeta) \bigg) \,  d_0(\zeta),
\label{eq:Tgcoll3}
\end{eqnarray}
where the function $d_0(\zeta)$ is
\begin{equation}
d_0(\zeta) = -i \int_q
\frac{\delta(\zeta - 2q.\tilde P/p_3.\tilde P)}
{[(\tfrac12 p_3+q)^2 - m_t^2  + i \epsilon]^{1+\delta_3} [(\tfrac12 p_3-q)^2-m_t^2  + i \epsilon]^{1+\delta_3} }.
 \label{eq:d0zetadef}
\end{equation}
The measure for the  integral over $q$ in Eq.~\eqref{eq:d0zetadef} is given by Eq.~\eqref{eq:intreg} with $\delta_2=\delta_3$ and $\delta_1=0$. The integral defines a Lorentz scalar function of $p_3$ and $\tilde P$ that is a homogeneous function of  $\tilde P$ with degree 0. Since such a function cannot be formed from the Lorentz scalars $p_3^2=0$, $\tilde P^2=0$, and $p_3.\tilde P$, the integral must actually be independent of $\tilde P$. The dimensionless function $d_0(\zeta)$ defined by Eq.~\eqref{eq:d0zetadef} depends only on $\zeta$ and on ratios of the mass $m_t$ and the regularization scales $\mu$ and $\nu$.

\subsection{Form factor}

A factorized expression for the gluon collinear contribution to the LP form factor in which the scales $Q$ and $M$ are separated can be obtained by contracting the tensor $\mathcal{T}_{g\, \text{coll}}^{\mu \nu}$ in Eq.~\eqref{eq:Tgcoll3}  with the tensor in Eq.~\eqref{eq:Fdef}:
\begin{eqnarray}
\mathcal{F}_{g\, \text{coll}}^\text{LP}(\hat s, m_t^2) = 
- g_s^2 y_t  \left[\frac{ \hat s+ i \epsilon}{2 e^{i \pi} \nu^2} \right]^{-\delta_1}\int_{-1}^{+1}\!\!\! d\zeta
\left( \frac{1}{(1+\zeta)^{1 + \delta_1} } + \frac{1}{(1-\zeta)^{1 + \delta_1} } \right)
d_0(\zeta),
\label{eq:Fgcollfact}
\end{eqnarray}
where $d_0(\zeta)$ is defined by the momentum integral in Eq.~\eqref{eq:d0zetadef}. This function can be obtained from the expression for $d(\zeta)$ in Eq.~\eqref{eq:d-zeta} by setting $r=0$ and replacing $\delta_1$ by $\delta_3$:
\begin{equation}
d_0(\zeta) = \frac{1}{32 \pi^2}
\left[ \frac{\mu^2}{m_t^2} \right]^{\epsilon} \left[ \frac{\nu^2}{m_t^2} \right]^{2\delta_3}
\frac{(1)_{\epsilon+2\delta_3}}{(1)_{\epsilon}(1)_{\delta_3}(1)_{\delta_3}} \frac{1}{\epsilon+2\delta_3}
\left( \frac{1-\zeta^2}{4} \right)^{\delta_3}.
 \label{eq:d0-zeta}
\end{equation}
The integral over $\zeta$ in Eq.~\eqref{eq:Fgcollfact} can be calculated analytically. The subsequent integral over $\zeta$ in Eq.~\eqref{eq:Fgcollfact} produces a pole in $\delta_1-\delta_3$. The result agrees with the expression for the gluon collinear contribution to the LP form factor in Eq.~\eqref{eq:F0gcoll} with  $\delta_2 = \delta_3$.

The scales $Q$ and $M$ are separated in Eq.~\eqref{eq:Fgcollfact}. All the dependence on $\hat s$ is in the prefactor $\hat s^{-\delta_1}$. All the dependence on $m_t$ is in the function $d_0(\zeta)$ in the integrand. The  expression in Eq.~\eqref{eq:Fgcollfact} has the schematic form $\widetilde{\mathcal{F}} [H + t \bar t_{8T}]  \otimes d[t \bar t_{8T} \to g]$, which corresponds to one of the terms in the factorization formula in Eq.~\eqref{eq:Ffact}. The notation $t \bar t_{8T}$ represents a collinear $t \bar t$ pair in the color-octet Lorentz-tensor channel. The symbol $\otimes$ represents the integral over $\zeta$ in Eq.~\eqref{eq:Tgcoll1}.

The factorized expression for the gluon collinear contribution to the LP form factor in Eq.~\eqref{eq:Fgcollfact} can be simplified by choosing $\delta_1= 0$. 
The rapidity divergence is now  a pole in $\delta_3$. The gluon collinear contribution to the LP form factor reduces to
\begin{eqnarray}
\mathcal{F}_{g\, \text{coll}}^\text{LP}(\hat s, m_t^2) &=& 
-2g_s^2 y_t \int_{-1}^{+1}\!\!\! d\zeta \,  \frac{d_0(\zeta)}{1-\zeta^2}.
\label{eq:FfactgcollH}
\end{eqnarray}
One advantage of this choice of  analytic regularization parameters is that the gluon collinear contribution no longer depends on $\hat s$. Another advantage is that the poles in $\delta_3$ and $\epsilon$ can be extracted before the integration over $\zeta$. The Laurent expansion in $\delta_3$ and $\epsilon$ can be obtained from that of $d(\zeta)/(1-\zeta^2)$ in Eq.~\eqref{eq:dLaurentr} by setting $r=0$ and replacing $\delta_1$ by $\delta_3$:
\begin{eqnarray}
\frac{d_0(\zeta)}{1-\zeta^2} =
\frac{1}{32 \pi^2} \left[ \frac{\mu^2}{m_t^2} \right]^{ \epsilon} \left[ \frac{\nu^2}{m_t^2} \right]^{2 \delta_3}
\Bigg\{
 \left( \frac{1}{\epsilon \delta_3} - \frac{2}{\epsilon^2} +  \frac{\pi^2}{3} \right)  \, \delta\big(1 - \zeta^2\big) 
+  \frac{1}{\epsilon} \, \frac{1}{(1 - \zeta^2)_+} \Bigg\}.
\label{eq:d0Laurent}
\end{eqnarray}
The plus distribution is defined in Eq.~\eqref{eq:intplus}. The integral over $\zeta$ in Eq.~\eqref{eq:FfactgcollH} can  be evaluated easily. The result agrees with the gluon collinear contribution in Eq.~\eqref{eq:F0gcoll1}.

\section{LP Form Factor using Rapidity Regularization}
\label{sec:RapidReg}

In this Section, we calculate the LP form factor using rapidity regularization in conjunction with dimensional regularization to separate the contributions from the various regions. We set $m_H=0$ in this section to simplify the calculations.

\subsection{Rapidity regularization and zero-bin subtraction}

In Sections~\ref{sec:AnalReg}, \ref{sec:FactorHcoll}, and \ref{sec:Factorgcoll}, we  used analytic regularization to separate the contributions to the LP form factor from the various regions. The factorized expressions for the Higgs collinear and gluon collinear contributions derived in Sections~\ref{sec:FactorHcoll} and \ref{sec:Factorgcoll} involve an integral over the relative longitudinal momentum fraction $\zeta$. The rapidity divergences were made explicit in the integrand by using different analytic regularization parameters in the Higgs collinear and gluon collinear contributions. A rather arbitrary prescription was used to separate the soft contribution  into two contributions with different regularization parameters in order to cancel the rapidity divergences in the collinear contributions. It could be very difficult to extend this prescription to higher orders of perturbation theory.

Analytic regularization has other drawbacks.  It violates gauge invariance, which is a severe complication in proofs of factorization to all orders in perturbation theory  \cite{Becher:2011dz}. This problem is especially serious in QCD, because soft contributions can be nonperturbative. While the process we consider here  is completely perturbative, the violation of gauge invarince could complicate the extension of our calculation to NLO. Another disadvantage of analytic regularization is that rapidity divergences appear naturally as infrared divergences. This makes it difficult to interpret the cancellation of rapidity divergences as a renormalization procedure.

In this Section, we separate the contributions to the LP form factor from the various regions using an alternative regularization method for rapidity divergences called {\it rapidity regularization}. Rapidity regularization in conjunction with zero-bin subtraction  was introduced as a method for regularizing rapidity divergences by Manohar and Stewart  \cite{Manohar:2006nz}. Rapidity regularization separates the contributions from collinear and soft regions by explicitly breaking the boost invariance. Zero-bin subtractions of collinear contributions are required to avoid double counting of soft  contributions. With rapidity regularization, the rapidity divergence from each region is an ultraviolet divergence. This allows the cancellation of rapidity divergences to be implemented as a renormalization procedure.

In order to specify the rapidity regularization factors, it is convenient to introduce light-like vectors $n$ and $\bar n$ such that the only components of $P^\mu$ and $p_3^\mu$ that are of order $Q$ are $P.n$ and $p_3.\bar n$. We choose the normalizations of $n$ and $\bar n$  so that $n . \bar n=2$, which implies $P.n\, p_3.\bar n = \hat s$. Dimensional regularization is used to separate the hard contribution from the sum of the remaining contributions. The integration measure of the loop momentum can be expressed as
\begin{equation}
\int_q \equiv \int \frac{d (q.n) d(q.\bar n)}{8 \pi^2} \int_{\bm{q}_\perp},
\label{eq:intq}
\end{equation}
where the measure of the dimensionally regularized  transverse momentum integral is
\begin{equation}
\int_{\bm{q}_\perp}  \equiv \, \mu^{2\epsilon}
\frac{(4 \pi)^{-\epsilon}}{\Gamma(1+\epsilon)} \int \frac{d^{2-2\epsilon}q_\perp}{(2 \pi)^{2-2\epsilon}}.
\label{eq:intqperp}
\end{equation}
We can use the 4-vectors $n$ and $\bar n$ to define regions of $q$. In the $n$ collinear region, $q.\bar n$ is order $Q$, $q^2$ is order $M^2$,  and $q.n$ is order $M^2/Q$. In the $\bar n$ collinear region, $q.n$ is order $Q$, $q^2$ is order $M^2$,  and $q.\bar n$ is order $M^2/Q$. In the soft region, $q.\bar n$, $q. n$, and $\bm{q}_\perp$ are all order $M$.

With rapidity regularization, different regularization factors are used in different regions. The specific forms of the regularization factors required for our problem were used in Ref.~\cite{Chiu:2011qc} and described more explicitly in Ref.~\cite{Chiu:2012ir}. The regularization factors in each of the regions of $q$ are
\begin{subequations}
\begin{eqnarray}
n~\text{collinear:} ~&&~  \big( |q.n|/\nu_+ \big)^{-\eta},
\label{eq:rapidreg-n}
\\\
\bar n~\text{collinear:} ~&&~ \big( |q.\bar n|/\nu_- \big)^{-\eta},
\label{eq:rapidreg-nbar}
\\\
\text{soft:}~~~~~~~~~ ~&&~ \big( |q.(n-\bar n)|/\nu \big)^{-\eta}.
\label{eq:rapidreg-soft}
\end{eqnarray}
\label{eq:rapidreg}%
\end{subequations}
where $\eta$ is the regularization parameter and $\nu_+$, $\nu_-$, and $\nu$ are regularization scales. The regularization scales are constrained by an equation that depends on the application. In most cases, the equation is either $\nu_+ \nu _- = \nu^2$ or $\nu_+ \nu _- = -\nu^2$.

\subsection{Hard contribution}

In the hard region of the loop momentum $q$, all its components of $q$ are order $\sqrt{\hat s}$. The hard contribution to the LP form factor is given by the integral in Eq.~\eqref{eq:F0hardint}. There are no  rapidity divergences from  this region, so there is no need for rapidity regularization. The analytic result is given in Eq.~\eqref{eq:F0hard}. A Laurent expansion in $\epsilon$  gives the final result in Eq.~\eqref{eq:F0hardeps}.

\subsection{Higgs collinear contribution}

In the Higgs collinear region of the loop momentum $q$, $p_3.q$ is order $Q^2$ but $q^2$ and $\tilde P.q$ are order $M^2$. The Higgs collinear contribution to the LP form factor with rapidity regularization but before any zero-bin subtractions is
\begin{eqnarray}
\label{eq:FHcollregint}
\mathcal{F}^{\text{LP}}_{H\,\text{coll,reg}}
=\frac{2i g_s^2 y_t}{D-2}
\int_q\frac{-4(p_3.q)^2/\tilde P.p_3 + 2(D-3)p_3.q + (D-2)\tilde P.p_3}{[(q+\tilde P)^2-m_t^2+i\epsilon]\, [q^2-m_t^2+i\epsilon]\, [-2p_3.q+i\epsilon]} 
\nonumber\\
\times   \bigg[\frac{|q.n|}{\nu_1} \bigg]^{-\eta}
 \bigg[ \frac{|(q+\tilde P).n|}{\nu_1} \bigg]^{-\eta}.
\end{eqnarray}
The measure for the integral over $q$ is given in Eq.~\eqref{eq:intq}. Since we set $m_H =0$ in this Section, we have replaced the 4-momentum $P$ of the Higgs by the light-like 4-vector $\tilde P$ whose 3-vector component is collinear to $\bm{P}$  and whose normalization is given by  $2\tilde{P}.p_3=\hat{s}$. The rapidity divergence from  the denominator $-2p_3.q$ is regularized by multiplying the integrand by the factor in Eq.~\eqref{eq:rapidreg-n} with the regularization scale $\nu_+$ replaced by $\nu_1$. In order to  maintain the symmetry between the two denominators with momenta $q$ and $q+\tilde P$, 
we have also multiplied the integrand by that same factor with $q$ replaced by $q+\tilde P$.

The only component of $p_3$ that is order $Q$ is $p_3.\bar n$. The leading-power contribution to Eq.~\eqref{eq:FHcollregint} can therefore be simplified by replacing $p_3.\tilde P$ by $p_3.\bar n\, \tilde P.n/2$ and   $p_3.q$ by $p_3.\bar n\, q.n/2$. The factors of $p_3.\bar n$ then cancel in Eq.~\eqref{eq:FHcollregint}, and it is evident that the only physical scales in the integral  are $m_t$ and $\tilde P.n$. The integral over $q.\bar{n}$ can be evaluated by contours. The integral over $q.n$ produces an  infrared  pole in the rapidity regularization parameter $\eta$. The dimensionally regularized integral over $\bm{q}_\perp$ produces an  ultraviolet  pole in $\epsilon$. The analytic result from  integrating over $q$ is
\begin{equation}
\label{eq:FHcollreg}
\mathcal{F}^{\text{LP}}_{H\,\text{coll,reg}}
=\frac{g_s^2 y_t}{16\pi^2}
\left[\frac{\mu^2}{m_t^2}\right]^\epsilon \bigg[ \frac{\tilde P.n}{\nu_1}\bigg]^{-2\eta}
\frac{1}{\epsilon} \frac{(1)_{-\eta}(1)_{-\eta}}{(1)_{-2\eta}}
\left( \frac{1}{\eta_\text{ir}} + \frac{2}{1-2\eta} \right).
\end{equation}
The subscript ir on the pole in $\eta$ indicates that the divergence has an infrared origin. 

Because the Higgs collinear region has an overlap with the soft region, the integral in Eq.~\eqref{eq:FHcollregint} requires a zero-bin subtraction. The subtraction integral is
\begin{eqnarray}
\label{eq:FHcollzbsint}
\mathcal{F}^{\text{LP}}_{H\,\text{coll,zbs}}
=2i g_s^2 y_t
\int_q\frac{\tilde P.p_3}{[2\tilde P.q+i\epsilon]\, [q^2-m_t^2+i\epsilon]\, [-2p_3.q+i\epsilon]}
 \bigg[\frac{|q.n|}{\nu_1} \bigg]^{-\eta}
 \bigg[ \frac{|(q+\tilde P).n|}{\nu_1} \bigg]^{-\eta}.~~~
\end{eqnarray}
The denominator with momentum $q+\tilde P$ in Eq.~\eqref{eq:FHcollregint} has been replaced by its soft limit. The integral over $q.\bar{n}$ can be evaluated by contours. The integral over $q.n$   gives  an infrared divergence and an ultraviolet  divergence, both of which are regularized by the parameter $\eta$. The dimensionally regularized integral over $\bm{q}_\perp$ produces an ultraviolet pole in $\epsilon$. The analytic result for the integral over $q$ is
\begin{equation}
\label{eq:FHcollzbs}
\mathcal{F}^{\text{LP}}_{H\,\text{coll,zbs}}
=\frac{g_s^2 y_t}{16\pi^2}
\left[\frac{\mu^2}{m_t^2}\right]^\epsilon \bigg[\frac{\tilde P.n}{\nu_1}\bigg]^{-2\eta}
\frac{1}{\epsilon}
\left\{\frac{1}{\eta_\text{ir}} \frac{(1)_{-\eta}(1)_{-\eta}}{(1)_{-2\eta}}
- \frac{1}{2\eta_\text{uv}} \frac{(1)_{-\eta}(1)_{2\eta}}{(1)_{\eta}}\right\}.
\end{equation}
The subscripts ir and uv indicate the origins of the divergences.

The complete contribution to the LP form factor from the Higgs collinear region is obtained by subtracting Eq.~\eqref{eq:FHcollzbs} from Eq.~\eqref{eq:FHcollreg}. The infrared poles in $\eta$  cancel, leaving an ultraviolet pole. After a Laurent expansion in $\eta$, the Higgs collinear contribution   reduces to
\begin{equation}
\label{eq:FHcollrap}
\mathcal{F}^{\text{LP}}_{H\,\text{coll}}(m_t^2,\tilde P.n)
= \frac{g_s^2 y_t}{16\pi^2}
\left[\frac{\mu^2}{m_t^2}\right]^\epsilon
\frac{1}{\epsilon}
\left(\frac{1}{2\eta_\text{uv}}-\log\frac{\tilde P.n}{\nu_1}+2 \right).
\end{equation}
It  depends logarithmically on $\tilde P.n$.

\subsection{Gluon collinear contribution}

In the gluon collinear region of the loop momentum $q$, $\tilde P.q$ is order $Q^2$ but $q^2$ and $p_3.q$ are order $M^2$. The gluon collinear contribution to the LP form factor with rapidity regularization before any zero-bin subtraction is
\begin{eqnarray}
\label{eq:Fgcollregint}
\mathcal{F}^{\text{LP}}_{g\,\text{coll,reg}}
=2i g_s^2 y_t
\int_q\frac{\tilde P.p_3}{[2 \tilde P.q+i\epsilon]\, [q^2-m_t^2+i\epsilon]\, [(q-p_3)^2-m_t^2+i\epsilon]} 
\nonumber\\
\times  \bigg[\frac{|q.\bar n|}{\nu_3} \bigg]^{-\eta}
 \bigg[\frac{|(q-p_3).\bar n|}{\nu_3} \bigg]^{-\eta}.
\end{eqnarray}
The measure for the integral over $q$ is given in Eq.~\eqref{eq:intq}. The rapidity divergence from  the denominator $2\tilde P.q$ is regularized by multiplying the integrand by the  factor in Eq.~\eqref{eq:rapidreg-nbar} with the regularization scale $\nu_-$ replaced by $\nu_3$. In order to  maintain the symmetry between the two denominators with momenta $q$ and $q-p_3$, we have also multiplied the integrand by that same factor with $q $ replaced by $q-p_3$.

The only component of $\tilde P$ that is order $Q$ is $\tilde P.n$. The leading-power contribution to Eq.~\eqref{eq:Fgcollregint} can therefore be simplified by replacing $\tilde P.p_3$ by $\tilde P.n\, p_3.\bar n/2$ and   $\tilde P.q$ by $\tilde P.n\, q. \bar n/2$. The factors of $\tilde P.n$ then cancel in Eq.~\eqref{eq:Fgcollregint}, and it is evident that the only physical scales in the integral  are $m_t$ and $p_3.\bar n$. The integral over $q.n$ can be evaluated by contours. The integral over $q.\bar n$ produces an infrared pole in $\eta$. The integral over $\bm{q}_\perp$ produces an ultraviolet pole in $\epsilon$. The analytic result for the integral over $q$ is 
\begin{equation}
\label{eq:Fgcollreg}
\mathcal{F}^{\text{LP}}_{g\,\text{coll,reg}}
=\frac{g_s^2 y_t}{16\pi^2}
\left[\frac{\mu^2}{m_t^2}\right]^\epsilon \left[\frac{p_3.\bar n}{\nu_3}\right]^{-2\eta}
\frac{1}{\epsilon\, \eta_\text{ir}}
\frac{(1)_{-\eta}(1)_{-\eta}}{(1)_{-2\eta}}.
\end{equation}
The subscript ir on the pole in $\eta$ indicates that the divergence has an infrared origin. 

Because the gluon collinear region has an overlap with the soft region, the integral in Eq.~\eqref{eq:Fgcollregint} requires a zero-bin subtraction. The subtraction integral is
\begin{eqnarray}
\label{eq:Fgcollzbsint}
\mathcal{F}^{\text{LP}}_{g\,\text{coll,zbs}}
=2i g_s^2 y_t
\int_q\frac{\tilde P.p_3}{[2\tilde P.q+i\epsilon]\, [q^2-m_t^2+i\epsilon]\, [-2p_3.q+i\epsilon]}
\bigg[\frac{|q.\bar n|}{\nu_3} \bigg]^{-\eta}
 \bigg[\frac{|(q-p_3).\bar n|}{\nu_3} \bigg]^{-\eta}.~~
 \end{eqnarray}
The denominator with momentum $q-p_3$ in Eq.~\eqref{eq:Fgcollregint} has been replaced by its soft limit. The integral over $q.n$ can be evaluated by contours. The  integral over $q.\bar n$ produces an infrared pole in $\eta$ and an ultraviolet pole in $\eta$. The  integral over $\bm{q}_\perp$ produces an ultraviolet pole in $\epsilon$. The analytic result for the integral over $q$ is 
\begin{equation}
\label{eq:Fgcollzbs}
\mathcal{F}^{\text{LP}}_{g\,\text{coll,zbs}}
=\frac{g_s^2 y_t}{16\pi^2}
\left[\frac{\mu^2}{m_t^2}\right]^\epsilon \left[\frac{p_3.\bar n}{\nu_3}\right]^{-2\eta}
\frac{1}{\epsilon}
\left\{\frac{1}{\eta_\text{ir}} \frac{(1)_{-\eta}(1)_{-\eta}}{(1)_{-2\eta}}
- \frac{1}{2\eta_\text{uv}} \frac{(1)_{-\eta}(1)_{2\eta}}{(1)_{\eta}}\right\}.
\end{equation}
The subscripts ir and uv indicate the origins of the divergences.

The complete contribution to the LP form factor from the gluon collinear region is obtained by subtracting Eq.~\eqref{eq:Fgcollzbs} from Eq.~\eqref{eq:Fgcollreg}. The infrared poles in $\eta$  cancel, leaving an ultraviolet pole. After a Laurent expansion in $\eta$,   the gluon collinear contribution   reduces to
\begin{equation}
\label{eq:Fgcollrap}
\mathcal{F}^{\text{LP}}_{g\,\text{coll}}(m_t^2,p_3.\bar n)
= \frac{g_s^2 y_t}{16\pi^2}
\left[\frac{\mu^2}{m_t^2}\right]^\epsilon
\frac{1}{\epsilon}
\left(\frac{1}{2\eta_\text{uv}}-\log\frac{p_3.\bar n}{\nu_3} \right).
\end{equation}
It depends logarithmically  on $p_3.\bar n$.

\subsection{Soft contribution}

In the soft region of the loop momentum $q$, all the components of $q$ are order $m_t$. The soft contribution to the LP form factor with rapidity regularization is
\begin{equation}
\label{eq:Fsoftrapint}
\mathcal{F}^{\text{LP}}_{\text{soft}}
=2i g_s^2 y_t 
\int_q\frac{\tilde P.p_3}{[2\tilde P.q +i\epsilon]\, [q^2-m_t^2+i\epsilon]\, [-2p_3.q+i\epsilon]}
 \left[\frac{|q.(n-\bar n)|}{\nu} \right]^{-2\eta}.
\end{equation}
The 4-momentum $P$ of the Higgs has been replaced by the light-like 4-vector $\tilde P$. The rapidity divergences from the two denominators $2\tilde P.q$ and $-2p_3.q$ have been regularized by multiplying the integrand by two identical copies of the factor in Eq.~\eqref{eq:rapidreg-soft}. The integral over $q$ in Eq.~\eqref{eq:Fsoftrapint} includes a Higgs collinear region in which $q.\bar n$ is small and a gluon collinear region in which $q.n$ is small. In the  Higgs collinear region, the soft regularization factor is proportional to $|q.n|^{-2\eta}$. It has the same form as the Higgs collinear regularization factor in Eq.~\eqref{eq:FHcollregint} in the ultraviolet limit. Thus the ultraviolet divergences from the Higgs collinear region in Eq.~\eqref{eq:Fsoftrapint} cancel against ultraviolet divergences from the zero-bin subtraction for the Higgs collinear contribution in Eq.~\eqref{eq:FHcollzbsint}. In the gluon collinear region, the soft regularization factor is proportional to $|q.\bar n|^{-2\eta}$. It has the same form as the gluon collinear regularization factor in Eq.~\eqref{eq:Fgcollregint} in the ultraviolet limit. Thus the ultraviolet divergences from the gluon collinear region in Eq.~\eqref{eq:Fsoftrapint} cancel against ultraviolet divergences from the zero-bin subtraction for the gluon collinear  contribution in Eq.~\eqref{eq:Fgcollzbsint}.

The integral  over $q$ in Eq.~\eqref{eq:Fsoftrapint} gives ultraviolet  poles in $\eta$ and in $\epsilon + \eta$:
\begin{eqnarray}
\label{eq:Fsoftrap2}
\mathcal{F}^{\text{LP}}_{\text{soft}}=
- \frac{g_s^2 y_t}{16\pi^2} \left[\frac{\mu^2}{m_t^2}\right]^\epsilon \left[ \frac{\nu}{2m} \right]^{2\eta}
\frac{(\tfrac12)_{-\eta} (1)_{\epsilon+\eta}}{(1)_\epsilon}
\frac{1}{\eta_\mathrm{uv} (\epsilon + \eta)}.
\end{eqnarray}
After a Laurent expansion in $\eta$,  the   soft contribution  reduces to
\begin{equation}
\label{eq:Fsoftrap}
\mathcal{F}^{\text{LP}}_{\text{soft}}(m_t^2) =
- \frac{g_s^2 y_t}{16\pi^2}
\left[\frac{\mu^2}{m_t^2}\right]^\epsilon
\left(\frac{1}{\epsilon\, \eta_\text{uv}} -\frac{1}{\epsilon^2}+\frac{1}{\epsilon}\log\frac{\nu^2}{m_t^2} +\frac{\pi^2}{6} \right).
\end{equation}
It depends logarithmically on  $m_t$. 

\subsection{LP form factor}

In the sum of the Higgs collinear contribution in Eq.~\eqref{eq:FHcollrap}, the gluon collinear contribution in Eq.~\eqref{eq:Fgcollrap}, and the soft contribution in Eq.~\eqref{eq:Fsoftrap}, the ultraviolet poles in $\eta$ cancel. The only divergences that remain are double and single poles in $\epsilon$:
\begin{equation}
\label{eq:Fcollsoftreg}
\mathcal{F}^{\text{LP}}_{H\,\text{coll}} +\mathcal{F}^{\text{LP}}_{g\,\text{coll}}
+\mathcal{F}^{\text{LP}}_{\text{soft}} =
\frac{g_s^2 y_t}{16\pi^2}
\left[\frac{\mu^2}{m_t^2}\right]^\epsilon
\left\{ \frac{1}{\epsilon^2} - \frac{1}{\epsilon} \left( \log\frac{\tilde P.n\, p_3.\bar n}{\nu_1\, \nu_3}+\log\frac{\nu^2}{m_t^2} - 2 \right)
-\frac{\pi^2}{6} \right\}.
\end{equation}
Comparing with the sum of the Higgs collinear contribution and the gluon collinear contribution using analytic regularization in Eq.~\eqref{eq:F0H+gcoll}, we see that they agree provided the rapidity regularization scales satisfy
\begin{equation}
\label{eq:rapidregscales}
\nu_1\, \nu_3 = e^{+i \pi}\,  \nu^2.
\end{equation}
We obtained this nontrivial constraint on the rapidity regularization scales by comparing with the result from analytic regularization. It would be preferable to derive it  more directly within the framework of rapidity regularization.

The complete LP form factor with rapidity regularization is obtained by adding the hard contribution in Eq.~\eqref{eq:F0hardeps} to the sum of the Higgs collinear, gluon collinear, and soft contributions in Eq.~\eqref{eq:Fcollsoftreg}. The double and single poles in $\epsilon$ are canceled. The final result for the LP form factor with $m_H=0$ agrees with the result in Eq.~\eqref{eq:F0LP}.

\section{Hard Form Factors and Distribution Amplitudes}
\label{sec:Fragmentation}

In this Section, we calculate the factors in the Higgs collinear and gluon collinear contributions to the LP form factor in a way that involves only the single scale $Q$ or $M$. The factors involving the hard scale $Q$ are form factors for $t \bar t_{1V} +g$ and $H+ t \bar t_{8T}$. The factors involving the soft scale $M$ are  distribution amplitudes for a $t \bar t$ pair in the Higgs and for a $t \bar t$ pair in a real gluon. We use rapidity regularization to define the distribution amplitudes. At the end of this section, we discuss the relation between our distribution amplitudes and double-parton fragmentation functions, which were recently introduced for heavy quarkonium production, and the relation between our distribution amplitudes and those used for exclusive processes.

\subsection{Hard form factor for $\bm{ t\bar{t}_{1V} +g}$}

In Section~\ref{sec:Factorgcoll}, the scales $Q$ and $M$ in the Higgs collinear contribution to the LP form factor  were separated by expressing it as an integral over the relative longitudinal momentum fraction $\zeta$:
\begin{equation}
\mathcal{F}_{H\, \text{coll}}^{\text{LP}} =
 \int_{-1}^{+1}\!\!\!\! d \zeta \,\mathcal{\widetilde F}_{t \bar t_{1V}+ g}(\zeta) \,
d_{t \bar t_{1V} \to H}(\zeta) .
\label{eq:FLPHcoll1V}
\end{equation}
The integrand is the product of a hard form factor $\mathcal{\widetilde F}_{t \bar t_{1V}+ g}$ for producing a gluon and a collinear $t \bar t$ pair in the color-singlet Lorentz-vector ($1V)$ channel and a distribution amplitude $d_{t \bar t_{1V} \to H}$ for a $t \bar t$ pair in the Higgs. The hard form factor depends only on the scale $Q$. The distribution amplitude depends on the scale $M$. With rapidity regularization, it also depends logarithmically on $P.n$. 

\begin{figure}
\includegraphics[width=10cm]{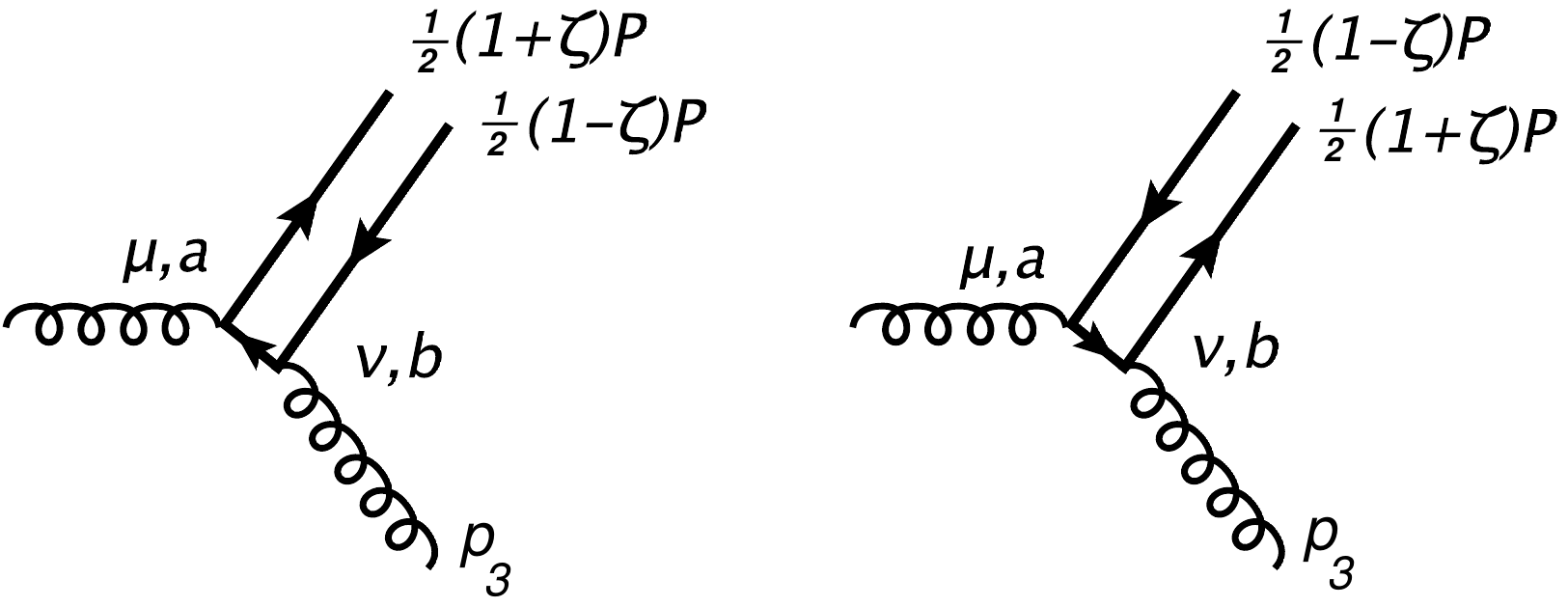}
\caption{Feynman diagrams for  the tensor amplitude $\mathcal{T}^{\mu\nu}$ for $g^*\to t\bar{t} +g$ at LO.
\label{fig:tt+g}}
\centering
\end{figure}

The amplitude $\mathcal{T}^{\mu a,  \nu b}$ for $g^* \to t \bar t + g$ is given by the sum of the two diagrams in Fig.~\ref{fig:tt+g}.  Since we only want the leading power, we can set the top-quark mass equal to zero. The amplitude for a virtual gluon with Lorentz index $\mu$ and color index $a$ to produce a real gluon with momentum $p_3$,  Lorentz index $\nu$, and color index $b$ and a color-singlet $t$ and $ \bar t$ pair with collinear momenta 
$\tfrac12(1+\zeta) \tilde P$ and $\tfrac12(1-\zeta) \tilde P$ is
\begin{eqnarray}
\mathcal{T}^{\mu a, \nu b} (P,p_3) &=& g_s^2  \frac{\text{tr}(T^a T^b)}{\sqrt{N_c}}
\left( \frac{\text{Tr}\big[  \gamma^\mu(\slash \!\!\!p_3 + \tfrac12(1-\zeta)\slash \!\!\!\!\tilde P) \gamma^\nu \, v\,\bar u \big] }
{(1-\zeta)\tilde P.p_3 } \right.
\nonumber\\
&& \hspace{3cm}
\left.  - \frac{\text{Tr}\big[  \gamma^\nu(\slash \!\!\!p_3 + \tfrac12(1+\zeta)\slash \!\!\!\!\tilde P) \gamma^\mu \, v\,\bar u \big] }
{(1+\zeta)\tilde P.p_3 }\right),
\label{eq:T-tt+g}
\end{eqnarray}
where $\bar u$ and $v$ are the Dirac spinors for the $t$ and $\bar t$. The factor $1/\sqrt{N_c}$, where $N_c$ is the number of quark colors, comes from projecting the  $t \bar t$ pair into a color-singlet state. The color trace  tr($T^a T^b$) can be absorbed into the prefactor of $\mathcal{T}^{\mu\nu}$ in Eq.~\eqref{eq:M}. The  $t \bar t$ pair can be projected onto the Lorentz-vector channel by replacing the spinor product $v \, \bar u$ by $ \slash \!\!\!\! \tilde P$.
 The $1V$ contribution to the tensor amplitude in Eq.~\eqref{eq:T-tt+g} is  
\begin{equation}
\mathcal{T}_{1V}^{\mu\nu} (P,p_3)= -\frac{4g_s^2}{\sqrt{N_c}} 
\left( \frac{\tilde P.p_3  g^{\mu \nu} - (\tilde P^\mu p_3^\nu +  p_3^\mu \tilde P^\nu ) 
- (1-\zeta) \tilde P^\mu  \tilde P^\nu }
{(1-\zeta)\tilde P.p_3 } - (\zeta \to -\zeta) \right).
\label{eq:T-tt1V+g}
\end{equation}

The hard form factor for $g^* \to t \bar t_{1V} + g$  can be obtained by contracting  the tensor $\mathcal{T}_{1V}^{\mu\nu}$ in Eq.~\eqref{eq:T-tt1V+g} with the tensor in Eq.~\eqref{eq:Fdef}, with $P$ replaced by $\tilde P$. We choose to move a factor $1/(1-\zeta^2)$ to the distribution amplitude to allow the poles in the regularization parameters to be made explicit. A canceling factor $1-\zeta^2$ must appear in the hard form factor. We also choose to move the factor $1/m_t$ from Eq.~\eqref{eq:Fdef} and the factor $1/\sqrt{N_c}$ from Eq.~\eqref{eq:T-tt1V+g} to the distribution amplitude to simplify the expressions for the hard form factor and the distribution amplitude. The resulting expression for the hard form factor is
\begin{equation}
\widetilde{\mathcal{F}}_{t \bar t_{1V}+ g}(\zeta) =
-\frac{g_s^2}{D-2}(1-\zeta^2) \left(\frac{D-1-\zeta}{1-\zeta} - \frac{D-1+\zeta}{1+\zeta}  \right).
\label{eq:FFtt1Vg}
\end{equation}
We have given the contributions from the two diagrams separately. The dependence on $D$ cancels in their sum.

\subsection{Distribution amplitude for $\bm{ t \bar{t}_{1V} \to H}$}

The   soft  factor in the  expression for the Higgs collinear contribution to the LP form factor in Eq.~\eqref{eq:FLPHcoll1V} is the distribution amplitude for $t \bar t_{1V} \to H$. The distribution amplitude is a function of the relative  longitudinal momentum fraction $\zeta$ that describes how the longitudinal momentum of the Higgs is  distributed between a virtual $t$ and a virtual $\bar t$. It can be calculated by using ingredients from the  Feynman rules for 
double-parton fragmentation functions in Ref.~\cite{Ma:2013yla}. A fragmentation function can be expressed as the sum of cut diagrams that are products of an amplitude and the complex conjugate of an amplitude. The amplitude for $t \bar t$ fragmentation into a specific final state is the amplitude for that final state to be produced by sources that create the $t$ and the $\bar t$ in a specified color and Lorentz channel with relative longitudinal momentum fraction $\zeta$. The sources are the endpoints of eikonal lines that extend to future infinity. The Feynman rule for the sources is the product of a color matrix, a Dirac matrix, and a delta function. The Feynman rule for sources that create the $t$ and $\bar t$ in the $1V$ channel with momenta $p $ and $\bar p$ is
\begin{equation}
\frac{\mathds{1}}{\sqrt{N_c} }~ \frac{\slash\!\!\!n}{4(p + \bar p).n} ~
\delta \big( \zeta - (p - \bar p).n/(p + \bar p).n \big),
\label{eq:Frulett1V}
\end{equation}
where $n$ is the light-like 4-vector that defines the longitudinal direction.

\begin{figure}
\includegraphics[width=4cm]{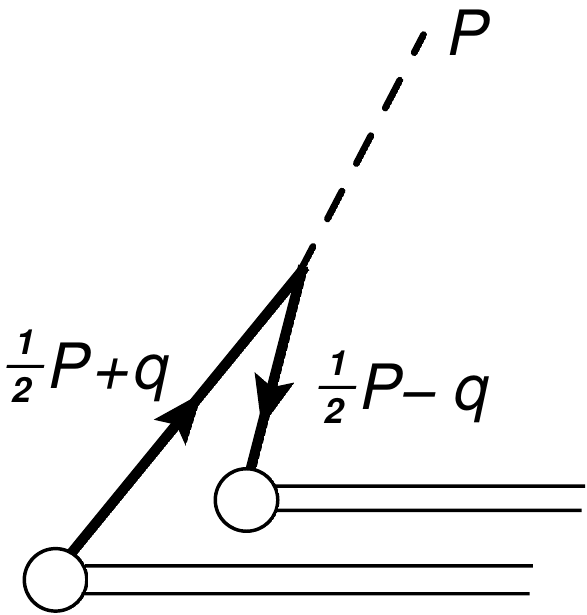}
\caption{Feynman diagram for the distribution amplitude for ${ t \bar{t}_{1V} \to H}$ at LO.
\label{fig:tt->H}}
\centering
\end{figure}

The leading-order diagram for the distribution amplitude for a $t \bar t$ pair  in the Higgs is shown  in Fig.~\ref{fig:tt->H}. The diagram has a factor of $-1$ for the closed fermion loop.  The expression for the distribution amplitude is
\begin{equation}
\frac{i \sqrt{N_c\, }y_t}{4P.n} \int_q  
\frac{ \delta(\zeta - 2q.n/P.n)\,
\text{Tr}\big[  (\tfrac12 \slash \!\!\!\!P - \slash \!\!\!q - m_t)  (\tfrac12 \slash \!\!\!\!P + \slash \!\!\!q + m_t)\slash \!\!\!n \big]}
{[(\tfrac12 P-q)^2 - m_t^2  + i \epsilon]\, [(\tfrac12 P+q)^2-m_t^2  + i \epsilon] }
=\sqrt{N_c} \,y_tm_t\, \zeta \, d(\zeta),
\label{eq:dtt1V->H}
\end{equation}
where the function $d(\zeta)$ is
\begin{equation}
d(\zeta) = -i  \int_q  
\frac{ \delta(\zeta - 2q.n/P.n) }
{[(\tfrac12 P+q)^2 - m_t^2  + i \epsilon]\, [(\tfrac12 P-q)^2-m_t^2  + i \epsilon] }.
\label{eq:drapid}
\end{equation}
In Eqs.~\eqref{eq:dtt1V->H} and \eqref{eq:drapid}, we have suppressed rapidity regularization factors and zero-bin subtractions for the integral over the loop momentum $q$. Multiplying by the factors $1/(\sqrt{N_c} \,m_t)$ and $1/(1-\zeta^2)$ that were removed from the form factor for $g^* \to t \bar t_{1V} + g$ in Eq.~\eqref{eq:FFtt1Vg}, we obtain the distribution amplitude 
\begin{equation}
d_{t \bar t_{1V} \to H}(\zeta) = y_t \, \zeta \, \frac{d(\zeta)}{1-\zeta^2} .
 \label{eq:dfragtt1V->H}
\end{equation}
The function $d(\zeta)$ is calculated with rapidity regularization and with appropriate zero-bin subtractions in Appendix~\ref{app:FragAmprapid}. The function $d(\zeta)/(1-\zeta^2)$ is given in Eq.~\eqref{eq:d-analreg}, with the ultraviolet poles in the regularization parameters $\epsilon$ and $\eta$ made explicit. The regularized distribution amplitude is
\begin{eqnarray}
d_{t \bar t_{1V} \to H}(\zeta) =
\frac{y_t}{32 \pi^2}
\left[\frac{\mu^2}{m_t^2}\right]^\epsilon \left[\frac{P.n}{\nu_1}\right]^{-2\eta}
\frac{1}{\epsilon}
\left\{-\frac{1}{2\eta_\text{uv}}\delta(1-\zeta^2)+\frac{1}{(1-\zeta^2)_+}\right\}
\nonumber\\
\times
\zeta \left[1-(1-\zeta^2)r^2-i\epsilon\right]^{-\epsilon}.
\label{eq:dfragtt1V->Hreg}
\end{eqnarray}
We have set the rapidity regularization scale to $\nu_1$.

\subsection{Hard form factor for $\bm{H+ t\bar{t}_{8T} }$}

In Section~\ref{sec:Factorgcoll}, the scales $Q$ and $M$ in the gluon collinear contribution to the LP form factor  were separated by expressing it as an integral over the momentum fraction variable $\zeta$:
\begin{equation}
\mathcal{F}_{g\, \text{coll}}^{\text{LP}} =
 \int_{-1}^{+1}\!\!\!\! d \zeta \,\mathcal{\widetilde F}_{H+t \bar t_{8T}}(\zeta) \,
d_{t \bar t_{8T} \to g}(\zeta) .
\label{eq:FLPgcoll8T}
\end{equation}
The integrand is the product of the hard form factor $\mathcal{\widetilde F}_{H+t \bar t_{8T}}$ for producing a Higgs and a collinear $t \bar t$ pair in the color-octet Lorentz-tensor ($8T)$ channel and the  distribution amplitude $d_{t \bar t_{8T} \to g}$ for a $t \bar t$ pair in a real gluon. The hard form factor depends only on the scale $Q$. The distribution amplitude depends on the scale $M$. With rapidity regularization, it also depends logarithmically on $p_3.\bar n$. 

\begin{figure}
\includegraphics[width=9cm]{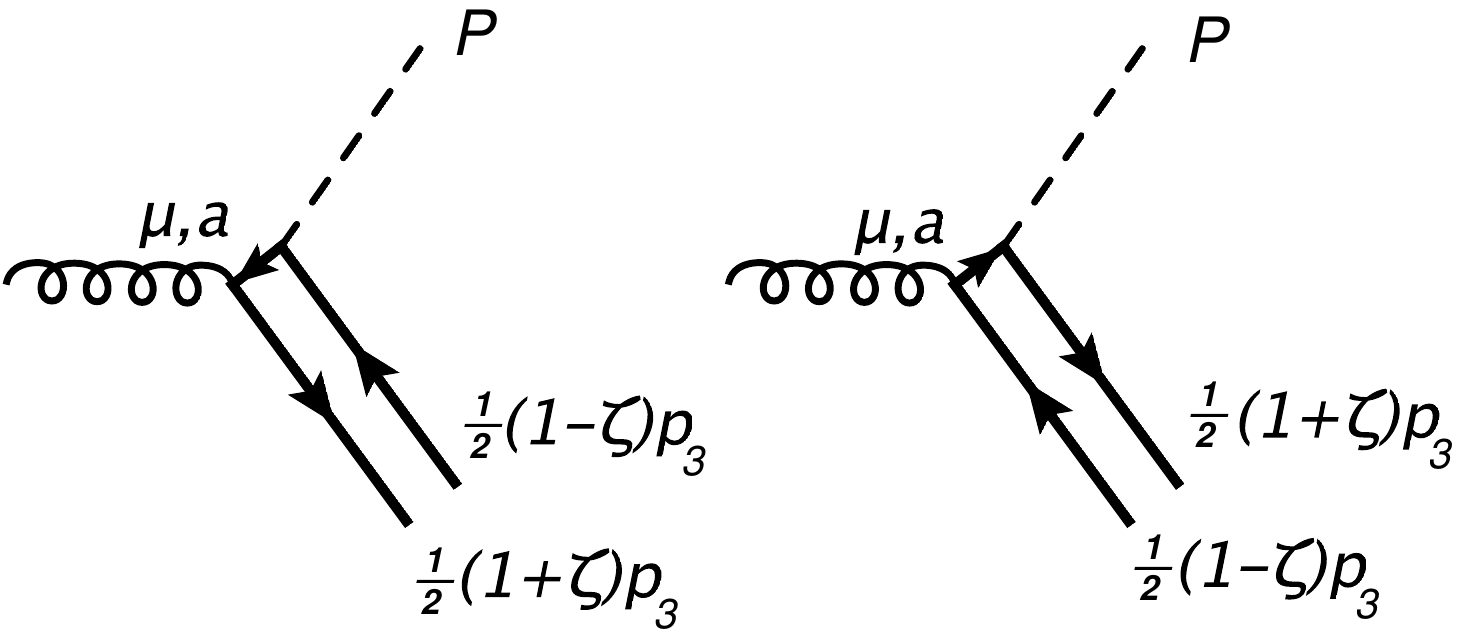}
\caption{Feynman diagrams for the tensor amplitude $\mathcal{T}^{\mu a,b}$ for $g^*\to H+ t\bar{t}$ at LO.
\label{fig:H+tt}}
\centering
\end{figure}

The amplitude $\mathcal{T}^{\mu a,b}$ for $g^* \to H+  t \bar t$ is given by the sum of the two diagrams in Fig.~\ref{fig:H+tt}. Since we only want the leading power, we can set the top quark mass equal to zero. The amplitude for a virtual gluon with Lorentz index $\mu$ and color index $a$ to produce a Higgs with momentum $\tilde P$ and a color-octet $t\bar t$ pair with collinear momenta $\tfrac12(1+\zeta)  p_3$ and $\tfrac12(1-\zeta) p_3$ and color index $b$ is 
\begin{eqnarray}
\mathcal{T}^{\mu a,b}(P,p_3) &=&  g_s y_t \left(\sqrt{2}\,  \text{tr}(T^a T^b)\right)
\left( \frac{\text{Tr}\big[ ( \slash \!\!\!\!\tilde P + \tfrac12(1+\zeta)\slash \!\!\!p_3)  \gamma^\mu \, v\,\bar u \big] }
{(1+\zeta)\tilde P.p_3 } \right.
\nonumber\\
&&\hspace{4cm} \left.
- \frac{\text{Tr}\big[ \gamma^\mu( \slash \!\!\!\!\tilde P + \tfrac12(1-\zeta)\slash \!\!\!p_3)   \, v\,\bar u \big] }
{(1-\zeta)\tilde P.p_3 } \right),
\label{eq:T-H+tt}
\end{eqnarray}
where $\bar u$ and $v$ are the Dirac spinors for the $t$ and $\bar t$. The factor of $\sqrt{2}$ comes from projecting the $t \bar t$ pair onto a color-octet state.
The $t \bar t$ pair can be projected onto the Lorentz-tensor channel with a Lorentz index $\nu$ by replacing the spinor product $v \, \bar u$ by $\slash \!\!\!\! p_3 \gamma_\perp^\nu$, where $ \gamma_\perp^\nu$ are Dirac matrices that are perpendicular to specified light-like 4-vectors $n$ and $\bar n$. They can be expressed as $\gamma_\perp^\nu = g_\perp^{\nu\alpha} \gamma_\alpha$, where the perpendicular  metric tensor is
\begin{equation}
 g_{\perp\alpha\beta} = 
 g_{\alpha\beta} - \frac{n_\alpha \bar n_\beta + \bar n_\alpha n_\beta}{n.\bar n}.
\label{eq:gperp}
\end{equation}
The color trace tr$(T^a T^b)$ can be absorbed into the prefactor of $\mathcal{T}^{\mu\nu}$ in Eq.~\eqref{eq:M}. The $8T$ contribution to the vector amplitude $\mathcal{T}^{\mu}$ in Eq.~\eqref{eq:T-H+tt} defines the tensor amplitude
\begin{equation}
\mathcal{T}_{8T}^{\mu\nu} (P,p_3)= - 4 \sqrt{2}g_s y_t g_\perp^{\mu\nu}
 \left(\frac{1}{1+\zeta} + \frac{1}{1-\zeta}  \right).
\label{eq:T-H+tt8T}
\end{equation}

The hard form factor for $g^* \to H+ t \bar t_{8T}$  can be obtained  by contracting the tensor $\mathcal{T}_{8T}^{\mu\nu}$ in Eq.~\eqref{eq:T-H+tt8T} with the tensor in Eq.~\eqref{eq:Fdef}, with $P$ replaced by $\tilde P$. We choose to move a factor $1/(1-\zeta^2)$ to the distribution amplitude to allow the poles in the regularization parameters to be made explicit. A canceling factor $1-\zeta^2$ must appear in the hard form factor. We also choose to move the factor $1/m_t$ from Eq.~\eqref{eq:Fdef} and the factor $\sqrt{2}$ from Eq.~\eqref{eq:T-H+tt8T} to the distribution amplitude to simplify the expressions for the hard form factor and the distribution amplitude. The resulting expression for the hard form factor is  
\begin{equation}
\widetilde{\mathcal{F}}_{H+t \bar t_{8T}}(\zeta) =
-g_s y_t(1-\zeta^2) \left(\frac{1}{1+\zeta} + \frac{1}{1-\zeta}  \right).
\label{eq:FFHtt8T}
\end{equation}
We have given the contributions from the two diagrams separately.

\subsection{Distribution amplitude for $\bm{ t \bar{t}_{8T} \to g}$}

\begin{figure}
\includegraphics[width=4cm]{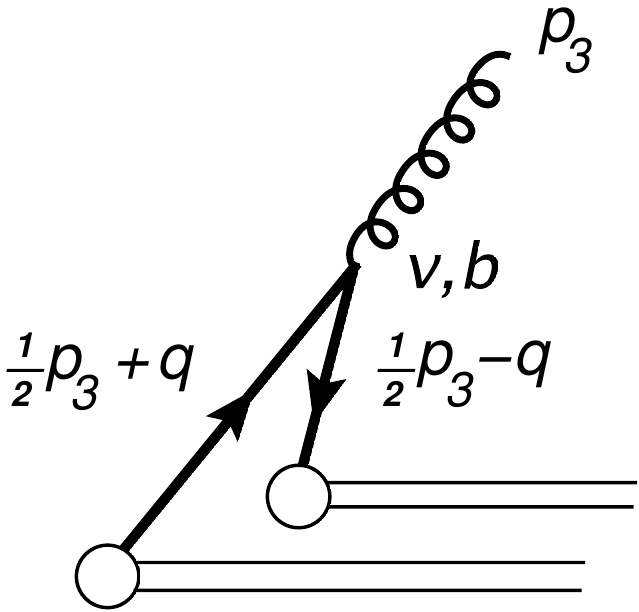}
\caption{Feynman diagram for the distribution amplitude for ${ t \bar{t}_{8T} \to g}$ at LO.
\label{fig:tt->g}}
\centering
\end{figure}

The collinear factor in the  expression for the gluon collinear contribution to the LP form factor in Eq.~\eqref{eq:FLPgcoll8T} is the distribution amplitude for $t \bar t_{8V} \to g$. The distribution amplitude is a function of the relative  longitudinal momentum fraction $\zeta$ that describes how the longitudinal momentum of the real gluon is  distributed between a virtual $t$ and a virtual $\bar t$. It can be calculated from the diagram  in Fig.~\ref{fig:tt->g} by using ingredients from the  Feynman rules for double-parton fragmentation functions in Ref.~\cite{Ma:2013yla}. The amplitude for $t \bar t$ fragmentation into a specific final state is the amplitude for that final state to be produced by sources that create the $t$ and the $\bar t$ in a specified color and Lorentz channel with relative longitudinal momentum fraction $\zeta$. The sources are the endpoints of eikonal lines that extend to future infinity. The Feynman rule for the sources is the product of a color matrix, a Dirac matrix, and a delta function. The Feynman rule for sources that create the $t$ and $\bar t$ in the $8T$ channel with momenta $p $ and $\bar p$ is
\begin{equation}
\sqrt{2}\,T^a~ \frac{\slash\!\!\!\bar n \gamma_\perp^\mu}{4(p + \bar p).\bar n} ~
\delta \big( \zeta - (p - \bar p).\bar n/(p + \bar p).\bar n \big),
\label{eq:Frulett8T}
\end{equation}
where $\bar n$ is the light-like 4-vector that defines the longitudinal direction, $\gamma_\perp^\mu = g_\perp^{\mu\alpha}\gamma_\alpha$, and the metric tensor $g_{\perp\mu\beta}$ is defined in Eq.~\eqref{eq:gperp}.

The leading-order diagram for the distribution amplitude for a $t \bar t$ pair in a real gluon is shown  in Fig.~\ref{fig:tt->g}. The diagram has a factor of $-1$ for the closed fermion loop. The amplitude for the source to  produce a real gluon with  polarization vector $\varepsilon_3$ and color index $a$  is
\begin{eqnarray}
- \frac{i \sqrt{2}\, g_s }{4 p_3.\bar n}\text{tr}(T^a T^b)\, \varepsilon^*_{3\nu}\int_q  
\frac{ \delta(\zeta - 2q.\bar n/p_3.\bar n)\,
\text{Tr}\big[  (\tfrac12 \slash \!\!\!p_3 - \slash \!\!\!q - m_t) \gamma^\nu (\tfrac12 \slash \!\!\!p_3 + \slash \!\!\!q + m_t)
\slash\!\!\!\bar n \gamma_\perp^\mu \big]}
{[(\tfrac12 p_3+q)^2 - m_t^2  + i \epsilon]\, [(\tfrac12 p_3-q)^2-m_t^2  + i \epsilon] }
\nonumber\\
=\frac{g_s m_t}{\sqrt{2}} \delta^{ab} (-g_\perp^{\mu \nu}) \epsilon_{3\nu}^* \, d_0(\zeta),
\label{eq:dtt8T->g}
\end{eqnarray}
where the function $d_0(\zeta)$ is
\begin{eqnarray}
d_0(\zeta) = -i \int_q  
\frac{ \delta(\zeta - 2q.\bar n/p_3.\bar n)}
{[(\tfrac12 p_3+q)^2 - m_t^2  + i \epsilon]\, [(\tfrac12 p_3-q)^2-m_t^2  + i \epsilon] }.
\label{eq:d0rapid}
\end{eqnarray}
In Eqs.~\eqref{eq:dtt8T->g} and \eqref{eq:d0rapid}, we have suppressed rapidity regularization factors and zero-bin subtractions in the integral over the loop momentum $q$. We can identify the distribution amplitude for a real gluon with transverse polarization vector in the same direction as the source and with the same color index as the source as the coefficient of $\delta^{ab} (-g_\perp^{\mu \nu}) \epsilon_{3\nu}^*$. Multiplying by the factors $\sqrt{2}/m_t$ and $1/(1-\zeta^2)$ that were removed from the form factor for $g^* \to H+t \bar t_{8T} $, we obtain the distribution amplitude 
\begin{equation}
d_{t \bar t_{8T} \to g}(\zeta) = g_s \, \frac{d_0(\zeta)}{1-\zeta^2} .
 \label{eq:dfragtt8T->g}
\end{equation}
The function $d_0(\zeta)/(1-\zeta^2)$  with rapidity regularization can be obtained from the function $d(\zeta)/(1-\zeta^2)$ in Eq.~\eqref{eq:d-analreg} by setting  $r=0$ and replacing $P.n$ with $p_3.\bar n$. The regularized distribution amplitude, with the ultraviolet poles in the regularization parameters $\epsilon$ and $\eta$ made explicit, is
\begin{eqnarray}
d_{t \bar t_{8T} \to g}(\zeta) =
\frac{g_s}{32 \pi^2}
\left[\frac{\mu^2}{m_t^2}\right]^\epsilon \left[\frac{p_3.\bar n}{\nu_3}\right]^{-2\eta}
\frac{1}{\epsilon} \left\{-\frac{1}{2\eta_\text{uv}}\delta(1-\zeta^2)+\frac{1}{(1-\zeta^2)_+}\right\}.
\label{eq:dfragtt8T->greg}
\end{eqnarray}
We have set the rapidity regularization scale to $\nu_3$.

\subsection{Relation to double-parton fragmentation functions}

Our factorization framework for the exclusive production of Higgs was inspired by recent progress in the QCD factorization of heavy quarkonium. We proceed to describe the connection between our distribution amplitude for $t \bar t  \to H$ and double-parton fragmentation functions for Higgs production. Factorization formulas for inclusive Higgs production with large transverse momentum $P_T$ in the Standard Model can be deduced from the corresponding factorization formulas for inclusive hadron production in QCD \cite{Braaten:2015ppa}. For an inclusive cross section $d \sigma/dP_T^2$, the leading  power is $1/P_T^4$ and the next-to-leading power is $1/P_T^6$. For inclusive hadron production, the leading power comes from a mechanism called {\it fragmentation}:  the production of a parton with larger transverse momentum followed by the decay of the virtual parton into states that include the hadron.
For inclusive Higgs production with a top quark, the leading power contribution to the differential cross section can be expressed in the form of the leading-power (LP) factorization formula:
\begin{equation}
d \tilde{\sigma}_{H+t+X}(P_T) +
\int_0^1 \!\!dz\, d \tilde{\sigma}_{t+X}(P_T/z)\,
D_{t \to H}(z),
\label{eq:dsigLP}
\end{equation}
where $d \tilde{\sigma}_{H+t+X}$ is the inclusive hard-scattering cross section for producing $H$ with transverse momentum $P_T$ and $d \tilde{\sigma}_{t+X}$ is the inclusive hard-scattering cross section for producing $t$ with larger transverse momentum $P_T/z$. The integral is over the fraction $z$ of the longitudinal momentum of $t$ carried by $H$. The fragmentation function $D_{t \to H}(z)$ is the probability distribution for $z$ from the decay of the virtual  $t$ into states that include $H$. The LP factorization formula in Eq.~\eqref{eq:dsigLP} separates the large scale $P_T$, which appears only in $d \tilde{\sigma}_{H+t+X}$ and $d \tilde{\sigma}_{t+X}$, from the smaller scale $M$ of the masses $m_t$ and $m_H$, which appear only in $D_{t \to H}$. The first term in Eq.~\eqref{eq:dsigLP} corresponds to the direct production of $H$ at short distances. This has no analog in the LP factorization formula for QCD: a color-singlet hadron cannot be produced directly at short distances at leading power.

A significant step forward  in QCD factorization was the extension of the factorization  formula to the next-to-leading power in $1/P_T^2$ for the case of heavy quarkonium. The next-to-leading power  (NLP) factorization formula was proven by Kang, Qiu, and Sterman using a diagrammatic analysis  \cite{Kang:2011mg} and derived by Fleming, Leibovich, Mehen, and Rothstein using Soft Collinear Effective Theory  \cite{Fleming:2012wy}. There are contributions at NLP that come from expanding the hard-scattering cross sections $d \tilde{\sigma}$ to first order in $M^2/P_T^2$, but there are additional contributions that come from a new mechanism called {\it double-parton fragmentation}:  the production of a heavy quark and antiquark with a larger total transverse momentum followed by the decay of the virtual quark-antiquark pair into states that include the heavy quarkonium. For inclusive Higgs production, the $t \bar t$ fragmentation contribution to the NLP factorization formula has the form
\begin{equation}
\int_0^1 \!\!dz\, \int_{-1}^{+1}\!\!\!\! d \zeta \, \int_{-1}^{+1}\!\!\!\! d \zeta'\, d \tilde{\sigma}_{t\bar t+X}(P_T/z,\zeta,\zeta')\,
D_{t \bar t \to H}(z,\zeta,\zeta'),
\label{eq:dsigNLP}
\end{equation}
where $d \tilde{\sigma}_{t\bar t+X}$ is the inclusive hard-scattering cross section for producing $t\bar t$ with total transverse momentum $P_T/z$. The integrals are over the fraction $z$ of the total longitudinal momentum of $t \bar t$ carried by $H$, the relative longitudinal momentum fraction $\zeta$ of  the $t$ and $\bar t$ in the amplitude, and the relative longitudinal momentum fraction $\zeta'$ of the $t$ and $\bar t$ in the complex conjugate of the amplitude. For given $\zeta$ and $\zeta'$, the fragmentation function $D_{t \to H}(z,\zeta,\zeta')$ is the distribution in $z$ from the decay of the virtual  $t \bar t$ pair into states that include $H$. The term in Eq.~\eqref{eq:dsigNLP} in the NLP factorization formula separates the large scale $P_T$, which appears only in $d \tilde{\sigma}_{t \bar t+X}$, from the smaller scale $M$ of the masses, which appear only in $D_{t \bar t \to H}$.

The production of Higgs at large $P_T$ with no final-state top quark has contributions from the LP factorization formula in Eq.~\eqref{eq:dsigLP} beginning at NLO in $\alpha_s$.  At LO in $\alpha_s$, the leading power in $1/P_T^2$ comes from the NLP factorization formula in Eq.~\eqref{eq:dsigNLP}. At this order, the fragmentation process is the annihilation of the virtual $t \bar t$ pair into a Higgs only. The entire longitudinal momentum of the $t \bar t$ pair is carried by the Higgs, so the fragmentation function has a factor of $\delta(1-z)$. The fragmentation function at LO is
\begin{eqnarray}
D_{t \bar t_{1V} \to H}(z,\zeta,\zeta')=
N_c m_t^2\,
\Big[ (1-\zeta^2)\,  d_{t \bar t_{1V} \to H}(\zeta) \Big]
\Big[(1-{\zeta'}^2)\, d_{t \bar t_{1V} \to H}(\zeta') \Big]^* \, \delta(1-z) ,
\label{eq:Dtt->H}
\end{eqnarray}
where  $d_{t \bar t_{1V} \to H}(\zeta)$ is the  regularized distribution amplitude in Eq.~\eqref{eq:dfragtt1V->Hreg}. A renormalized fragmentation function can be obtained by the minimal subtraction of the poles in $\eta$ and in $\epsilon$.

\subsection{Relation to distribution amplitudes for exclusive processes}

For an exclusive process in QCD in which hadrons are scattered with a large momentum transfer $Q$, the matrix element can be expressed as a factorization formula in which the hard scale $Q$ is separated from the soft hadronic scale $\Lambda$ \cite{Lepage:1980fj}. The hard factor is an amplitude for the hard-scattering of  collinear constituents of each of the hadrons. The soft factor for each hadron is a {\it distribution amplitude} that gives the amplitude for the constituents of the hadron to have specified longitudinal momentum fractions. In the case of a $q \bar q$ meson with large momentum $p$, the distribution amplitude $\phi(x)$ is the amplitude for its constituents to be $q $ and $\bar q$ with momenta $x p$ and $(1-x)p$. The longitudinal momentum fraction $x$ has the range $0\le x \le 1$. The distribution amplitude of the meson can be defined in terms of its light-front wavefunction $\psi(x,\bm{k}_\perp)$ in the light-front gauge  \cite{Lepage:1980fj}:
\begin{equation}
\phi(x) =
\int \frac{d^2k_\perp}{(2 \pi)^2} \psi(x,\bm{k}_\perp) .
\label{eq:phi-psi}
\end{equation}

The distribution amplitude for $t \bar t _{1V} \to H$ in our factorization formula can be interpreted as the conventional distribution amplitude for exclusive processes involving the $t \bar t$ component of the Higgs up to a normalization factor and a factor of $1-\zeta^2$:
\begin{equation}
d_{t \bar t_{1	V} \to H}(\zeta) = \sqrt{N_c}\,m_t \, (1-\zeta^2)\, \phi \big( x=\tfrac12(1+\zeta) \big) .
\label{eq:dtt1VH-psi}
\end{equation}
We have defined the distribution amplitude diagrammatically as the amplitude for producing  the Higgs only from $t$ and $ \bar t$ sources with the Feynman rule  in Eq.~\eqref{eq:Frulett1V} and with eikonal lines extending to future infinity. This definition could be expressed formally as the matrix element of local operators multiplied by Wilson lines. Since the $t$ and $\bar t$ created by the sources are in a color-singlet state and the Higgs is a color singlet, the product of the Wilson lines at future infinity must also be color singlet. The product of the color-triplet  Wilson line and the color-antitriplet  Wilson line therefore 
behaves like a trivial color-singlet  Wilson line as the time approaches future infinity. This ensures that the distribution amplitude is gauge invariant.

The distribution amplitude for $t \bar t _{8T} \to g$ in our factorization formula can be interpreted as a distribution amplitude for exclusive processes involving the $t \bar t$ component of a real gluon. The light-front wavefunction for $t$ and $\bar t$ in a real gluon with polarization vector $\varepsilon$ perpendicular to $n$ and $\bar n$ has a term of the form $\psi(x,\bm{k}_\perp)\, \bm{k}_\perp.\bm{\varepsilon}_\perp$. The $8T$ distribution amplitude can be expressed as an integral of $\psi(x,\bm{k}_\perp)$ over $\bm{k}_\perp$ analogous to that in Eq.~\eqref{eq:phi-psi}. We have defined the distribution amplitude diagrammatically as the amplitude for producing  the gluon only from $t$ and $ \bar t$ sources with the Feynman rule  in Eq.~\eqref{eq:Frulett8T} and with eikonal lines extending to future infinity. This definition could be expressed formally as the matrix element of local operators multiplied by Wilson lines. The product of the color-triplet  Wilson line and the color-antitriplet  Wilson line behaves like a color-octet  Wilson line as the time approaches future infinity. The distribution amplitude is  not gauge invariant. However, as long as the same gauge is used to calculate each piece in the factorization formula in Eq.~\eqref{eq:Ffact}, the gauge dependence will cancel after all pieces are added.

\section{Renormalized Factorization Formula}
\label{sec:Factorization}

The divergences  in the contributions to the LP form factor from the hard, Higgs collinear, gluon collinear, and soft regions  cancel between the different regions. In this Section, we define renormalized contributions to the LP form factor by the minimal subtraction of the poles from dimensional regularization and from the regularization of rapidity divergences. This renormalization procedure is equivalent  to canceling the divergences by moving the divergent terms between different regions. The renormalized contribution from each region depends on the renormalization scheme, but the sum over all regions is scheme independent. The renormalized contributions are combined into a renormalized factorization formula for the LP form factor in which there are no divergences.

\subsection{LP form factor}

The factorization formula for the LP form factor was given in a schematic form in Eq.~\eqref{eq:Ffact}. The explicit form of  the renormalized factorization formula for the LP form factor is 
\begin{eqnarray}
\mathcal{F}^{\text{LP}}(\hat s, m_t^2,m_H^2) &=& \widetilde{\mathcal{F}}_{H+g}(\hat s)
+ \int_{-1}^{+1}\!\!\!\!d \zeta\,  \widetilde{\mathcal{F}}_{t \bar t_{1V}+g}(\zeta) \, d_{t \bar t_{1V} \to H}(\zeta;m_t^2,m_H^2,P.n) 
\nonumber\\
&&\hspace{1.5cm}
+ \int_{-1}^{+1}\!\!\!\!d \zeta\,   \widetilde{\mathcal{F}}_{H+t \bar t_{8T}}(\zeta) \, d_{t \bar t_{8T} \to g}(\zeta;m_t^2,p_3.\bar n) 
+\mathcal{F}_{\text{endpt}}(m_t^2) .
\label{eq:Ffactren}
\end{eqnarray}
All the dependences on physical scales are indicated explicitly by the  arguments  in Eq.~\eqref{eq:Ffactren}. Each of the individual pieces in the factorization formula is given below.

The regularized hard contribution to the LP form factor is given in Eq.~\eqref{eq:F0hardeps}. We define the renormalized contribution from direct production of $H+g$ by minimal subtraction of the poles in $\epsilon$:
\begin{equation}
\widetilde{\mathcal{F}}_{H+g}(\hat s) = 
\frac{g_s^2 y_t}{16 \pi^2} 
\left(  -\frac12 \log^2\frac{-\hat s - i \epsilon}{\mu^2}
 +2\log\frac{-\hat s - i \epsilon}{\mu^2} + \frac{\pi^2}{6}   - 6 \right).
\label{eq:F0hardren}
\end{equation}
With the measure of the dimensionally regularized momentum integral defined in Eq.~\eqref{eq:intq}, the minimal subtraction of the poles in $\epsilon$ corresponds to the modified minimal subtraction ($\overline{\text{MS}}$) renormalization scheme. The renormalized  hard contribution  depends logarithmically on $\hat s$.

The Higgs collinear contribution to the LP form factor is given by the integral over the momentum fraction variable $\zeta$ in Eq.~\eqref{eq:FLPHcoll1V}. The hard form factor for $g^* \to t \bar t_{1V} + g$  is given in Eq.~\eqref{eq:FFtt1Vg}. It reduces to
\begin{equation}
\widetilde{\mathcal{F}}_{t \bar t_{1V}+ g}(\zeta) = - 2 g_s^2 \zeta.
\label{eq:FFtt1Vgren}
\end{equation}
The  distribution amplitude with rapidity regularization is given in Eq.~\eqref{eq:dfragtt1V->Hreg}. We define a renormalized distribution amplitude by minimal subtraction of the ultraviolet poles  in $\eta$ and in $\epsilon$:
\begin{eqnarray}
d_{t \bar t_{1V} \to H}(\zeta) =
\frac{y_t}{32 \pi^2} \zeta
\left[ \log\frac{\mu^2}{m_t^2} \left(\log\frac{P.n}{\nu_1}\,\delta(1-\zeta^2)
+ \frac{1}{(1-\zeta^2)_+} \right)
 -\frac{\log\big(1-(1-\zeta^2)r^2 \big)}{1-\zeta^2}\right].~~~~~~
 \label{eq:dfragtt1V->Hren}
\end{eqnarray}
This distribution amplitude   depends logarithmically on $m_t$ and on $P.n$.

The gluon collinear contribution to the LP form factor is given by the  integral over the momentum fraction variable $\zeta$ in Eq.~\eqref{eq:FLPgcoll8T}. The form factor for $g^* \to H +  t \bar t_{8T}$  is given in Eq.~\eqref{eq:FFHtt8T}. It reduces to
\begin{equation}
\widetilde{\mathcal{F}}_{H+t \bar t_{8T}}(\zeta) = - 2 g_s y_t.
\label{eq:FFHtt8Tren}
\end{equation}
The  distribution amplitude with rapidity regularization is given in Eq.~\eqref{eq:dfragtt8T->greg}. We define a renormalized distribution amplitude by minimal subtraction of the ultraviolet poles  in $\eta$ and in $\epsilon$:
\begin{equation}
d_{t \bar t_{8T} \to g}(\zeta) = 
\frac{g_s}{32 \pi^2}
\log\frac{\mu^2}{m_t^2}
\left( \log\frac{p_3.\bar n}{\nu_3} \delta(1-\zeta^2)+\frac{1}{(1-\zeta^2)_+}\right).
 \label{eq:dfragtt8T->gren}
\end{equation}
This distribution amplitude  depends logarithmically on $m_t$ and on $p_3.\bar n$.

The soft contribution to the LP form factor using rapidity regularization is given in Eq.~\eqref{eq:Fsoftrap}. We define the renormalized endpoint contribution by minimal subtraction of the ultraviolet poles  in $\eta$ and in $\epsilon$:
\begin{equation}
\label{eq:Fsoftren}
\mathcal{F}_{\text{endpt}}(m_t^2) =
\frac{g_s^2 y_t}{16\pi^2}
\left( \frac12 \log^2\frac{\mu^2}{m_t^2} - \log\frac{\mu^2}{m_t^2}\log\frac{\nu^2}{m_t^2} - \frac{\pi^2}{6} \right).
\end{equation}
The endpoint contribution depends logarithmically on $m_t$.

The integrals over $\zeta$ in the factorization formula in Eq.~\eqref{eq:Ffactren} are
\begin{subequations}
\begin{eqnarray}
 \int_{-1}^{+1}\!\!\!\! d \zeta \,\widetilde{\mathcal{F}}_{t \bar t_{1V}+g}(\zeta) \,
d_{t \bar t_{1V} \to H}(\zeta) &=& 
\frac{g_s^2 y_t}{16\pi^2}
\left(  -\log\frac{\mu^2}{m_t^2}\log\frac{P.n}{\nu_1}
+ 2 \log\frac{\mu^2}{m_t^2} \right. 
 \nonumber \\
&&\hspace{1.5cm} \left.
-2\arcsin^2 r -\frac{4\sqrt{1-r^2}}{r} \arcsin r + 4 \right),
\label{eq:FLPHcollren}
\\ \int_{-1}^{+1}\!\!\!\! d \zeta \,\widetilde{\mathcal{F}}_{H+t \bar t_{8T}}(\zeta) \,
d_{t \bar t_{8T} \to g}(\zeta) &=&
\frac{g_s^2 y_t}{16 \pi^2}
\left( - \log\frac{\mu^2}{m_t^2} \log\frac{p_3.\bar n}{\nu_3} \right).
\label{eq:FLPgcollren}
\end{eqnarray}
\end{subequations}
The logarithms of $P.n$ and $p_3.\bar n$ in these two terms combine to give a logarithm of $\hat s$. The last three terms  in the factorization formula in  Eq.~\eqref{eq:Ffactren} depend on the rapidity regularization scales $\nu_1$, $\nu_3$, and $\nu$. The dependence on these scales cancels upon using the  relation between $\nu_1$, $\nu_3$, and $\nu$ in  Eq.~\eqref{eq:rapidregscales}. All four terms  in Eq.~\eqref{eq:Ffactren}  depend on the dimensional regularization scale $\mu$. The dependence cancels when all the terms are added. The sum of the four terms in Eq.~\eqref{eq:Ffactren} reproduces the LP form factor in Eq.~\eqref{eq:FFLP}.

\subsection{Improved Mass Dependence}
\label{sec:massdep}

The LP form factor is an approximation to the full form factor with errors of order $m_t^2/Q^2$ and $m_H^2/Q^2$. It is relatively easy to modify the renormalized factorization formula in Eq.~\eqref{eq:Ffactren} to decrease the errors to order $m_H^2/Q^2$. Since the top quark mass threshold $2 m_t$ is significantly larger than $m_H$, one may be able to improve the accuracy by keeping the leading terms of an expansion in $m_H^2/Q^2$ without expanding  in $m_t^2/Q^2$. This will not change the parametric dependence of the error, which still decreases as $1/Q^2$. However, since the ratio $r = m_H/(2 m_t)$ satisfies  $r^2 \approx 0.13$, one  might hope for an order-of-magnitude decrease in the numerical size of the error. For the subprocess $q \bar q \to H t \bar t$ considered in Ref.~\cite{Braaten:2015ppa}, this hope was not realized.  The error in the leading power in $m_H^2/Q^2$ had the opposite sign as the error in the leading power in $M^2/Q^2$ but approximately the same magnitude. We will show below that for the subprocess $q \bar q \to H g$, there is in fact a significant decrease in the numerical size of the error.

In the factorization formula for the LP form factor in Eq.~\eqref{eq:Ffactren}, the hard form factors are independent of the masses $m_t$ and $m_H$. It is not essential that the hard form factors be independent of $m_t$ and $m_H$, but they must be {\it infrared safe}, which means that they can not have any mass singularities. One can include the top quark mass dependence by taking the hard scale to be $Q\sim P_T, \sqrt{\hat{s}}$ and the soft scale to be $M\sim m_H$, but allowing $m_t$ to be an arbitrary scale that could be order $M$ or order $Q$ or an intermediate scale. Since $m_t$ could be order $Q$, the form factor cannot be expanded in powers of $m_t^2/\hat s$. Since $m_t$ could be order $M$, the form factor cannot be expanded in powers of $m_H^2/m_t^2$. The leading term in an expansion of the form factor in powers of $m_H^2/Q^2$ has an error of order $m_H^2/Q^2$. We denote this approximation to the form factor by $\mathcal{F}^\text{LPH}(\hat s, m_t^2,m_H^2)$. We will show that it can be expressed in the same form as the factorization formula in Eq.~\eqref{eq:Ffactren}, with the only change being in the hard form factor $\widetilde{\mathcal{F}}_{H+g}(\hat s)$. The modified hard form factor depends on $m_t$ and we denote it by $\widetilde{\mathcal{F}}^{(t)}_{H+g}(\hat s, m_t^2)$.

Using the schematic factorization formula in Eq.~\eqref{eq:Ffact}, the hard form factor $\widetilde{\mathcal{F}}_{H+g}(\hat s)$  can be expressed as 
\begin{eqnarray}
 \widetilde{\mathcal{F}}[H+g]  &=& \mathcal{F}^\text{LP}[H+g]  
- \widetilde{\mathcal{F}}[t \bar t_{1V}+g]  \otimes d[t \bar t_{1V} \to H]
\nonumber\\
&& \hspace{2.2cm}
- \widetilde{\mathcal{F}} [H + t \bar t_{8T}]  \otimes d[t \bar t_{8T} \to g]
- \mathcal{F}_\text{endpt}[H+g].
\label{eq:Fhardm=0}
\end{eqnarray}
Since the left side is independent of $m_t$ and $m_H$, we can take the simultaneous  limits $m_t \to 0$ and $m_H \to 0$ on the right side.  All the mass singularities must cancel on the right side to make these simultaneous limits well defined.  The mass singularities also cancel between the LP form factor $\mathcal{F}^\text{LP}$ and the full form factor $\mathcal{F}$, which has the complete dependence on $m_t$ and $m_H$. We can therefore replace $\mathcal{F}^\text{LP}$ inside the limits by $\mathcal{F}$. The resulting expression for the hard form factor is
\begin{eqnarray}
 \widetilde{\mathcal{F}}[H+g]  &=& \Big[ \mathcal{F}[H+g]  
- \widetilde{\mathcal{F}}[t \bar t_{1V}+g]  \otimes d[t \bar t_{1V} \to H]
\nonumber\\
&& \hspace{2.2cm}
- \widetilde{\mathcal{F}} [H + t \bar t_{8T}]  \otimes d[t \bar t_{8T} \to g]
- \mathcal{F}_\text{endpt}[H+g] \Big]_{\substack{m_t\to 0 \\ m_H\to 0}}.
\label{eq:hardFFbb}
\end{eqnarray}
We define the $m_t$-dependent hard form factor $ \widetilde{\mathcal{F}}^{(t)}[H+g]$ simply by removing the limit $m_t \to 0$ from the right side of Eq.~\eqref{eq:hardFFbb}:
\begin{eqnarray}
 \widetilde{\mathcal{F}}^{(t)}[H+g]  &\equiv& \Big[ \mathcal{F}[H+g]
 - \widetilde{\mathcal{F}}[t \bar t_{1V}+g]  \otimes d[t \bar t_{1V} \to H]
\nonumber\\
&& \hspace{2cm}
- \widetilde{\mathcal{F}} [H + t \bar t_{8T}]  \otimes d[t \bar t_{8T} \to g]
- \mathcal{F}_\text{endpt}[H+g]\Big]_{m_H= 0}.
\label{eq:hardcontribution3}
\end{eqnarray}
The only terms  on the right side that depend on $m_H$ are the full form factor $\mathcal{F}$ and the distribution amplitude for $t \bar t_{1V} \to H$. We define the LPH form factor by replacing the hard form factor $\widetilde{\mathcal{F}}[H+g]$ in the schematic  factorization formula in Eq.~\eqref{eq:Ffact} by the $m_t$-dependent hard form factor $\widetilde{\mathcal{F}}^{(t)}[H+g]$ in Eq.~\eqref{eq:hardcontribution3}:
\begin{eqnarray}
\mathcal{F}^\text{LPH}[H+g]   &\equiv&  \widetilde{\mathcal{F}}^{(t)}[H+g] 
+ \widetilde{\mathcal{F}}[t \bar t_{1V}+g]  \otimes d[t \bar t_{1V} \to H]
\nonumber\\
&& \hspace{2.2cm}
+ \widetilde{\mathcal{F}} [H + t \bar t_{8T}]  \otimes d[t \bar t_{8T} \to g]
+ \mathcal{F}_\text{endpt}[H+g].
\label{eq:Fhard(H)}
\end{eqnarray}

We proceed to show that the errors in the LPH form factor defined by Eq.~\eqref{eq:Fhard(H)} are order $m_H^2/\hat s$. The difference between the LPH form factor and the LP form factor  in Eq.~\eqref{eq:Ffact} is $\widetilde{\mathcal{F}}^{(t)}[H+g] -\widetilde{\mathcal{F}}[H+g]$, which is order $m_t^2/\hat s$. Since the error in the LP form factor decreases as $1/\hat s$, the error in the  LPH form factor also decreases as $1/\hat s$. By inserting the expression for $\widetilde{\mathcal{F}}^{(t)}[H+g]$ in Eq.~\eqref{eq:hardcontribution3} into the expression for $\mathcal{F}^\text{LPH}[H+g]$ in Eq.~\eqref{eq:Fhard(H)}, we find that the difference between the LPH form factor and the full form factor can be expressed as
\begin{eqnarray}
\mathcal{F}^\text{LPH}[H+g]  - \mathcal{F}[H+g] &=&  
\left(\mathcal{F}[H+g]\big|_{m_H= 0}  - \mathcal{F}[H+g] \right)
\nonumber\\
&& \hspace{0cm}
+ \widetilde{\mathcal{F}}[t \bar t_{1V}+g]  \otimes \left(d[t \bar t_{1V} \to H] - d[t \bar t_{1V} \to H]\big|_{m_H= 0} \right).
\label{eq:FFdiff}
\end{eqnarray}
The right side is 0 for $m_H=0$.  Thus the error in the LPH form factor is order $m_H^2/\hat s$.

The expression for the $m_t$-dependent hard form factor $\widetilde{\mathcal{F}}^{(t)}$ in Eq.~\eqref{eq:hardcontribution3} seems to require calculating the full form factor $\mathcal{F}$ and then taking the limit $m_H \to 0$. If this were true, the LPH form factor would have no calculational advantage over the full form factor. It would require a calculation involving all three scales $\hat s$, $m_t$, and $m_H$. However the $m_t$-dependent hard form factor can be calculated more easily by not taking the limit $m_H \to 0$, but instead setting $m_H=0$ from the beginning. The two terms on the right side of Eq.~\eqref{eq:hardcontribution3} that depend on $m_H$ are finite if $m_H=0$. Thus $\widetilde{\mathcal{F}}^{(t)}$ can be obtained by calculations that  involve only the two scales $\hat s$ and $m_t$. For some other processes, such as  double-Higgs  production through a virtual Higgs, the limit  $m_H\to 0$ in the equation analogous to Eq.~\eqref{eq:hardcontribution3} produces additional infrared divergences. The calculation can still be carried out with fewer scales by setting $m_H=0$ from the beginning and using dimensional regularization to regularize the additional infrared divergences. After the subtractions analogous to those in Eq.~\eqref{eq:hardcontribution3},  these additional infrared divergence must cancel.

In the schematic expression for the $m_t$-dependent hard form factor $\widetilde{\mathcal{F}}^{(t)}$ in Eq.~\eqref{eq:hardcontribution3}, the first term on the right side  is the $m_H=0$ form factor, which is given in Eq.~\eqref{eq:FmH=0}. The three subtraction terms in Eq.~\eqref{eq:hardcontribution3} are the Higgs collinear contribution  with $m_H=0$, the gluon collinear contribution, and the soft contribution. The sum of the three regularized contributions using rapidity regularization is given in Eq.~\eqref{eq:Fcollsoftreg}. Thus the regularized $m_t$-dependent hard form factor is the difference between Eqs.~\eqref{eq:FmH=0} and \eqref{eq:Fcollsoftreg}. The renormalized $m_t$-dependent hard form factor can be defined by the minimal subtraction of the poles in $\epsilon$:
\begin{eqnarray}
\widetilde{\mathcal{F}}_{H+g}^{(t)}(\hat s,m_t^2) &=& 
 \frac{g_s^2 y_t}{16 \pi^2} 
\left\{   
 2\,  \frac{\hat s + 4 m_t^2}{\hat s}  \arcsin^2 z + 4 \frac{\sqrt{1-z^2}}{z} \arcsin z   \right.
\nonumber\\
&&\hspace{1.5cm} \left.
- \frac{1}{2}\log^2\frac{\mu^2}{m_t^2}
+ \log\frac{\mu^2}{m_t^2}\left(\log\frac{-s-i\epsilon}{m_t^2}-2\right)
+ \frac{\pi^2}{6}  -  6 \right\},
\label{eq:F0hardrenmt}
\end{eqnarray}
where $z=[(\hat s + i \epsilon)/4 m_t^2]^{1/2}$.

The explicit form of the   factorization formula for the LPH form factor in Eq.~\eqref{eq:Ffactren} is
\begin{eqnarray}
\mathcal{F}^{\text{LPH}}(\hat s, m_t^2,m_H^2) &\equiv& \widetilde{\mathcal{F}}_{H+g}^{(t)}(\hat s,m_t^2)
+ \int_{-1}^{+1}\!\!\!\!d \zeta\,  \widetilde{\mathcal{F}}_{t \bar t_{1V}+g}(\zeta) \, d_{t \bar t_{1V} \to H}(\zeta;m_t^2,m_H^2,P.n) 
\nonumber\\
&&\hspace{1.5cm}
+ \int_{-1}^{+1}\!\!\!\!d \zeta\,   \widetilde{\mathcal{F}}_{H+t \bar t_{8T}}(\zeta) \, d_{t \bar t_{8T} \to g}(\zeta;m_t^2,p_3.\bar n) 
+\mathcal{F}_{\text{endpt}}(m_t^2) .~
\label{eq:FfactLPH}
\end{eqnarray}
This can be expressed as the sum of the LP form factor   in Eq.~\eqref{eq:FFLP} and the difference between the hard form factors $\widetilde{\mathcal{F}}_{H+g}^{(t)}$ in Eq.~\eqref{eq:F0hardrenmt} and $\widetilde{\mathcal{F}}_{H+g}$  in Eq.~\eqref{eq:F0hardren}. The explicit result for the LPH form factor is
\begin{eqnarray}
\mathcal{F}^{\text{LPH}}(\hat s,m_t^2,m_H^2)
&=& \frac{g_s^2 y_t}{16\pi^2}
\left\{
2 \,  \frac{\hat s + 4 m_t^2}{\hat s} \arcsin^2 z + 4 \frac{\sqrt{1-z^2}}{z} \arcsin z   \right.
\nonumber\\
&&\hspace{2cm}
\left. -2 \arcsin^2 r-\frac{4\sqrt{1-r^2}}{r}\arcsin r-2\right\},
\label{eq:FFLPmt}
\end{eqnarray}
where $z=[(\hat s + i \epsilon)/4 m_t^2]^{1/2}$.

\subsection{Comparison with full form factor}
\label{sec:compare}

\begin{figure}
\includegraphics[width=16cm]{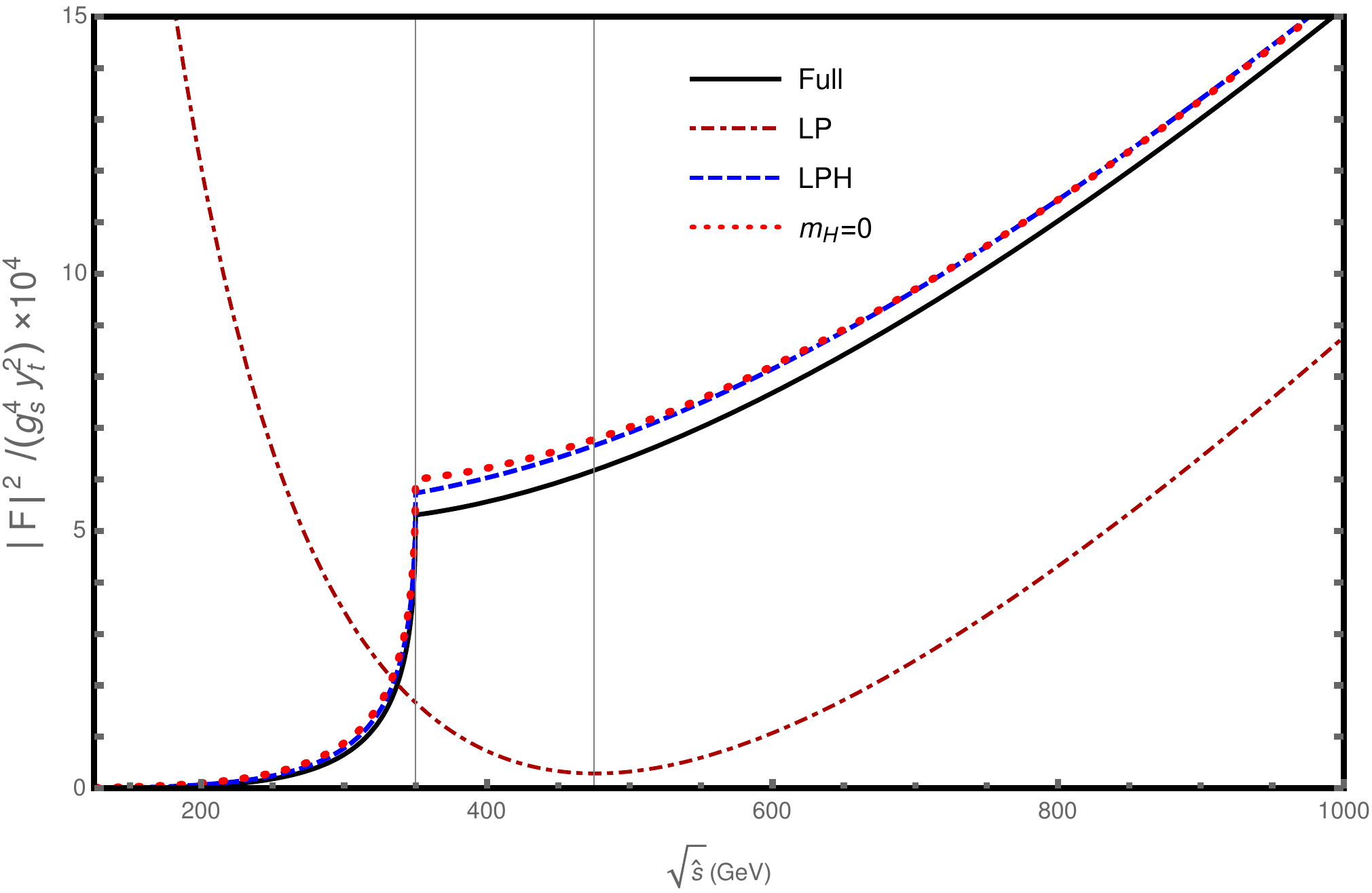}
\caption{Form factors for $q\bar{q}\to H+g$ as functions of the center-of-mass energy $\sqrt{\hat{s}}$: the full form factor $|\mathcal{F}|^2$ (solid curve), the $m_H=0$ form factor (dotted curve), the LP form factor (dot-dashed curve), and the LPH form factor (dashed curve). The two vertical lines mark the $t\bar{t}$ threshold $2m_t$ and the $t\bar{t}H$ threshold $2m_t+m_H$.
\label{fig:FullvsLP}}
\centering
\end{figure}

In Fig.~\ref{fig:FullvsLP}, we compare three approximations to the form factor for $q\bar{q}\to H+g$ at LO. The full form factor $\mathcal{F}(\hat s,m_t^2,m_H^2)$ is given in Refs.~\cite{Ellis:1987xu, Baur:1989cm}. The approximations are 
\begin{itemize}
\item
the $m_H=0$ form factor $\mathcal{F}(\hat s,m_t^2,0)$  in Eq.~\eqref{eq:FmH=0}, which is obtained by setting $m_H=0$ in the full form factor,
\item
the LP form factor $\mathcal{F}^\text{LP}(\hat s,m_t^2,m_H^2)$ in Eq.~\eqref{eq:FFLP}, which is leading power in $m_t^2/\hat s$ and $m_H^2/\hat s$,
\item
the LPH form factor $\mathcal{F}^{\text{LPH}}(\hat s,m_t^2,m_H^2)$  in Eq.~\eqref{eq:FFLPmt}, which is leading power in $m_H^2/\hat s$ only.
\end{itemize}
We set $m_H=125$~GeV and $m_t=175$~GeV. The squares of the absolute values of the form factors are shown as functions of the center-of-mass energy $\sqrt{\hat{s}}$, which ranges  from the threshold $m_H$ for producing the Higgs to 1~TeV. The $m_H=0$ form factor  and the LPH form factor have the same qualitative behavior as the full form factor. The  LPH form factor seems to provide a little better approximation to the full form factor
than the $m_H=0$ form factor. At the $t \bar t $ threshold, the percentage errors in the absolute squares of the form factors are about 8\% for $|\mathcal{F}^\text{LPH}|^2$ and about 14\% for  $|\mathcal{F}^{m_H=0}|^2$. The error in $|\mathcal{F}^{m_H=0}|^2$ becomes smaller than the error in $|\mathcal{F}^\text{LPH}|^2$ when $\sqrt{\hat s}$ increases above 0.8~TeV. The LP form factor has a completely different qualitative behavior from the full form factor and it provides a very poor approximation in the range of $\sqrt{\hat{s}}$ shown in  Fig.~\ref{fig:FullvsLP}.

\begin{figure}
\includegraphics[width=8cm]{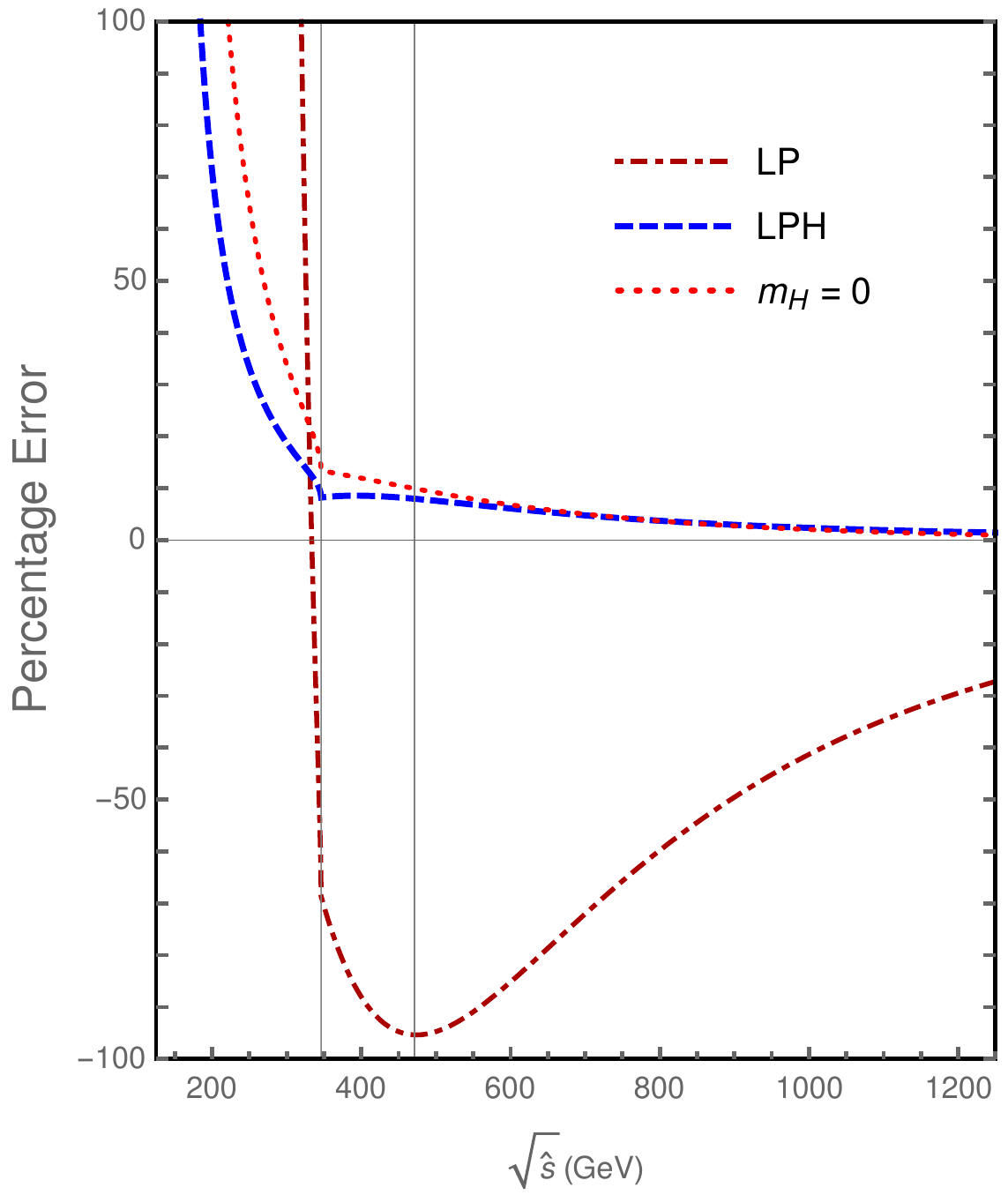}
\includegraphics[width=7.75cm]{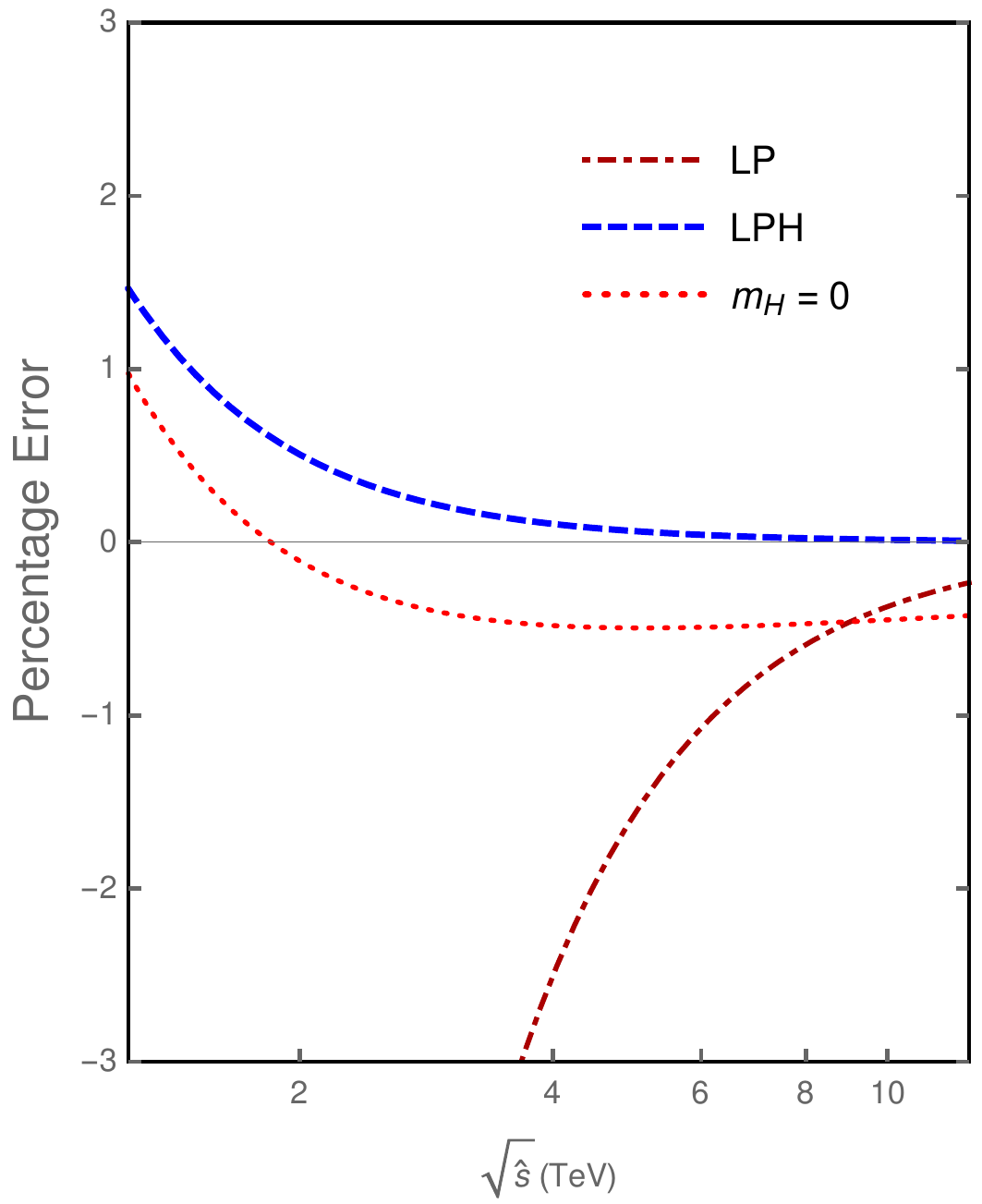}
\caption{Percentage errors in form factors for $q\bar{q}\to H+g$ as functions of the center-of-mass energy $\sqrt{\hat{s}}$: the $m_H=0$ form factor (dotted curve), the LP form factor (dot-dashed curve), and the LPH form factor (dashed curve). The ranges of $\sqrt{\hat{s}}$ are $m_H$ to $10m_H$ on a linear scale (left panel) and from $10m_H$ to $100m_H$ on a log scale (right panel). The two vertical lines mark the $t\bar{t}$ threshold $2m_t$ and the $t\bar{t}H$ threshold $2m_t+m_H$.
\\~
\label{fig:err}}
\centering
\end{figure}

In Fig.~\ref{fig:err}, we compare the percentage errors in the three approximations to the full form factor. The percentage error in the  absolute square of a form factor is defined as the difference from $|\mathcal{F}|^2$ divided by $|\mathcal{F}|^2$. The percentage errors are shown as functions of the center-of-mass energy $\sqrt{\hat{s}}$, which ranges  from $m_H$ to $100\,m_H$. The right panel of Fig.~\ref{fig:err} shows that the error in $|\mathcal{F}^{m_H=0}|^2$ changes sign near 1.8 TeV. It becomes larger than the error in $|\mathcal{F}^\text{LPH}|^2$ above 2.5 TeV. That there is a range of $\hat s$ in which the error in $|\mathcal{F}^{m_H=0}|^2$ is smaller than the error in $|\mathcal{F}^\text{LPH}|^2$ is just a fortuitous consequence of the change in sign of the error in $|\mathcal{F}^{m_H=0}|^2$. The right panel of Fig.~\ref{fig:err} shows that the percentage error in $|\mathcal{F}^{m_H=0}|^2$ decreases to a minimum of $-0.49$\% at $\sqrt{\hat{s}}= 4.8$~TeV, and then approaches  zero very slowly. The reason for  the slow approach to 0 is that the absolute error is a constant at large $\hat{s}$, while the full form factor $|\mathcal{F}|^2$ increases as $\log^4(\hat{s}/m_t^2)$ when $\hat{s}$ is large. The right panel of Fig.~\ref{fig:err} shows that both $\mathcal{F}^\text{LP}$ and $\mathcal{F}^\text{LPH}$ approach the full form factor at large $\sqrt{\hat{s}}$, which is expected since their errors decrease as $1/\hat{s}$. However, $\mathcal{F}^\text{LPH}$ approaches the full form factor much more rapidly. The percentage error in $|\mathcal{F}^\text{LPH}|^2$ drops to $5\%$ at $\sqrt{\hat{s}}\sim 0.68$ TeV. The percentage error in $|\mathcal{F}^\text{LP}|^2$ drops to $5\%$ at $\sqrt{\hat{s}}\sim 2.9$ TeV.

Figs.~\ref{fig:FullvsLP} and \ref{fig:err} seem to indicate that the $m_H=0$ form factor is almost as good an approximation as the LPH form factor. Since it is significantly easier to calculate the $m_H=0$ form factor, one might question whether the additional effort is worthwhile. It is important to emphasize that the error in the $m_H=0$ form factor approaches a constant at large $\sqrt{\hat s}$:
\begin{equation}
\mathcal{F}(\hat s,m_t^2,0) - \mathcal{F}(\hat s,m_t^2,m_H^2) 
\longrightarrow \frac{g_s^2 y_t}{16\pi^2} \left\{ 2\arcsin^2 r + \frac{4\sqrt{1-r^2}}{r}\arcsin r-4\right\}.
\label{eq:err_mHeq0}
\end{equation}
This absolute error  is order $r^2$, which is approximately 0.13. The percentage error at large $\sqrt{\hat s}$ in the right panel of Fig.~\ref{fig:err} is about $0.5\%$. The reason this is so small is that the full form factor in the denominator of the percentage error increases as $\log^2({\hat s}/m_t^2)$. 

There may be other processes for which the error of order $r^2$ from the $m_H=0$ approximation is disastrous. One such case is when there is interference between amplitudes that makes the differential cross section small. Bear in mind that the most important motivation for accurate calculations of Higgs production at large transverse momentum is the search for new physics beyond the standard model. An error of order $r^2$ could overwhelm a small signal  for new physics. Moreover, there are terms dropped in the $m_H=0$ approximation, such as $(m_H^2/m_t^2)\log(\hat{s}/m_t^2)$, that depend on kinematic variables. Dropping such terms may change the shape of the differential cross section. Finally there are processes, such as  double-Higgs  production through a virtual Higgs, in which the $m_H=0$ approximation gives a divergent cross section. The LPH approximation is much more reliable, because the absolute error  of order $m_H^2/\hat{s}$ approaches  zero rapidly as $\hat{s}$ increases. Even when multiplied with powers of logarithms, such as $(m_H^2/\hat{s})\log(\hat{s}/m_t^2)$, the power suppression is strong enough to make the omitted terms  small at large $\hat{s}$. The LPH approximation can even be applied to cross sections that diverge in the $m_H\to 0$ limit, such as  double-Higgs production through a virtual Higgs.

\section{Summary and Outlook}
\label{sec:Discuss}

In this work, we applied factorization methods  developed for exclusive production of hadrons in QCD to high-energy  exclusive production of the Higgs boson. This factorization approach can also be applied to  high-energy  exclusive production of other elementary particles, such as the weak gauge bosons $W^\pm$ and $Z^0$. The formalism for exclusive production of hadrons in QCD  is well developed. It can be readily generalized to high-energy exclusive  production of an elementary particle, but there are important differences. One difference is that the elementary particle can be produced directly by hard interactions, but there is no analogous contribution to the exclusive production of a hadron. Another difference is that  all the pieces in the factorization formula for the high-energy exclusive production of an elementary particle are perturbatively calculable. As a result, an all-order proof of a factorization formula is not essential in order to apply it to the exclusive production of an elementary particle. In this sense, the factorization formalism  is simpler than for exclusive production of hadrons in QCD. 

We applied factorization  to Higgs production at large transverse momentum through a top-quark loop. Production of the Higgs at large $P_T$ is complicated by the multiple energy scales: the hard kinematic scales $P_T,\hat s^{1/2} \sim Q$ and the soft mass scales $m_t,m_H \sim M$. Factorization can be used to separate the scales $M$ and $Q$ and expand in powers of $M^2/Q^2$. To illustrate the factorization approach, we applied it to the subprocess $q \bar q \to H g$ at LO in $\alpha_s$ and at the leading power in $M^2/Q^2$. The matrix element for this subprocess is determined by the form factor $\mathcal{F}(\hat s, m_t^2,m_H^2)$ defined in Eq.~\eqref{eq:Fdef}. We defined the {\it leading-power} (LP) form factor $\mathcal{F}^\text{LP}$ as the leading terms in the expansion of $\mathcal{F}$ in powers of $M^2/\hat{s}$. A factorization formula for the LP form factor in which the scales $Q$ and $M$ are separated  
is given schematically in Eq.~\eqref{eq:Ffact}. The explicit renormalized form of the factorization formula is given in Eq.~\eqref{eq:Ffactren}. We also defined the LPH form factor $\mathcal{F}^\text{LPH}$ as the leading terms in the expansion of $\mathcal{F}$ in powers of $m_H^2/\hat{s}$, keeping all dependence on $m_t$ that is not suppressed by $m_H^2/\hat{s}$. The goal of this paper was to obtain these approximations to the full form factor through diagrammatic calculations that each involves fewer scales than the calculation of the full form factor.

The LP form factor can be calculated using the method of regions. The relevant regions and the corresponding contributions to the LP form factor were labeled {\it hard}, {\it Higgs collinear}, {\it gluon collinear}, and {\it soft}. The method of regions introduces rapidity divergences in addition to the infrared and ultraviolet divergences that can be regularized by dimensional regularization.  We regularized the rapidity divergences using analytic regularization in Section~\ref{sec:AnalReg} and using rapidity regularization in Section~\ref{sec:RapidReg}. With analytic regularization, the rapidity divergences appear naturally as infrared divergences. With  rapidity regularization, the rapidity divergences appear naturally as ultraviolet divergences after zero-bin subtractions. With analytic regularization, the only kinematic variable the contribution from each region can depend on is $\hat s$. With  rapidity regularization, 
the Higgs collinear contribution depends logarithmically on $P.n$, where $P$ is the momentum of the Higgs, and the gluon collinear contribution depends  logarithmically  on $p_3.\bar n$, where $p_3$ is the momentum of the gluon. These logarithms combine to give a logarithm of $\hat s$. One complication of  rapidity regularization is that it requires a constraint on the rapidity regularization scales in the various regions that is given in Eq.~\eqref{eq:rapidregscales}. This constraint was derived by comparing with results using analytic regularization. It would be preferable to deduce this constraint directly using only rapidity regularization.

The factorization formula given schematically in Eq.~\eqref{eq:Ffact} separates the hard scales $Q$ and the soft scales $M$. The hard contribution to the LP form factor depends only on the hard scale $Q$. It can be regularized with dimensional regularization only, and it is given in Eq.~\eqref{eq:F0hardeps}. With rapidity regularization, the soft contribution to the LP form factor depends only on the scale $M$ and it  is given in Eq.~\eqref{eq:Fsoftrap}. The Higgs collinear and gluon collinear contributions depend on both the hard scale $Q$ and the  soft scale $M$. In Sections~\ref{sec:Factorgcoll} and Section~\ref{sec:FactorHcoll}, we separated the scales $Q$ and $M$ in the Higgs collinear and  gluon collinear contributions. Each collinear contribution can be factorized into the integral of the product of a hard form factor and a distribution amplitude. In Section~\ref{sec:Fragmentation}, we showed how the hard form factors and the distribution amplitudes could be obtained through separate diagrammatic calculations. In the Higgs collinear contribution, the hard form factor for $t \bar t_{1V}+g$ is given in Eq.~\eqref{eq:FFtt1Vg}, and the distribution amplitude  for $t \bar t_{1V} \to H$ with rapidity regularization is given in Eq.~\eqref{eq:dfragtt1V->H}. In the gluon collinear contribution, the hard form factor for $H+t \bar t_{8T}$ is given in Eq.~\eqref{eq:FFHtt8T}, and the distribution amplitude for $t \bar t_{8T} \to g$ with rapidity regularization is given in Eq.~\eqref{eq:dfragtt8T->g}.

In the schematic factorization formula  in Eq.~\eqref{eq:Ffact}, the pieces that have poles in the regularization parameters are  the hard form factor $\widetilde{\mathcal{F}}[H+g]$, the  distribution amplitude for $d[t \bar t_{1V} \to H]$, the  distribution amplitude for $d[t \bar t_{8T} \to g]$, and the endpoint form factor $\mathcal{F}_\text{endpt}[H+g]$. The poles in the dimensional regularization parameter $\epsilon$ and the rapidity regularization parameter $\eta$ 
cancel in the sum of all four terms in the factorization formula. Given the cancellation of the poles, they can alternatively be eliminated by subtractions applied to each of the divergent pieces of the factorization formula. Minimal subtraction of the poles in $\eta$ and the poles in $\epsilon$ was used to define the finite pieces in the renormalized  factorization formula in Eq.~\eqref{eq:Ffactren}. The hard form factor  $\widetilde{\mathcal{F}}_{H+g}$  is given in Eq.~\eqref{eq:F0hardren}. The distribution amplitudes  for $t \bar t_{1V} \to H$ and $t \bar t_{8T} \to g$ are given in Eqs.~\eqref{eq:dfragtt1V->Hren} and \eqref{eq:dfragtt8T->gren}. The endpoint form factor  $\mathcal{F}_\text{endpt}$ is given in Eq.~\eqref{eq:Fsoftren}. With  rapidity regularization, the poles in $\eta$ are ultraviolet divergences. The subtraction of the poles in each of the regularized pieces of the factorization formula can therefore be interpreted as a renormalization procedure. It could be expressed in terms of the renormalization of an operator in an effective field theory that resembles  soft collinear effective field theory in QCD. We made no attempt to develop the effective-field-theory formalism.

The LP form factor $\mathcal{F}^\text{LP}$ is a good approximation to the full form factor only at extremely large $\hat{s}$. The error is of order $M^2/\hat{s}$, where $M\sim m_t, m_H$, so the error decreases to 0 as $\hat{s}$ increases. As shown in Fig.~\ref{fig:err}, the rapid decrease in the error in $|\mathcal{F}^\text{LP}|^2$ does not begin until $\sqrt{\hat{s}}$ is well above the $t\bar{t}H$ threshold. The percentage error does not decrease to less than 5\% until $\sqrt{\hat s} > 3$~TeV. Thus the LP form factor has no practical use at LHC  energies. The LPH form factor $\mathcal{F}^\text{LPH}$ was obtained  by a simple modification of the factorization formula that requires   additional  calculations with $m_H=0$. Thus it can also be obtained through calculations that involve fewer scales than  the  full form factor. The error in $|\mathcal{F}^\text{LPH}|^2$ is order $m_H^2/\hat{s}$. As shown in Fig.~\ref{fig:err}, the LPH form factor has the same qualitative behavior as the full form factor. The percentage error is only 8\% already at the $t \bar t$ threshold 0.35~TeV, and it decreases to less  than $5\%$  at 0.7~TeV. 

We illustrated our factorization approach by applying it to the form factor $\mathcal{F}$ for the subprocess $q \bar q \to H g$, which is a function of a positive Mandelstam variable $\hat s$. The form factors for the  subprocesses $g\, q \to H+q$ and $g\, \bar q  \to H+\bar q$ are given by the same function $\mathcal{F}$ with $\hat s$ analytically continued to a negative Mandelstam variable $\hat{t}$. The factorization formula involves a resolved-gluon amplitude for $g \to t \bar t$ instead of the distribution amplitude for  $t \bar t \to g$. The matrix elements for the subprocesses $g g \to H+g$ at LO are determined by four form factors that are functions of three Mandelstam variables. Three of the four form factors are given by the same function with different permutations of the three Mandelstam variables. It should be possible to express the LP contributions to these form factors as factorization formulas analogous to Eq.~\eqref{eq:Ffactren} in which all the pieces are calculated analytically.

Our factorization approach can be used to simplify calculations of the Higgs $P_T$ distribution at higher orders in $\alpha_s$. The NLO calculation of the form factor for $q \bar q \to H+g$ would require calculating each of the pieces in the factorization formula in Eq.~\eqref{eq:Ffactren} to NLO. The NLO calculations of the hard form factors $\widetilde{\mathcal{F}}_{H+g}$, $\widetilde{\mathcal{F}}_{t \bar t_{1V} +g}$, and $\widetilde{\mathcal{F}}_{H+t \bar t_{8T}}$ require straightforward perturbative QCD calculations with massless quarks. The NLO calculation of the endpoint form factor $\mathcal{F}_\text{endpt}$ with rapidity regularization may be more difficult, since it may have nontrivial dependence on the scale $Q$ through the hard form factor $\widetilde{\mathcal{F}}_{t + \bar t}$ for producing $t+\bar t$. The NLO calculation of the distribution amplitudes  for $t \bar t_{1V} \to H$ and  for $t \bar t_{8T} \to g$ may be the most challenging steps in the NLO calculation of the LP form factor. At NLO, there may be additional terms in the factorization formula associated with other double-parton channels, such as $t \bar t_{1S}$, $t \bar t_{1T}$, $t \bar t_{8S}$, and $t \bar t_{8V}$. These additional terms would require only LO calculations.

One advantage of the factorization approach is that it is in principle systematically improvable. HEFT can be used to systematically improve predictions for Higgs production at $P_T < 2 m_t$ by including operators of  dimension 7 and higher in the HEFT Lagrangian. The factorization approach could in principle be used to systematically improve predictions for Higgs production at large $P_T $  by including higher powers in the expansion in $M^2/Q^2$. The straightforward factorization methods used to obtain the LP form factor provide improvements for $P_T > 2 m_t$. The improvement used to obtain the LPH form factor expand the region of validity to $P_T >m_H$. There is an overlap region of $P_T$ between $m_H$ and $2 m_t$ where HEFT and the LPH factorization approach both apply. By combining these two approaches, we should be able to obtain systematically improved approximations to the $P_T$ distribution over the entire range of $p_T$.

We derived our factorization formula diagrammatically. It could be derived more formally using effective field theory methods analogous to those used in soft collinear effective field theory in QCD. The individual pieces in the factorization formula could all be expressed in terms of matrix elements of operators in the effective field theory. These formal definitions could be useful in the calculation of the form factor to higher orders in $\alpha_s$. They would also facilitate the all-order resummation of potentially large logarithms by solving renormalization group equations. The LP and LPH form factors for $q\bar{q}\to H+g$ involve single and double logarithms of $\hat s/m_t^2$. The resummation of these logarithms  is not important at the LHC, but it could be necessary  at a future 100~TeV proton-proton collider. The resummation of logarithms of $\hat s/m_b^2$ could be important for Higgs production through a bottom quark loop at the LHC.

     
\acknowledgments
The work of EB and HZ was supported in part by the Department of Energy under grant DE-SC0011726. JWZ was supported in part by the Natural Science Foundation of China under Grant No.11347024, the Natural Science Foundation Project of CQCSTC under Grant No. 2014jcyjA00030, the Scientific and Technological Research Program of Chongqing Municipal Education Commission under Grant No. KJ1401313, the Research Foundation of Chongqing University of Science and Technology under Grant No. CK2016Z03. HZ would like to thank Jian-Wei Qiu, Xiaohui Liu and Yan-Qing Ma for beneficial discussions. We acknowledge the use of FeynCalc \cite{Mertig:1990an,Shtabovenko:2016sxi} in this research.


\appendix

\section{Calculations of Distribution Amplitude}
\label{app:FragAmp}

In this Appendix, we calculate the function  $d(\zeta)$ that appears in the distribution amplitude for $t \bar t_{1V} \to H$ using analytic regularization and using rapidity regularization. We also give expressions for  $d(\zeta)/(1-\zeta^2)$ in  which the poles in the regularization parameters are explicit. 

\subsection{Analytic regularization}
\label{app:FragAmpanal}

The function $d(\zeta)$ is defined using analytic regularization in Eq.~\eqref{eq:dzetadef}:
\begin{equation}
d(\zeta) = -i \int_q
\frac{\delta(\zeta - 2q.n/P.n)}
{[(\tfrac12 P+q)^2 - m_t^2  + i \epsilon]^{1+\delta_1} [(\tfrac12 P-q)^2-m_t^2  + i \epsilon]^{1+\delta_1} },
 \label{eq:ddef}
\end{equation}
where $P$ is the 4-momentum of the Higgs, $n$ is an arbitrary light-like four-vector, and $\delta_1$ is the analytic regularization parameter. The measure for the  integral over $q$ is given by Eq.~\eqref{eq:intreg} with $\delta_2 = \delta_1$ and $\delta_3=0$.  

To calculate the integral  in Eq.~\eqref{eq:ddef}, we begin by combining the denominators using a Feynman parameter:
\begin{equation}
d(\zeta) =
\frac{\Gamma(2+2\delta_1)}{\Gamma^2(1+\delta_1)}\int_0^1 \!\!dx \,x^{\delta_1}\, (1-x)^{\delta_1}\, (-i) \!\!\int_q
\frac{\delta(\zeta - 2q.n/P.n)}
{[q^2  +(2x-1) P.q -m_t^2 + m_H^2/4 + i \epsilon]^{2+2\delta_1} }.
 \label{eq:dint1}
\end{equation}
After the shift $q \to q -(x - \frac12) P$ in the loop momentum, this reduces to
\begin{equation}
d(\zeta) =
\frac{\Gamma(2+2\delta_1)}{\Gamma^2(1+\delta_1)}\int_0^1 \!\!dx \,x^{\delta_1}\, (1-x)^{\delta_1}\, (-i) \!\!\int_q
\frac{\delta(\zeta - 1+2x-2q.n/P.n)}
{[q^2   -m_t^2 + x(1-x)m_H^2 + i \epsilon]^{2+2\delta_1} }.
 \label{eq:dint2}
\end{equation}
We will show below that we can set $q.n =0$ in the argument of the delta  function, after which  the delta  function can  be pulled outside the momentum integral. The momentum integral can then be evaluated analytically. Finally the delta function can be used to evaluate the integral over $x$. The result is
\begin{equation}
d(\zeta) = \frac{1}{32 \pi^2}
\left[ \frac{\mu^2}{m_t^2} \right]^{\epsilon} \left[ \frac{\nu^2}{m_t^2} \right]^{2\delta_1}
\frac{(1)_{\epsilon+2\delta_1}}{(1)_{\epsilon}(1)_{\delta_1}(1)_{\delta_1}} \frac{1}{\epsilon+2\delta_1}
\left( \frac{1-\zeta^2}{4} \right)^{\delta_1}
\big[ 1 - (1-\zeta^2) (r^2 + i \epsilon) \big]^{-\epsilon-2\delta_1},
 \label{eq:dint3}
\end{equation}
where $r = m_H/2m_t$. There is an implied constraint  $-1 \le \zeta \le +1$ that comes from the integral over $x$.

We now verify that we can set $q.n =0$ in the argument of the delta  function in Eq.~\eqref{eq:dint2}. This is a special case of the more general identity
\begin{equation}
\int_q \frac{f(q.n)} {[q^2   -M^2 + i \epsilon]^p }
= f(0) \int_q \frac{1} {[q^2   -M^2 + i \epsilon]^p }.
 \label{eq:intid}
\end{equation}
It is convenient to use light-cone variables $q_+$, $q_-$, and $\bm{q}_\perp$ for the 4-momentum $q$, where $q.n = q_+$. The integral over $q_-$ has the form
\begin{equation}
\int dq_- \frac{f(q_+)} {[q_+ q_-  - q_\perp^2  -M^2 + i \epsilon]^p}
= A (q_\perp)\, f(0) \, \delta(q_+),
 \label{eq:intq-}
\end{equation}
where $A (q_\perp)$ is a function of $q_\perp$. If $q_+ \neq 0$, the integral over $q_-$ on the left side can be shown to vanish by closing the integration contour in a half-plane that is determined by $\epsilon$ and depends on the sign of $q_+$. If $q_+ = 0$, the integral over $q_-$ is infinite, because the integrand does not depend on $q_-$. The integral is actually proportional to a delta function of $q_+$, as indicated on the right side of Eq.~\eqref{eq:intq-}.
Thus the factor of $f(q_+)$ on the left side  of Eq.~\eqref{eq:intq-} can be pulled outside the integral as a prefactor $f(0)$. The coefficient $A (q_\perp)$ can then be determined by setting $f(q_+) = 1$ in Eq.~\eqref{eq:intq-} and integrating both sides over $q_+$:
\begin{equation}
A (q_\perp) = \int dq_+ \int dq_- \frac{1} {[q_+ q_-  - q_\perp^2  -M^2 + i \epsilon]^p}.
 \label{eq:A-int}
\end{equation}
Integrating both sides of  Eq.~\eqref{eq:intq-} over $q_+$, inserting the expression for $A (q_\perp)$ in   Eq.~\eqref{eq:A-int},  and then integrating both sides of the equation over $\bm{q}_\perp$ gives the identity in Eq.~\eqref{eq:intid}.

\subsection{Explicit poles in the regularization parameters}

If the function $d(\zeta)$ in Eq.~\eqref{eq:dint3} is divided by $1-\zeta^2$, the poles in the regularization parameters $\delta_1$ and $\epsilon$ can be made explicit. The function $d(\zeta)$ in Eq.~\eqref{eq:dint3} has a factor of $[(1 - \zeta^2)/4]^{\delta_1}$. To make the poles explicit, we use the expansion
\begin{eqnarray}
\frac{1}{1 - \zeta^2} \bigg( \frac{1 - \zeta^2}{4} \bigg)^{\!\!\delta_1} &=&
\frac{1}{\delta_1} \frac{(1)_{\delta_1} (1)_{\delta_1}}{(1)_{2\delta_1}} \, \delta\big(1 - \zeta^2\big) + \frac{1}{(1-\zeta^2)_+} 
\nonumber\\
&& \hspace{1cm}
+ \delta_1 \bigg( \frac{\log(1-\zeta^2)-2\log 2}{1-\zeta^2}\bigg)_{\!\!\!+} + O(\delta_1^2).
 \label{eq:expandplus}
\end{eqnarray}
The distributions in $\zeta$ on the right side of Eq.~\eqref{eq:expandplus} can be defined by specifying the integral of the product of the distribution and a smooth function $f(\zeta)$ over the closed interval   $-1 \le \zeta \le +1$. The Dirac delta function can be defined by
\begin{equation}
\int_{-1}^{+1}\!\!d \zeta \, \delta\big(1 - \zeta^2\big) \, f(\zeta) \equiv
\frac{f(1)+f(-1)}{2}.
 \label{eq:intdelta}
\end{equation}
The integral is 0 if $f(\zeta)$ is an odd function of $\zeta$.  The plus distributions are defined by
\begin{equation}
\int_{-1}^{+1}\!\!d \zeta \, g(\zeta)_+ \, f(\zeta) \equiv
\int_{-1}^{+1}\!\!d \zeta \, g(\zeta)  \,  \frac{f(\zeta)+f(-\zeta)-f(1)-f(-1)}{2}.
 \label{eq:intplus}
\end{equation}
The integral is 0 if $f(\zeta)$ is a constant or an odd function of $\zeta$. All the higher order terms in the Laurent expansion in Eq.~\eqref{eq:expandplus} are plus distributions. The prefactor of the Dirac delta function in Eq.~\eqref{eq:expandplus} can be verified by integrating both sides of the equation over $\zeta$ and using the fact that the integrals of the plus distributions are 0.

After dividing the function $d(\zeta)$ in Eq.~\eqref{eq:dint3} by $1 - \zeta^2$, the expansion in Eq.~\eqref{eq:expandplus} can be inserted. The Laurent expansion in $\delta_1$ followed by the Laurent expansion in $\epsilon$  gives
\begin{eqnarray}
\frac{d(\zeta)}{1-\zeta^2} =
\frac{1}{32 \pi^2} \left[ \frac{\mu^2}{m_t^2} \right]^{\epsilon} \left[ \frac{\nu^2}{m_t^2} \right]^{2 \delta_1}
\Bigg\{
 \left(\frac{1}{\epsilon \delta_1}  - \frac{2}{\epsilon^2} +  \frac{\pi^2}{3} \right)  \, \delta\big(1 - \zeta^2\big) 
 + \frac{1}{\epsilon} \frac{1}{(1 - \zeta^2)_+}
\nonumber \\
- \frac{\log\big(1- (1 - \zeta^2) r^2 \big)}{1 - \zeta^2} \Bigg\}.
\label{eq:dLaurent}
\end{eqnarray}
In the last term inside the braces in Eq.~\eqref{eq:dLaurent}, the distribution $1/(1-\zeta^2)_+$ has been replaced by the function $1/(1-\zeta^2)$, because the logarithm vanishes when $\zeta^2=1$.

\subsection{Rapidity regularization}
\label{app:FragAmprapid}

The function  $d(\zeta)$ with rapidity regularization is defined by the integral over $q$ in Eq.~\eqref{eq:drapid}, with the integrand multiplied by appropriate regularization factors and with zero-bin subtractions. The rapidity regularization factor  is the product of two factors like that in Eq.~\eqref{eq:rapidreg-n} with $q$ replaced by $\tfrac12P+q$ and with $q$ replaced by  $\tfrac12P-q$. The zero-bin subtractions remove contributions from  the region where $\tfrac12P+q$ is soft and the region where $\tfrac12P-q$ is soft. It is convenient to divide $d(\zeta)$ by $1-\zeta^2$  in order to  facilitate the explicit extraction of the poles in the regularization parameter $\eta$.

In the collinear region, $P.n$ and $q.n$ are order $Q$, but $q^2$, $P.q$, and $P^2$ are order $M^2$. The integral over the entire collinear region before any zero-bin subtractions  is
\begin{eqnarray}
\left[ \frac{d(\zeta)}{1-\zeta^2} \right]_{\text{coll}} =
\frac{-i}{1-\zeta^2} \int_q
\frac{\delta(\zeta - 2q.n/P.n)}
{[(\tfrac12 P+q)^2 - m_t^2  + i \epsilon]\,  [(\tfrac12 P-q)^2-m_t^2  + i \epsilon] }
\nonumber\\
\times \bigg[\frac{\big|(\tfrac12 P+q).n \big|}{\nu} \bigg]^{-\eta}
\bigg[\frac{\big|(\tfrac12 P-q).n\big|}{\nu} \bigg]^{-\eta}.
\label{eq:d-coll1}
\end{eqnarray}
The measure for the integral over $q$ is given in Eq.~\eqref{eq:intq}. The integral  over $q$ in Eq.~\eqref{eq:d-coll1}  can be evaluated analytically:
\begin{eqnarray}
\left[ \frac{d(\zeta)}{1-\zeta^2} \right]_{\text{coll}} =
\frac{1}{32\pi^2\epsilon}
\left[\frac{\mu^2}{m_t^2}\right]^\epsilon \left[\frac{P.n}{\nu}\right]^{-2\eta}
\frac{1}{1-\zeta^2} 
\left(\frac{1-\zeta^2}{4}\right)^{-\eta}
\left[1-(1-\zeta^2)r^2-i\epsilon\right]^{-\epsilon}.~~~
\label{eq:d-coll2}
\end{eqnarray}
The infrared pole in $\eta$ can be made explicit by using a Laurent expansion like that in Eq.~\eqref{eq:expandplus}:
\begin{eqnarray}
\left[ \frac{d(\zeta)}{1-\zeta^2} \right]_{\text{coll}} =
\frac{1}{32\pi^2}
\left[\frac{\mu^2}{m_t^2}\right]^\epsilon \left[\frac{P.n}{\nu}\right]^{-2\eta}
\frac{1}{\epsilon}
\left\{-\frac{1}{\eta_\text{ir}}\delta(1-\zeta^2)+\frac{1}{(1-\zeta^2)_+}\right\}
\nonumber\\
\times
\left[1-(1-\zeta^2)r^2-i\epsilon\right]^{-\epsilon}.
\label{eq:d-coll3}
\end{eqnarray}

Two zero-bin subtractions are required to  remove the contributions from the soft regions. To calculate the zero-bin subtractions, it is convenient to pull the factor $1/(1-\zeta^2)$  inside the integral, expressing it as a function of $q.n$:
\begin{eqnarray}
&&\frac{-i}{1-\zeta^2} \int_q
\frac{\delta(\zeta - 2q.n/P.n)}
{[(\tfrac12 P+q)^2 - m_t^2  + i \epsilon]\,  [(\tfrac12 P-q)^2-m_t^2  + i \epsilon] }
\nonumber\\
&& = -i\int_q \frac{1}{1 - (2q.n/P.n)^2} \, 
\frac{\delta(\zeta - 2q.n/P.n)}
{[(\tfrac12 P+q)^2 - m_t^2  + i \epsilon]\,  [(\tfrac12 P-q)^2-m_t^2  + i \epsilon] }.
 \label{eq:ddefrap}
\end{eqnarray}
The zero-bin subtraction for the region where $\tfrac12 P+q$ is soft can be obtained from the right side of  Eq.~\eqref{eq:d-coll1} with the modification in Eq.~\eqref{eq:ddefrap} by first  making the substitution $q \to k - \tfrac12 P$ and then making soft approximations for $k$:
\begin{eqnarray}
\left[ \frac{d(\zeta)}{1-\zeta^2} \right]_{\text{zbs,+}} =
-\frac{i}{4} \, \delta(1+\zeta)
\int_k \frac{P.n/k.n}{ [k^2-m_t^2+i\epsilon]\, [-2k.P+i\epsilon]}
 \bigg[\frac{\big|k.n \big|}{\nu} \bigg]^{-\eta}
  \bigg[\frac{\big|(P-k).n\big|}{\nu} \bigg]^{-\eta}.~~~~~~~
\label{eq:d-subint+}
\end{eqnarray}
The delta function of $1+\zeta$ comes from making a soft approximation in the argument of the delta function in Eq.~\eqref{eq:ddefrap}. The factor of $P.n/k.n$ comes from the factor of $1+2q.n/P.n$ in the denominator in Eq.~\eqref{eq:ddefrap}. The integral can be evaluated analytically:
\begin{eqnarray}
\left[ \frac{d(\zeta)}{1-\zeta^2} \right]_{\text{zbs,+}} =
\frac{1}{64 \pi^2\epsilon}\delta(1+\zeta)
\left[\frac{\mu^2}{m_t^2}\right]^\epsilon \left[\frac{P.n}{\nu}\right]^{-2\eta}
\left\{ -\frac{1}{\eta_\text{ir}} \frac{(1)_{-\eta}(1)_{-\eta}}{(1)_{-2\eta}}
+\frac{1}{2\eta_\text{uv}} \frac{(1)_{-\eta}(1)_{2\eta}}{(1)_{\eta}}\right\}.~~~~~~~
\label{eq:d-unsubint}
\label{eq:d-sub+}
\end{eqnarray}
The zero-bin subtraction for the region where $\tfrac12 P-q$ is soft can be evaluated in a similar way, and the only difference is in the argument of  the delta function:
\begin{eqnarray}
\left[ \frac{d(\zeta)}{1-\zeta^2} \right]_{\text{zbs},-} =
\frac{1}{64\pi^2\epsilon}\delta(1-\zeta)
\left[\frac{\mu^2}{m_t^2}\right]^\epsilon \left[\frac{P.n}{\nu}\right]^{-2\eta}
\left\{ -\frac{1}{\eta_\text{ir}} \frac{(1)_{-\eta}(1)_{-\eta}}{(1)_{-2\eta}}
+\frac{1}{2\eta_\text{uv}} \frac{(1)_{-\eta}(1)_{2\eta}}{(1)_{\eta}}\right\}.~~~~~~~
\label{eq:d-sub-}
\end{eqnarray}

The complete expression for the function $d(\zeta)/(1-\zeta^2)$ with rapidity regularization is obtained by subtracting the zero-bin subtractions in Eqs.~\eqref{eq:d-sub+} and \eqref{eq:d-sub-} from the integral over the entire collinear region in Eq.~\eqref{eq:d-coll3}. The infrared poles in $\eta$ cancel, leaving only ultraviolet poles. The net result of the zero-bin subtractions is to replace the infrared pole $1/\eta_\text{ir}$ in Eq.~\eqref{eq:d-coll3} by the ultraviolet pole $1/(2\eta_\text{uv})$. Our final result for the regularized function is 
\begin{eqnarray}
 \frac{d(\zeta)}{1-\zeta^2}  =
\frac{1}{32 \pi^2}
\left[\frac{\mu^2}{m_t^2}\right]^\epsilon \left[\frac{P.n}{\nu}\right]^{-2\eta}
\frac{1}{\epsilon}
\left\{-\frac{1}{2\eta_\text{uv}}\delta(1-\zeta^2)+\frac{1}{(1-\zeta^2)_+}\right\}
\nonumber\\
\times
\left[1-(1-\zeta^2)r^2-i\epsilon\right]^{-\epsilon}.
\label{eq:d-analreg}
\end{eqnarray}


\providecommand{\href}[2]{#2}\begingroup\raggedright

\end{document}